\def\version{November 25, 2014}

\documentclass[12pt]{article}
\def\macrosPb{} % 12pt
\def\macrosHarxiv{} % arxiv hyperref
\usepackage[psamsfonts]{amsfonts}
\usepackage{amsmath,amssymb,amsthm}
\usepackage[dvips]{graphicx}
\usepackage{appendix}
\usepackage{bbm} % used in \newcommand{\1}{\mathbbm{1}}
\usepackage{amsbsy}
\usepackage{enumerate}
\usepackage{cite}

\def\macrosH{}

\InputIfFileExists{./macros_local.tex}{}{}

\ifdefined\macrosPa
  \usepackage[textwidth=465pt,textheight=650pt,centering]{geometry} % 11pt
\else\ifdefined\macrosPb
  \usepackage[textwidth=500pt,textheight=650pt,centering]{geometry} % 12pt
\fi\fi

\ifdefined\macrosS
  \makeatletter
  \def\boldsymbol{\pmb}
  \makeatother

  \usepackage{mathptmx}
  \DeclareMathAlphabet{\mathcal}{OMS}{cmsy}{m}{n}
\fi

\def\UseSection{%%
        \numberwithin{equation}{section}
	\theoremstyle{plain}% default theorem style 
        \newtheorem{theorem}    {Theorem}[section]
        \DefineTheorems % Use this to define other environments to be 
}

\def\DefineTheorems{%%
	
	\newtheorem{lemma}      [theorem] {Lemma}
	
	\newtheorem{prop}       [theorem] {Proposition}
	
	\newtheorem{cor}        [theorem] {Corollary}

	\theoremstyle{definition}% ``defn'' theorem style 
	\newtheorem{defn}       [theorem] {Definition}

	\newtheorem{rk} 	[theorem] {Remark}
	\theoremstyle{definition}% ``remark'' theorem style 

}

\newcommand{\bt}   {\begin{theorem}}
\newcommand{\et}   {\end  {theorem}}
\newcommand{\bl}   {\begin{lemma}}
\newcommand{\el}   {\end  {lemma}}
\newcommand{\bp}   {\begin{prop}}
\newcommand{\ep}   {\end  {prop}}
\newcommand{\bc}   {\begin{cor}}
\newcommand{\ec}   {\end  {cor}}
\newcommand{\bd}   {\begin{defn}}
\newcommand{\ed}   {\end  {defn}}

\newcommand{\ba}   {\begin{array}}
\newcommand{\ea}   {\end  {array}}
\newcommand{\be}   {\begin{enumerate}}
\newcommand{\ee}   {\end  {enumerate}}
\newcommand{\bi}   {\begin{itemize}}
\newcommand{\ei}   {\end  {itemize}}

\def\eq#1\en{\begin{equation}#1\end{equation}}  
\def\eqsplit#1\ensplit{
	\begin{equation}\begin{split}#1\end{split}\end{equation}
	}
\def\eqalign#1\enalign{
	\begin{align}#1\end{align}
	}
\def\eqmul#1\enmul{
	\begin{multline}#1\end{multline}
	}
\newcommand{\eqarrstar} {\begin{eqnarray*}} 
\newcommand{\enarrstar} {\end{eqnarray*}} 
\newcommand{\eqarray}   {\begin{eqnarray}} 
\newcommand{\enarray}   {\end{eqnarray}} 
\newcommand{\nnb}	{\nonumber \\} 

\newcommand{\lbeq}[1]  {\label{e:#1}}
\newcommand{\refeq}[1] {\eqref{e:#1}}    % AMS-LaTeX trick!
\makeatletter
\newcommand{\labelcounter}[2]{{%
	\stepcounter{#1}%	First, increase the ``countC'' by one.
	\protected@write\@auxout{}%
	{\string\newlabel{#2}{{\csname the#1\endcsname}{\thepage}}}%
	{\ref{#2}}%	Finally, make sure to refer to this label, 
	}}
\makeatother
\newcommand{\Ebold} {{\mathbb E}}

\newcommand{\Rbold} {{\mathbb R}}

\newcommand{\Zbold} {{\mathbb Z}}

\newcommand{\Bcal}   {\mathcal{B}} 
\newcommand{\Ccal}   {\mathcal{C}} 
\newcommand{\Dcal}   {\mathcal{D}} 
\newcommand{\Ecal}   {\mathcal{E}} 
 
\newcommand{\Gcal}   {\mathcal{G}} 
 
\newcommand{\Ical}   {\mathcal{I}}

\newcommand{\Lcal}   {\mathcal{L}} 
 
\newcommand{\Ncal}   {\mathcal{N}} 
 
\newcommand{\Pcal}   {\mathcal{P}}
\newcommand{\Qcal}   {\mathcal{Q}}

\newcommand{\Scal}   {\mathcal{S}} 
 
\newcommand{\Ucal}   {\mathcal{U}} 
\newcommand{\Vcal}   {\mathcal{V}}

\newcommand{\Ihat} {{\hat{I} }}

\newcommand{\Rd}    {{ {\Rbold}^d}}
\newcommand{\Zd}    {{ {\Zbold}^d }}

\newcommand{\spose}[1] {{\hbox to 0pt{#1\hss}} }
\newcommand{\ltapprox} {\mathrel{\spose{\lower 3pt\hbox{$\mathchar"218$}}
 \raise 2.0pt\hbox{$\mathchar"13C$}}}
\newcommand{\gtapprox} {\mathrel{\spose{\lower 3pt\hbox{$\mathchar"218$}}
 \raise 2.0pt\hbox{$\mathchar"13E$}}}

\UseSection   %Necessary to define Numbering scheme for Theorem, etc.
\usepackage[usenames]{color}

\definecolor{at}{rgb}{0.0, 0.5, 0.0} % green, darker than the standard green.

\newcommand{\LT}{{\rm Loc}  }

\newcommand{\DV}{\Dcal}

\renewcommand{\to} {\rightarrow}
\renewcommand{\qed}{\hfill\rule{2mm}{2mm}\bigskip}

\newcommand{\sumtwo}[2]{\sum_{ \mbox{ \scriptsize
    $\begin{array}{c}
                        {#1} \\ {#2}
                        \end{array} $ }
    }
}

\newcommand{\R}{\Rbold}
\newcommand{\Z}{\Zbold}

\newcommand{\C}{\mathbb{C}}

\newcommand{\Lambdabold}{\boldsymbol{\Lambda}}

\newcommand{\1}{\mathbbm{1}}

\newcommand{\psib}{\bar\psi}

\newcommand{\jm}{j_\Omega}

\newcommand{\Ex}{\mathbb{E}}

\newcommand{\Econst}{\alpha_{\Ebold}}

\newcommand{\Fconst}{f}

\newcommand{\chicCov}{{\chi}}
\newcommand{\ellconst}{\mathfrak{c}}

\newcommand{\Qcalnabla}{\Qcal}

\newcommand{\Gtilp}{\gamma}

\newcommand{\pair}[1]{\langle #1 \rangle}

\newcommand{\Phipoltil}{\widetilde{\Pi}}

\newcommand{\units}{\Ucal}

\newcommand{\cgam}{\gamma}

\newcommand{\Itilde}{\tilde{I}}

\newcommand{\Ttimes}{T}

\newcommand{\hldg}{h_{\rm lead}}
\newcommand{\cldg}{c_{\rm lead}}

\newcommand{\pt}{{\rm pt}}
\newcommand{\Ipt}{I_{\rm pt}}
\newcommand{\Ipttil}{\tilde{I}_{\rm pt}}

\newcommand{\Vpt}{V_{\rm pt}}
\newcommand{\gpt}{g_{\mathrm{pt}}}

\newcommand{\lambdaa}{\lambda^{\pp}}
\newcommand{\lambdab}{\lambda^{\qq}}

\newcommand{\qa}{q^{\pp}}
\newcommand{\qb}{q^{\qq}}

\newcommand{\h}{\mathfrak{h}}

\newcommand{\q}{q}

\newcommand{\Lcallr}{\stackrel{\leftrightarrow}{\Lcal}}

\newcommand{\gbar}{\bar{g}}

\newcommand{\ggen}{\tilde{g}}

\newcommand{\pp}{a}
\newcommand{\qq}{b}

\newcommand{\epW}{\epsilon_{W}}

\newcommand{\epV}{\epsilon_{V}}

\newcommand{\epVbar}{\epsilon_{g\tau^{2}}}

\newcommand{\epdV}{\bar{\epsilon}}

\newcommand{\gh}{\epV}

\newcommand{\phib}{\bar\phi}

\newcommand{\hred}{h_{\rm red}}

\ifdefined\macrosH
  \usepackage{xr-hyper}
  \usepackage{hyperref}
  \hypersetup{hypertexnames=false}
  \hypersetup{colorlinks,citecolor=blue,linkcolor=blue}  % Comment out for printing!

\else\ifdefined\macrosHarxiv
  \usepackage{xr-hyper}
  \usepackage{hyperref}
  \hypersetup{hypertexnames=false}

\else
  \newcommand{\texorpdfstring}[2]{#1}
  \usepackage{xr}
\fi\fi

\title  {
       A renormalisation group method.
       IV. Stability analysis
        }

\author{
David C. Brydges\thanks{Department of Mathematics,
University of British Columbia,
Vancouver, BC, Canada V6T 1Z2.
E-mail: {\tt db5d@math.ubc.ca}, {\tt slade@math.ubc.ca}.}\;
 and Gordon Slade$^*$}

\date\version

\begin{document}

\maketitle

\begin{abstract}
This paper is the fourth in a series devoted to the development of a
rigorous renormalisation group method for lattice field theories
involving boson fields, fermion fields, or both.
The third paper in the series presents a perturbative analysis
of a supersymmetric field theory which represents the continuous-time
weakly self-avoiding walk on $\Zd$.  We now present an analysis of the
relevant interaction functional of the supersymmetric
field theory, which permits a nonperturbative
analysis to be carried out
in the critical dimension $d = 4$.  The results in this paper
include: proof of stability
of the interaction, estimates which enable control of Gaussian
expectations involving both boson and fermion fields,
estimates which bound the errors in the perturbative
analysis, and a crucial contraction estimate
to handle irrelevant directions in the flow of the renormalisation group.
These results
are essential for the analysis of the general renormalisation group
step in the fifth paper in the series.
\end{abstract}

%%%%%%%%%%%%%%%%%%%%%%%%%%%%%%%%%%%%%%%%%%%%%%%%%%%%%%%%%%%%%%%%%%%
%%%%%%%%%%%%%%%%%%%%%%%%%%%%%%%%%%%%%%%%%%%%%%%%%%%%%%%%%%%%%%%%%%%
\section{Introduction}
\label{sec:ie}

This paper is the fourth in a series devoted to the development of a
rigorous renormalisation group method.  The method has been applied
to analyse the critical behaviour of the continuous-time
weakly self-avoiding walk  \cite{BBS-saw4-log,BBS-saw4},
and the $n$-component $|\varphi|^4$ spin model
\cite{BBS-phi4-log}, in the critical dimension $d = 4$.
In both cases, logarithmic corrections to mean-field scaling are established
using our method.

In part~I  \cite{BS-rg-norm}
of the series, we
presented elements of the theory of Gaussian integration and developed
norms and norm estimates for performing analysis with Gaussian
integrals involving both boson and fermion fields.
In part~II \cite{BS-rg-loc}, we defined and analysed a
localisation operator whose purpose is to extract relevant and
marginal directions in the dynamical system defined by the
renormalisation group.
In part~III \cite{BBS-rg-pt}, we began to
apply the formalism
of parts~I and II to a specific supersymmetric field theory that
arises as a representation of the continuous-time weakly self-avoiding
walk \cite{BIS09,BBS-saw4-log}, by studying the flow of coupling constants
in a perturbative analysis.
We now prove several nonperturbative estimates for the supersymmetric
field theory studied in part~III.
These estimates are essential
inputs for our analysis of a general renormalisation group step
in part~V \cite{BS-rg-step}, and therefore
for the analysis of the critical
behaviour of the continuous-time weakly self-avoiding walk
in dimension $d = 4$ in \cite{BBS-saw4-log,BBS-saw4}.

The results in this paper include:  proof of stability
of the interaction, estimates which enable control of Gaussian
expectations involving both boson and fermion fields,
estimates which bound the errors in the perturbative
analysis of part~III and thereby confirm that the perturbative
analysis does indeed isolate leading contributions,
and a crucial contraction estimate
to handle irrelevant directions in the flow of the renormalisation group.
All these results are needed in our analysis of a general renormalisation
group step in part~V.
The methods and results developed in this paper are
of wider applicability, but for the sake of concreteness, and for
the purposes of our specific application in \cite{BBS-saw4-log,BBS-saw4}, we
formulate the discussion in the context of the supersymmetric
field theory studied in part~III.
Supersymmetry is helpful: it ensures that the partition function is equal
to $1$, so it need never be estimated.

Several mathematically rigorous approaches to renormalisation in statistical
mechanics and quantum field theory have
been proposed in recent decades, e.g., the books \cite{BG95,Riva91,Salm99}.
Characteristic features of the approach we develop are:
(i) there is no
partition of unity in field space to separate large and small fields, and (ii)
fluctuation fields have finite range of dependence.
The avoidance of partitions of unity
is important for us because it is easier to maintain supersymmetry
without them.  The use of finite-range fluctuation fields bears some similarity
to the wavelet program
reviewed in
\cite{Fede87},
but has better translation invariance properties.
An attractive feature of (ii) is
that independence of Gaussian fields replaces cluster expansions.
 The
price to be paid for avoiding partitions of unity
is that norms must control the size of the basic objects
in all of field space, including large fields.
The goal of the present paper is to acquire this control.

Our analysis has antecedents in \cite{Abde07,BDH95,MS08}, though our
setting includes fermions as well as bosons.  A systematic development
of appropriate norms is given in \cite{BS-rg-norm}.  Part of the need
for these norms is to define complete spaces in order to apply the
dynamical system analysis of \cite{BBS-rg-flow} (discussion of past
errors related to completeness can be found in \cite{Abde07}).  The
norms include a notion that we call ``regulators'' because they
control (regulate) growth when fields are large.  These are always a
delicate part in this approach and important ideas that guide their
choice appear in \cite{DH00,Falc12}.  For our field theory, the choice
of regulators is less delicate because the $\phi^4$ term suppresses
large fields.  Another feature of our analysis is the inclusion of
observable fields to permit control of correlation functions; somewhat
related ideas were introduced in \cite{DH92}.  Different methods
to construct the correlations in the infinite volume have been
developed in \cite{Falc13}.

The renormalisation group can be defined directly in infinite
volume, but until \cite{Dimo09} and \cite{Falc13} it was not
demonstrated that the infinite volume theory defined in this way
coincides with the infinite volume defined by taking limits of
correlation functions and pressures defined in finite volume.  Our
analysis also prepares the way for results about this question for
the weakly self-avoiding walk.

In the remainder of Section~\ref{sec:ie}, we give the fundamental definitions
and provide an informal overview of the results of this paper.
The main results are then stated precisely in Section~\ref{sec:IE}.
Proofs are given in Sections~\ref{app:sp}--\ref{sec:ipcl}.
In Appendix~\ref{sec:Lp}, we prove a lattice Sobolev inequality
that lies at the heart of our stability estimates.
Finally, Appendix~\ref{sec:further-ie} concerns estimates of a more
specialised nature that are required for the analysis of a single renormalisation
group step in part~V \cite{BS-rg-step}.
Our focus throughout the paper is on the case $d=4$.

\subsection{Object of study}
\label{sec:study}

We begin with several definitions needed to formulate our results.
Many of these definitions are recalled from parts~III and I.
We begin by introducing the covariance decomposition which provides the
basis for a multi-scale analysis.
We then introduce the space of boson and fermion fields, and define the interaction
functional $I$.  We also recall the definition of the renormalised polynomial $\Vpt$
from part~III, and the definitions of the norms and regulators from part~I.

\subsubsection{Covariance decomposition}
\label{sec:cd}

Let $d \ge 4$ and let $\Lambda = \Zd/L^N\Z$ denote the discrete $d$-dimensional
torus of side $L^N$, with $L>1$ fixed (large).
We are interested in results which remain useful in the
infinite volume limit $N \to \infty$.
There are several places in this paper where $L$ must be taken
large, depending on unimportant parameters such as the
dimension $d$, or combinatorial constants.  We do not comment
explicitly on each occasion where $L$ must be taken to be large,
but instead \emph{we assume throughout the
paper that $L$ is large enough to satisfy each such requirement
that is encountered}.

For $e$ in the set $\units$ of $2d$ nearest neighbours of the origin
in $\Zd$, we define the finite difference operator $\nabla^e \phi_x =
\phi_{x+e}-\phi_x$, and the Laplacian $\Delta_\Zd = -\frac{1}{2}\sum_{e \in
\units}\nabla^{-e} \nabla^{e}$.
Let $C= (-\Delta_\Lambda + m^2)^{-1}$, where
$m^2>0$ is a positive parameter and $\Delta_\Lambda$ denotes
the discrete Laplacian on $\Lambda$.
We fix $N$ large and $m^2$ small, and wish to perform an analysis
which applies uniformly in $N,m^2$.

We require decompositions of
the covariances $(\Delta_\Zd + m^2)^{-1}$ and $C=(-\Delta_\Lambda +m^2)^{-1}$.
For the former, the massless Green function is well-defined for $d>2$ and we may
consider $m^2 \ge 0$, but for the latter we must take $m^2>0$.
In \cite[Section~\ref{pt-sec:Cdecomp}]{BBS-rg-pt},
there is a detailed discussion of decompositions
we use for each of these covariances, based on \cite{Baue13a}
(see also Section~\ref{sec:frp} below).
In particular, in \cite[Section~\ref{pt-sec:Cdecomp}]{BBS-rg-pt}
a sequence
$(C_j)_{1 \le j < \infty}$ (depending on $m^2 \ge 0$)
of positive definite covariances on $\Zd$ is defined,
such that
\begin{equation}
\lbeq{ZdCj}
    (\Delta_\Zd + m^2)^{-1} = \sum_{j=1}^\infty C_j
    \quad
    \quad
    (m^2 \ge 0).
\end{equation}
The covariances $C_j$ are translation invariant
and have the \emph{finite-range} property
\begin{equation}
\label{e:fin-range}
      C_{j;x,y} = 0 \quad \text{if \; $|x-y| \geq \frac{1}{2} L^j$}.
\end{equation}
For $j<N$, the covariances $C_j$ can therefore be identified with
covariances on $\Lambda$, and we use both interpretations.
We define
\begin{align}
\lbeq{wjdef}
    w_j &= \sum_{i=1}^j C_i \quad\quad (1 \le j < \infty),
\end{align}
and note that $w_j$ also obeys \refeq{fin-range}.

There is also a covariance $C_{N,N}$ on $\Lambda$ such that
\begin{equation}
\lbeq{NCj}
    C=(-\Delta_\Lambda + m^2)^{-1} = \sum_{j=1}^{N-1} C_j + C_{N,N}
    \quad
    \quad
    (m^2 > 0).
\end{equation}
Thus the finite volume decomposition agrees with the infinite volume
decomposition except for the last term in the finite volume decomposition,
which is the single term that accounts for the torus.

The expectation $\Ex_C$ denotes the combined bosonic-fermionic Gaussian
integration on $\Ncal$, with covariance $C$, defined in
\cite[Section~\ref{norm-sec:Grass}]{BS-rg-norm}.
The integral is performed successively, using
\begin{equation}
\lbeq{progexp}
    \Ex_C = \Ex_{C_N} \circ \Ex_{C_{N-1}} \theta \circ
    \cdots \circ \Ex_{C_1}\theta,
\end{equation}
where
$\theta$ defines a type of convolution and is discussed further below.

\subsubsection{Fields and field polynomials}

We study a field theory which consists of a complex boson field
$\phi : \Lambda \to \C$ with its complex conjugate $\bar\phi$,
a pair of conjugate fermion fields
$\psi,\bar\psi$, and a \emph{constant} complex
observable boson field $\sigma \in \C$ with its complex conjugate
$\bar\sigma$.
The fermion field is given in terms of the 1-forms $d\phi_x$ by
$\psi_x = \frac{1}{\sqrt{2\pi i}} d\phi_x$
and $\bar\psi_x = \frac{1}{\sqrt{2\pi i}}
d\bar\phi_x$, where we fix some square root of $2\pi i$.
This is the supersymmetric choice discussed in more detail in
\cite[Sections~\ref{norm-sec:df}--\ref{norm-sec:supersymmetry}]{BS-rg-norm}
and used in \cite{BBS-rg-pt}.

Let two points $\pp,\qq \in \Lambda$ be fixed.
We work with an
algebra $\Ncal$ which is defined in terms of a direct sum decomposition
\begin{equation}
\label{e:Ncaldecomp}
    \Ncal = \Ncal^\varnothing \oplus \Ncal^a \oplus \Ncal^b \oplus \Ncal^{ab}.
\end{equation}
Elements of $\Ncal^\varnothing$ are given by finite linear combinations
of products of an even number of fermion fields with coefficients that are functions
of the boson fields.  This restriction to forms of even degree results
in a commutative algebra.
Elements of $\Ncal^a, \Ncal^b , \Ncal^{ab}$
are respectively given by elements of $\Ncal^\varnothing$ multiplied
by $\sigma$, by $\bar\sigma$, and by $\sigma\bar\sigma$.
For example, $\phi_x \bar\phi_y \psi_x \bar\psi_x
\in \Ncal^\varnothing$, and $\sigma \bar\phi_x \in \Ncal^a$.
There are canonical projections $\pi_\alpha: \Ncal \to \Ncal^\alpha$ for
$\alpha \in \{\varnothing, a, b, ab\}$.  We use the abbreviation
$\pi_*=1-\pi_\varnothing = \pi_a+\pi_b+\pi_{ab}$.
The algebra $\Ncal$ is discussed further around
\cite[\eqref{loc-e:1Ncaldecomp}]{BS-rg-loc}
(there $\Ncal$ is written $\Ncal/\Ical$ but to
simplify the notation we write $\Ncal$ here instead).
The parameter $p_\Ncal$ which appears in its definition
is a measure of the smoothness of elements of $\Ncal$ (see
\cite[Section~\ref{norm-sec:Ncal}]{BS-rg-norm}); its precise value
is unimportant and can be fixed to be larger than the degree
of polynomials encountered in practice in the application of
the stability bounds.
Constants in estimates may depend on its
value, in an unimportant way.

We define the forms
\begin{equation}
    \label{e:addDelta}
    \tau_x = \phi_x \bar\phi_x + \psi_x \bar\psi_x,
    \quad\quad
   \tau_{\nabla \nabla,x}  =
   \frac 12
   \sum_{e \in \units}
   \left(
   (\nabla^e \phi)_x (\nabla^e \bar\phi)_x +
   (\nabla^e \psi)_x (\nabla^e \bar\psi)_x
   \right)
,
\end{equation}
\begin{align}
       \tau_{\Delta,x} &= \frac 12 \left(
    (-\Delta \phi)_{x} \bar{\phi}_{y} +
    \phi_{x} (-\Delta \bar{\phi})_{y} +
    (-\Delta \psi)_{x} \bar{\psi}_{y} +
    \psi_{x} (-\Delta \bar{\psi})_{y}
    \right)
    .
\end{align}
Let $\Qcalnabla$ denote the vector space of polynomials of the form
\begin{gather}
    V
=
    V_{\varnothing} + V_{\pp} + V_{\qq} + V_{\pp\qq},
\end{gather}
where
\begin{gather}
    V_{\varnothing}
    =
    g \tau^{2} + \nu \tau +
    z \tau_{\Delta} +
    y \tau_{\nabla \nabla},
\quad
    V_{\pp}
=
    \lambda_{\pp} \sigma \bar{\phi},
\quad
    V_{\qq}
=
    \lambda_{\qq}\bar{\sigma} \phi,
\quad
    V_{\pp \qq}
=
    q_{\pp\qq} \bar{\sigma}\sigma
    \label{e:Vx}
,
\end{gather}
\begin{align}
\label{e:lambda-defs}
&
    \lambda_{\pp}
    =
    -\lambdaa \,\1_{\pp},
\quad\quad
    \lambda_{\qq}
    =
    -\lambdab \,\1_{\qq},
\quad\quad
    q_{\pp\qq}
    =
    -\frac{1}{2} (\qa\1_{\pp} + \qb\1_{\qq})
    ,
\end{align}
$g,\nu,y,z,\lambdaa,\lambdab,\qa,\qb \in \C$, and the indicator
functions are defined by the Kronecker delta $\1_{a,x}=\delta_{a,x}$.
For $X \subset \Lambda$, we write
\begin{equation}
\label{e:VXdef}
    V(X)=\sum_{x\in X}V_x.
\end{equation}
There is an important scale, called the \emph{coalescence scale},
defined by
\begin{equation}
   \label{e:Phi-def-jc}
    j_{\pp \qq}
    =
    \big\lfloor
   \log_{L} (2 |\pp - \qq|)
   \big\rfloor
   .
\end{equation}
We assume that $\pi_{ab}V=0$ for $j<j_{ab}$; note that if the
coefficient $q$ is initially equal to zero, then under the flow
\cite[\eqref{pt-e:qpt2}]{BBS-rg-pt} it remains zero below the coalescence scale due to the
assumption \eqref{e:fin-range}.

The goal of our analysis is to understand
the Gaussian integral $\Ex_C e^{-V(\Lambda)}$.
Given a positive-definite
matrix $C$ whose rows and columns are indexed by $\Lambda$,
we define the \emph{Laplacian}
\begin{equation}
\label{e:LapC}
    \Lcal_C = \frac 12 \Delta_{C}
=
    \sum_{u,v \in \Lambda}
    C_{u,v}
    \left(
    \frac{\partial}{\partial \phi_{u}}
    \frac{\partial}{\partial \bar\phi_{v}}
    +
    \frac{\partial}{\partial \psi_{u}}
    \frac{\partial}{\partial \bar\psi_{v}}
    \right)
\end{equation}
(see \cite[\eqref{norm-e:Lapss}]{BS-rg-norm}).
The Laplacian is intimately related to Gaussian integration.
To explain this, suppose we are
given an additional boson field $\xi,\bar\xi$ and
an additional fermion field $\eta, \bar\eta$, with
$\eta = \frac{1}{\sqrt{2\pi i}}d\xi$,
$\bar\eta = \frac{1}{\sqrt{2\pi i}}d\bar\xi$,
and consider the ``doubled'' algebra $\Ncal(\Lambdabold\sqcup
\Lambdabold')$ containing the original
fields and also these
additional fields.
We define a
map $\theta : \Ncal(\Lambdabold) \to \Ncal(\Lambdabold\sqcup
\Lambdabold')$ by making the replacement
in an element of $\Ncal$ of $\phi$ by $\phi+\xi$,
$\bar\phi$ by $\bar\phi+\bar\xi$, $\psi$ by $\psi+\eta$, and
$\bar\psi$ by $\bar\psi+\bar\xi$.
According to
\cite[Proposition~\ref{norm-prop:conv}]{BS-rg-norm}, for a
\emph{polynomial} $A$
in the fields, the Gaussian expectation with covariance $C$ can
be evaluated using the Laplacian operator
via
\begin{equation}
\label{e:EWick}
    \Ex_C \theta A
     = e^{\Lcal_C} A,
\end{equation}
where the fields $\xi,\bar\xi,\eta,\bar\eta$ are integrated out by
$\Ex_C$, with $\phi, \bar\phi, \psi, \bar\psi$ kept fixed, and where
$e^{\Lcal_C}$ is defined by its power series.

\subsubsection{Form of interaction}
\label{sec:formint}

In \cite[Section~\ref{pt-sec:WPjobs}]{BBS-rg-pt},
we discussed reasons to define an interaction
\begin{equation}
    I_{j}(V,\Lambda)= e^{-V(\Lambda)} (1+ W_{j}(V,\Lambda)),
\end{equation}
where $W_{j}$ is a certain non-local polynomial in the fields whose
definition is recalled below.
Our main object of study in this paper is a modified version of
$I_{j}$ which is defined on subsets of $\Lambda$.

We recall the relevant definitions from \cite{BBS-rg-pt}.  For
polynomials $V',V''$ in the fields, we define bilinear functions
of $V'$ and $V''$ by
\begin{align}
\label{e:FCAB}
    F_{C}(V',V'') &
    = e^{\Lcal_C}
    \big(e^{-\Lcal_C}V'\big)
    \big(e^{-\Lcal_C}V'' \big) - V'V'',
    \\
\label{e:Fpi}
    F_{\pi ,C}(V',V'')
    &=
    F_{C}(V',\pi_\varnothing V'')
    + F_{C}(\pi_* V',V'').
\end{align}
By definition, when $V'$ is expanded in $F_{C} (V',V'')$ as
$V'=\pi_{\varnothing} V' + \pi_{*} V'$, there are
cross-terms
$F_C(\pi_\varnothing V', \pi_* V'') + F(\pi_* V',\pi_\varnothing V'')$,
and \eqref{e:Fpi} is obtained
from \eqref{e:FCAB} by  replacing these cross-terms by
$2 F_{C}(\pi_{*}V' , \pi_{\varnothing}V'')$.
This unusual bookkeeping is appropriate (indeed necessary) in the proof
of Proposition~\ref{prop:Wbounds}.

For nonempty $X \subset \Lambda$, the space $\Ncal (X)$ is defined in
\cite[\eqref{norm-e:NXdef}]{BS-rg-norm} as consisting of elements of
$\Ncal$ which depend on $\phi_x,\bar\phi_x,\psi_x,\bar\psi_x$ only
with $x \in X$.  Recall from \cite{BS-rg-loc} that we defined $F \in
\Ncal_{X}$ to mean that there exists a coordinate patch $\Lambda '$
such that $F \in \Ncal (\Lambda')$ and $X \subset \Lambda'$, and we
defined the condition $F \in \Ncal_X$ to guarantee that neither $X$
nor $F$ ``wrap around" the torus.  The operator $\LT_X : \Ncal_X \to
\Vcal(X)$ is defined in
\cite[Definition~\ref{loc-def:LTsym}]{BS-rg-loc}, and the particular
specification we use is that described in
\cite[Section~\ref{pt-sec:loc-specs}]{BBS-rg-pt}.  In particular, the
\emph{field dimensions} are $[\phi]=[\bar\phi]=[\psi]=[\bar\psi]
=\frac{d-2}{2}$, and we set $d_+ = d$ on $\Ncal^\varnothing$.  On
$\Ncal^{ab}$, we take $d_+=0$.  When $\LT$ acts at scale $k$ (in the
sense discussed in \cite[Section~\ref{pt-sec:loc-specs}]{BBS-rg-pt}),
on $\Ncal^a$ and $\Ncal^b$ we take $d_+=[\phi]=\frac{d-2}{2}=1$ if
$k<j_{\pp\qq}$, and $d_+=0$ for $k \ge j_{\pp\qq}$.

For $x\in \Lambda$, with $w_j$ given by \refeq{wjdef} we define
\begin{align}
\label{e:Wwdef}
    W_j(V,x) &=
    \frac 12 (1-\LT_{x}) F_{\pi,w_j}(V_x,V(\Lambda)) \quad\quad (j<N).
\end{align}
For $j<N$, the above application of $\LT_x$ is well-defined since
$F_{\pi,w_j}(V_x,V(\Lambda)) \in \Ncal_x$ due to the finite-range property
of $w_j$.
For $X \subset \Lambda$, we then define
\begin{equation}
\label{e:WLTF}
    W_j(V,X)= \sum_{x \in X} W_j(V,x).
\end{equation}
By definition, $w_0=0$ and $W_0=0$.

We consider
the natural paving of $\Lambda$ by disjoint blocks of side length $L^j$,
for $j=0,\ldots, N$.
The set of all scale-$j$ blocks is denoted $\Bcal_j$, and $\Pcal_j$ denotes
the set whose elements are finite unions of blocks in $\Bcal_j$.
We refer to elements of $\Pcal_j$ as scale-$j$ \emph{polymers}.
Given a polynomial $V\in \Vcal$, and $X \subset \Lambda$, let
\begin{equation}
\label{e:Icaldef}
    \Ical(V,X) = e^{-V(X)}.
\end{equation}
The interaction is defined, for $B \in \Bcal_j$ and $X \in \Pcal_j$,
by
\begin{equation}
\label{e:Fsoptb}
    I_j(V,B)
    =
    \Ical(V,B)
    \left( 1+W_j(V,B) \right) ,
    \quad
    \quad
    I_j(V,X)
    =
    \prod_{B \in \Bcal_{j}(X)}
    I_j(V,B).
\end{equation}
Due to the finite-range property \refeq{fin-range}, $I_j(V,B)\in \Ncal(B^+)$,
where $B^+$ denotes the union of $B$ with every block $B'$ such that $B \cup B'$
is connected.
We often write $I_j(V,X) = I_j^X(V)$.
We also consider the interaction defined,
for $b \in \Bcal_{j-1}$ and $X \in \Pcal_{j-1}$,
by
\begin{equation}
\label{e:Itildef}
    \tilde I_{j}(V,b) =
    \Ical(V,b)(1+W_j(V,b)),
    \quad \quad
    \Itilde_j(V,X)
    =
    \prod_{b \in \Bcal_{j-1} (X)}
    \Itilde_j(V,b).
\end{equation}
Thus $\tilde I_j$ is defined on blocks and polymers of scale $j-1$,
whereas $I_j$ is defined on blocks and polymers of scale $j$.

An element $F \in \Ncal$ is said to be
\emph{gauge invariant} if it is invariant under
the gauge flow
$q \mapsto e^{-2\pi i t}q$,
$\bar q \mapsto e^{+2\pi i t}\bar q$;
for all $q = \phi_{x} ,\psi_{x}, \sigma$;
$\bar q = \bar\phi_{x} ,
\psib_{x}, \bar\sigma$; and
$x \in \Lambda$.
The basic objects we study, including $V,F,W,I,\tilde{I}$,
are all gauge invariant.  Also, since we assume $V_{ab}=0$ for $j<j_{ab}$,
it follows that none of these basic objects has a nonzero component
in $\Ncal_{ab}$ unless $j \ge j_{ab}$.

\subsubsection{The renormalised field polynomial}
\label{sec:Vpt}

To simplify the notation, we write $\Lcal_{j} = \Lcal_{C_{j}}$.
Given $V\in \Qcalnabla$, as in \cite[\eqref{pt-e:PdefF}]{BBS-rg-pt} we define
\begin{align}
\label{e:PWdef}
    P_{j}(V,x)
    &=
    \LT_{x}\left(
    e^{\Lcal_{j+1}} W_{j}(V,x)
    + \frac{1}{2}
    F_{\pi,C_{j+1}}(e^{\Lcal_{j+1}}V_x,e^{\Lcal_{j+1}} V(\Lambda))
    \right)
    \quad (j+1<N),
\end{align}
and we write $P_j(V,X)=\sum_{x\in X}P_j(V,x)$ for $X \subset \Lambda$.
The local polynomial $\Vpt$ is defined, as in
\cite[\eqref{pt-e:Vptdef}]{BBS-rg-pt}, by
\begin{equation}
\label{e:Vptdef}
    V_{\pt,j+1}(V,x) = e^{\Lcal_{j+1}}  V_x - P_j(V,x)
    \quad (j+1<N)
    .
\end{equation}
By definition, $V_{\pt,j+1}(B)$ depends on fields and their derivatives at
sites in $B$, in contrast to $I_j(V,B)$ which depends on fields in the
larger region $B^{+}$ because of $W_j(V,B)$.
By \cite[Lemma~\ref{pt-lem:EV}]{BBS-rg-pt} we have $e^{\Lcal_{j+1}}V
=V+ 2gC_{j+1;0,0}\tau$, so
\begin{equation}
\lbeq{VptE}
    V_{\pt,j+1} = V+ 2gC_{j+1;0,0}\tau -P_j(V)
    \quad (j+1<N)
    .
\end{equation}

For $j<N$, an explicit formula $V_{\pt,j} = \varphi_{\pt,j-1}$ is given in
\cite[Proposition~\ref{pt-prop:Vptg}]{BBS-rg-pt}.  In particular, $P \in \Qcal$.
The definition
of $\Vpt$ is motivated by the fact
(shown in \cite[Section~\ref{pt-sec:WPjobs}]{BBS-rg-pt}) that
the definitions of $W$ and $\Vpt$ cooperate to arrange that,
as formal power series,
\begin{equation}
\lbeq{fps}
    \Ex \theta I_j(V,\Lambda) \approx I_{j+1}(\Vpt,\Lambda) +O(V^3).
\end{equation}
For $B \in\Bcal_j$, we make the abbreviation
\begin{equation}
\label{e:Ipttildef}
    \Ipttil(B) = \tilde{I}_{j+1}(\Vpt,B ),
\end{equation}

\subsubsection{The final scale}
\label{sec:finalscale}

The above definitions have been given for scales below but not
including the final scale $N$.  At scale $N$, the torus consists of a single
block $\Lambda \in \Bcal_N$, the periodicity of the
torus becomes preponderant, the definition of
$\LT_x F_{\pi,w_{N,N}}(V_x,V(\Lambda))$ breaks down due to lack
of a coordinate patch, and the definitions of $W$ and $P$ in
\refeq{Wwdef} and \refeq{PWdef} can no longer be used.
Initially this may appear problematic, since we
are ultimately interested in performing the last expectation
and computing $I_N$.  However, any apparent difficulty is only
superficial.  There is only one problematic scale out of an
unbounded number of scales.  Moveover, the covariance $C_{N,N}$ is extremely
small for large $m^2L^{2N}$
(see \refeq{scaling-estimate-Omega} below),
and we do take the limit $N \to \infty$
before $m^2 \downarrow 0$, so the last expectation is insignificant.
Nevertheless it is necessary to make appropriate definitions of
$\Vpt$ and $W$ at scale $N$.  We do this in such a way that the
analysis at scale $N$ differs minimally from that at previous scales.

For $\Vpt$, the natural choice $V_{\pt,N}=\varphi_{\pt,N-1}$ is made in
\cite[Definition~\ref{pt-def:VptZd}]{BBS-rg-pt};
this choice defines $V_{\pt,N}$ to
be equal to what it would be if the torus side length were at a higher
scale than scale $N$.
In terms of this choice, we define $P_{N-1}$ so that \refeq{Vptdef}
remains valid for scale $N$, namely
\begin{align}
\label{e:PNdef}
    P_{N-1}(V) & =
    -V_{\pt,N}(V) + \Ex_{C_{N}}\theta V
   .
\end{align}
There is no $P_N$, the last $P_j$ is $P_{N-1}$ since the last $\Vpt$ is
$V_{\pt,N}$.  Thus we have arranged the definitions at the last scale in
such a way that $V_{\pt,N}$ agrees with what it would be on a torus of scale
greater than $N$ (the use of $\Ex_{C_N}$ rather than $\Ex_{C_{N,N}}$
is intentional and for this reason).

For $W_N$, our choice is inspired by a key identity obeyed by
$W_j$ for $j<N$, proved in Lemma~\ref{lem:EW}.
The identity implies in particular that
\begin{align}
    W_{j} (V,x)
    &=
   e^{\Lcal_j} W_{j-1} (e^{-\Lcal_j}V,x)
   - P_{j-1}(e^{-\Lcal_j}V,x)
   +
   \frac{1}{2} F_{\pi,C_j }(  V_x,V(\Lambda))
   \quad (j<N)
   .
\end{align}
The above identity is instrumental in the proof that the perturbative
analysis of \cite{BBS-rg-pt} is accurate beyond formal power series,
and thus plays a fundamental role.
We define $W_N$ to maintain this identity.  Thus, with $P_{N-1}$
given by \refeq{PNdef}, we define
\begin{align}
\label{e:WNdef}
    W_{N}(V,x) & =
    e^{\Lcal_{N,N}} W_{N-1}(e^{-\Lcal_{N,N}}V,x)
    -P_{N-1}(e^{-\Lcal_{N,N}}V,x)
    + \frac 12  F_{\pi,C_{N,N}}(V_x,V(\Lambda))
     .
\end{align}

\subsubsection{Norms and field regulators}
\label{sec:reg}

Our estimates are typically
expressed in terms of the $T_\phi$ semi-norm and
two important functions of $\phi$ that we
refer to as \emph{field regulators}.
We now recall the relevant definitions.

\subsubsection*{The $T_\phi$ semi-norm}

We make heavy use of the $\Phi_j(\h_j)$ norm on test functions
and the $T_{\phi,j}(\h_j)$ semi-norm on $\Ncal$.
The definition of the $\Phi_j(\h_j)$ norm on test functions
is given in \cite[Example~\ref{norm-ex:h}]{BS-rg-norm}
in terms of a parameter
$p_\Phi \ge
d+1-\frac{d-2}{2}=\frac{d+4}{2}$ (consistent with the requirement
above the statement of
\cite[Proposition~\ref{loc-prop:LTKbound}]{BS-rg-loc}),
and here we take $R=L^j$ in \cite[Example~\ref{norm-ex:h}]{BS-rg-norm} where $j$ is
the scale.  The value of $p_\Phi$ is fixed but unimportant,
and constants in estimates may depend on it.
The space $\Phi(\h)$ consists of test functions
$g : \vec\Lambdabold^* \to \C$.
The definition of the norm requires the specification of its
``sheets'' and the values of the components of $\h_j$ for
each sheet (particular choices are made in Section~\ref{sec:hex} below).
We assume that in the definition of the norm there are
sheets for each of the fields
$\phi,\bar\phi,\psi,\bar\psi,\sigma,\bar\sigma$.
The boson and fermion fields have a common component of $\h_j$,
and we sometimes abuse notation by writing $\h_j$ for this
particular component value.
Also, the fields $\sigma,\bar\sigma$ have a common value
$\h_{\sigma,j}$.

The $T_\phi(\h)$ semi-norm is defined in
\cite[Definition~\ref{norm-def:Tphi-norm}]{BS-rg-norm}, and provides
a family of semi-norms indexed by
the vector $\h$.  We often keep $\h$ as a parameter
in our results, as our applications ultimately use more than one
choice.  Properties of the $T_\phi$ semi-norm are derived in
\cite{BS-rg-norm}; prominent among them is the product property of
\cite[Proposition~\ref{norm-prop:prod}]{BS-rg-norm} which asserts that
$\|FG\|_{T_\phi} \le \|F\|_{T_\phi}\|G\|_{T_\phi}$ for all $F,G \in
\Ncal$.

\subsubsection*{Fluctuation-field regulator}
\label{sec:ffreg}

A special case of the $\Phi(\h)$ norm
is obtained by regarding the boson field as a test function:
given $\h_j>0$ its $\Phi_j=\Phi_j(\h_j)$ norm is
\eq
\lbeq{phinorm}
    \|\phi\|_{\Phi_j(\h_j)}
    =
    \h_j^{-1}
    \sup_{x\in \Lambda}
    \sup_{|\alpha|_1  \le p_\Phi}
    L^{j|\alpha|_1}
    |\nabla^{\alpha} \phi_x|.
\en
The estimates given in \cite[Proposition~\ref{pt-prop:Cdecomp}]{BBS-rg-pt}
(see \cite[\eqref{pt-e:scaling-estimate-Omega}]{BBS-rg-pt})
for the covariance decomposition show, in particular, that
      \begin{equation}
      \label{e:scaling-estimate}
        |\nabla_x^\alpha \nabla_y^\beta C_{j;x,y}|
        \leq
        cL^{-(j-1)(2[\phi]+(|\alpha|_1+|\beta|_1))}.
      \end{equation}
with $[\phi]$ the \emph{field dimension}
\begin{equation}
    [\phi]=\frac{d-2}{2}
\end{equation}
and
where $c$ is independent of $j,L$ and $m^2\in[0,\delta]$ for $j<N$,
while in the special case $C_j=C_{N,N}$,  $c$ is independent of $N,L,m^2$ as long as
$m^2 \in [\varepsilon L^{-2(N-1)},\delta]$
with the constant $c$ now depending on $\varepsilon>0$.
This suggests that under the expectation
$\Ex_{C_j}$, $|\nabla^{\alpha} \phi_x|$ is typically
$O(L^{-(j-1)([\phi]+(|\alpha|_1))})$.  We choose a value $\ell_j$ for $\h_j$
which makes the norm
$\|\phi\|_{\Phi_j(\ell_j)}$ be small for typical $\phi$, i.e.,
we choose for $\h_j$ the value
\begin{equation}
\lbeq{elldef}
    \ell_j = \ell_0 L^{-j[\phi]},
\end{equation}
with an $L$-dependent (large)
constant $\ell_0$ whose value gets fixed
at \refeq{CLbd} below.

As in \cite[(\ref{norm-e:PhiXdef})]{BS-rg-norm}, for $X \subset
\Lambda$ we define a local norm of the boson field $\phi$  by
\begin{align}
\label{e:PhiXdef}
    \|\phi\|_{\Phi_j(X)}
    &=
    \inf \{ \|\phi -f\|_{\Phi_j} :
    \text{$f \in \C^\Lambda$ such that $f_{x} = 0$
    $\forall x\in X$}\}.
\end{align}
This definition localises the norm to $X$ by minimising over all
extensions to the complement of $X$.
A \emph{small set} is defined to be a
connected polymer $X \in \Pcal_j$ consisting of at most $2^d$ blocks
(the specific number
$2^d$ plays a role only in the combinatorial geometry
of \cite[Section~\ref{step-sec:gl}]{BS-rg-step}
and it is only important in this paper that it be a finite constant independent of $L$).
The set of small sets is denoted $\Scal_j \subset \Pcal_j$.
The \emph{small set neighbourhood} of $X \subset \Lambda $ is
the enlargement of $X$ defined by
\begin{equation}
\label{e:ssn}
    X^{\Box}
=
    \bigcup_{Y\in \Scal_{j}:X\cap Y \not =\varnothing } Y.
\end{equation}
Given $X \subset \Lambda$ and $\phi \in \C^{\Lambda}$,
we recall from \cite[Definition~\ref{norm-def:ffregulator}]{BS-rg-norm}
that the
\emph{fluctuation-field regulator} $G_j$ is defined
by
\begin{align}
\label{e:GPhidef}
    G_j(X,\phi)
    =
    \prod_{x \in X} \exp
    \left(|B_{x}|^{-1}\|\phi\|_{\Phi_j (B_{x}^\Box,\ell_j )}^2 \right)
    ,
\end{align}
where $B_{x}\in \Bcal_j$ is the unique block that contains $x$,
and hence $|B_x| = L^{dj}$.

\subsubsection*{Large-field regulator}
\label{sec:lfr}

For $j<N$ (and $L$ large), and for $B \in \Bcal_j$, the diameter of
$B^\Box$ is less than the period of the torus.  We can therefore
identify $B^\Box$ with a subset of $\Zd$ and use this identification
to define polynomial functions from $B^\Box$ to $\C$.  More generally,
for $X$ with diameter less than the period of the torus, we define
\begin{equation}
\label{e:Phipoltildef}
    \Phipoltil (X)
    =
    \left\{ f \in \C^{\Lambda} \mid
    \text{$f$ restricted to $X$
    is a linear polynomial }\right\}.
\end{equation}
Then, for $\phi \in \C^{\Lambda}$,
we define the semi-norm
\begin{equation}
\label{e:Phitilnorm}
    \| \phi \|_{\tilde{\Phi} (X)}
=
    \inf \{ \| \phi -f\|_{\Phi}  : f \in \Phipoltil (X)
    \}.
\end{equation}
We recall from \cite[Definition~\ref{norm-def:regulator}]{BS-rg-norm} that the
\emph{large-field regulator} $\tilde G_j$ is defined by
\begin{align}
\label{e:9Gdef}
    \tilde G_j  (X,\phi)
    =
    \prod_{x \in X}
    \exp \left(
    \frac 12 |B_{x}|^{-1}\|\phi\|_{\tilde\Phi_j (B_{x}^\Box,\ell_j)}^2
    \right)
    .
\end{align}
The definition \refeq{9Gdef} is only used for $j<N$, since the norm on its
right-hand side is not defined at the final scale $j=N$.
Since $\| \phi \|_{\tilde{\Phi}  (B^\Box)} \le \| \phi \|_{\Phi (B^\Box)}$
by definition, $\tilde G_j(X,\phi) \le G_j(X,\phi)^{1/2}$.
The $\frac 12$ in the exponent of \refeq{9Gdef} is
a convenience that was used in \cite[Proposition~\ref{norm-prop:KKK}]{BS-rg-norm}.
The role of $\tilde G_j$ is discussed in Section~\ref{sec:lfp} below.

\subsubsection*{Regulator norms}

The two regulators lead us to the following definition.

\begin{defn}
\label{def:Gnorms}
Norms on $\Ncal (X^{\Box})$ are defined, for $F \in \Ncal (X^{\Box})$
and $\Gtilp \in (0,1]$, by
\begin{align}
\label{e:Gnormdef1}
    \| F\|_{G_j}
    &=
    \sup_{\phi \in \C^\Lambda}
    \frac{\|F\|_{T_{\phi,j}}}{G_{j}(X,\phi)}
    \quad
    j \le N,
\\
\label{e:Gnormdef2}
    \|F\|_{\tilde G_j^{\Gtilp}}
    &=
    \sup_{\phi \in \C^\Lambda}
    \frac{\|F \|_{T_{\phi,j}}}{\tilde{G}^{\Gtilp}_{j}(X,\phi)}
    \quad
    j<N.
\end{align}
\end{defn}

The norms depend on the choice of $\h_j$ used in the $T_{\phi,j}(\h_j)$
semi-norm
on the right-hand sides.
We write $\|F\|_j$ for the left-hand sides of \refeq{Gnormdef1}--\refeq{Gnormdef2}
in statements that apply to both the $G$
and $\tilde G$ norms.
Note that the norm $\| F\|_{G_j}$ is defined for all scales $j\le N$ whereas
we $\| F\|_{\tilde G_j}$ is undefined at the last scale.
At scale $N$, statements about the norm $\|F\|_j$ are to be understood
as applying \emph{only} to the $G$ norm.

A fundamental property of the norms \refeq{Gnormdef1}--\refeq{Gnormdef2}
is that each obeys the
\emph{product property}
\begin{equation}
\label{e:norm-fac}
    \|F G \|_j
\le
    \|F \|_j \|G \|_j
\qquad
\text{when $F\in \Ncal(X), G \in \Ncal(Y)$ for \emph{disjoint} $X,Y\in \Pcal_{j}$}.
\end{equation}
This is an immediate consequence of
the above mentioned product property which states that
$\|FG\|_{T_\phi} \leq \|F\|_{T_\phi}\|G\|_{T_\phi}$
for \emph{any} $F,G \in \Ncal$,
 together with
the fact that by definition
$G_j(X\cup Y,\phi)=G_j(X,\phi)G_j(Y,\phi)$ for disjoint
$X,Y$, and similarly for $\tilde G_j$.

\subsection{Overview of results}

Our goal in this paper is to obtain a thorough understanding
of the interaction functional
$I =I_{j}$.
The main results are stated in Section~\ref{sec:IE}, with proofs deferred to
Sections~\ref{app:sp}--\ref{sec:ipcl}.
The results include proof of
stability bounds for $I$,
estimates on Gaussian expectations involving both boson and fermion fields,
estimates verifying the accuracy of the perturbative
calculations in \cite{BBS-rg-pt},
and proof of the crucial contraction property needed to
control irrelevant directions in the flow of the renormalisation
group.  These all play a role in the analysis of a
single renormalisation group step in \cite{BS-rg-step}.
Before making precise statements
in Section~\ref{sec:IE}, in this section
we provide an informal overview of and motivation for
the results.

\subsubsection{Stability, expectation and the large-field problem}
\label{sec:lfp}

In Section~\ref{sec:stab}, we state a series of \emph{stability estimates}.
In particular, Proposition~\ref{prop:Iupper} provides the bound
\begin{equation}
\lbeq{intro-Ibd}
    \|I_j(V,B)F(B)\|_{T_\phi(\ell_j)} \le 2 \|F(B)\|_{T_0(\ell_j)}G_j(B,\phi)
\end{equation}
for $B \in \Bcal_j$, and for a polynomial $F(B)$ in the fields in $B$
of degree at most the parameter $p_\Ncal$ in the definition of the space $\Ncal$,
under suitable hypotheses expressing a
smallness condition on the coupling constants in $V$.
Since $G_j(B,\phi)= \exp[\|\phi\|_{\Phi(B^\Box, \ell_j)}^2]$, \refeq{intro-Ibd}
provides information on the growth of the left-hand side for large fields $\phi$.
This estimate does not take advantage of the quartic decay provided by $e^{-g\tau^2}$
to compensate for the quadratic part $e^{-\nu\tau}$ in $e^{-V}$
(with $\nu$ possibly negative).  This is
reflected by the quadratic growth in the exponent on the right-hand side
of \refeq{intro-Ibd}.

The renormalisation group method is based on iterated expectation to progressively
take into account fluctuations on increasingly larger scales.
One difficulty with \refeq{intro-Ibd} is that it degenerates under
expectation and change of scale, as we discuss next.
These ideas play a role in the proof of Proposition~\ref{prop:ip},
which is our main estimate on Gaussian expectation.
We make the abbreviation
$\Ex_j = \Ex_{C_j}$ for the Gaussian expectation with covariance $C_j$.
Since the expectation involves both boson and fermion fields
(see \cite{BIS09,BS-rg-norm}), it would more
accurately be termed ``super-expectation'' but we use the term ``expectation'' for brevity.
It is shown in
\cite[Proposition~\ref{norm-prop:EK}]{BS-rg-norm},
that for any $K \in \Ncal$,
\begin{equation}
\lbeq{intro-EI00}
        \|\Ex_{j+1}\theta K\|_{T_\phi(\h_j)}
        \le
        \Ex_{j+1}\|K\|_{T_{\phi\sqcup \xi}(\h_j \sqcup \ell_j)}.
\end{equation}
In more detail, in \cite[Proposition~\ref{norm-prop:EK}]{BS-rg-norm}
we choose $w=\h_j$ and $w'=\ell_j$, and the hypothesis $\|C_{j+1}\|_{\Phi_{j+1}(\ell_{j+1})} \le 1$
is verified at \refeq{CLbd} below.
The integrand on the
right-hand side of \refeq{intro-EI00}
is a function only of the boson field, so the super-expectation
reduces to a standard Gaussian expectation with covariance $C_{j+1}$
(see \cite[Section~\ref{norm-sec:cbf}]{BS-rg-norm}).
The fermion field ceases to play a
significant role in the analysis once this inequality has been applied,
and this is a beneficial aspect of our method.

By \refeq{intro-Ibd}--\refeq{intro-EI00} and \refeq{GPhidef},
and by the inequality $\|\phi+\xi\|^2 \le 2(\|\phi\|^2+\|\xi\|^2)$,
\begin{equation}
\lbeq{intro-EI0}
        \|\Ex_{j+1}\theta I_j(V,B)\|_{T_\phi(\ell_j)}
        \le
        \Ex_{j+1}\|I_j(V,B)\|_{T_{\phi\sqcup \xi}(\ell_j\sqcup \ell_j)}
        \le
        2 G_j(B,\phi)^2 \Ex_{j+1} G_j(B,\xi)^2.
\end{equation}
According to \cite[Proposition~\ref{norm-prop:EG2}]{BS-rg-norm},
$ \Ex_{j+1} G_j(B,\xi)^2 \le 2$.
Therefore,
\begin{equation}
\lbeq{intro-EI1}
        \|\Ex_{j+1}\theta I_j(V,B)\|_{T_\phi(\ell_j)}
        \le
        4 G_j(B,\phi)^2.
\end{equation}
The left-hand side can only become smaller when the semi-norm is changed
from scale $j$ to scale $j+1$ (this useful monotonicity property is
proved in Lemma~\ref{lem:Imono} below).
To see the effect of a change of scale on the right-hand side,
consider the particular case $\phi_x = a$ for all $x$,
where $a$ is a constant. In this case, by definition,
\begin{align}
    L^{-dj}\|\phi\|^2_{\Phi_j(B_{x,j}^\Box, \ell_j)}
    &=
    L^{-dj} \ell_j^{-2} a^2
    =
    L^{2} L^{-d (j+1)} \ell_{j+1}^{-2} \, a^2
    =
    L^{2} L^{-d (j+1)}
    \|\phi\|^2_{\Phi_{j+1}(B_{x,j}^\Box, \ell_{j+1})}
    \nnb & =
    L^{2} L^{-d(j+1)}
    \|\phi\|^2_{\Phi_{j+1}(B_{x,j+1}^\Box, \ell_{j+1})}
    ,
\lbeq{martGfail}
\end{align}
so for this case $G_{j}(B,a)=G_{j+1}^{L^2}(B,a)$.  Thus the estimate after
expectation and change of scale is substantially worse than
\refeq{intro-Ibd} (it is the growth in $\phi$ that is problematic,
the constant $4$ in \refeq{intro-EI1} is not).
It is in this way that the so-called \emph{large-field problem}
enters our analysis.  We postpone the problem by setting
$\phi=0$, so that the regulator plays no role in \refeq{intro-EI1}.
With $\phi=0$, \refeq{intro-EI0} becomes
\begin{equation}
\lbeq{intro-EI0a}
        \|\Ex_{j+1}\theta I_j(V,B)\|_{T_0(\ell_j)}
        \le
        \Ex_{j+1}\|I_j(V,B)\|_{T_{0\sqcup \xi}(\ell_j\sqcup \ell_j)}
        \le
        4
        .
\end{equation}
From this, we see that control of $I_j$ is needed for \emph{all}
field values in order
to estimate the expectation of the fluctuation field $\xi$, even when $\phi=0$.
Thus we are able to obtain a useful estimate in the $T_0$ semi-norm at scale $j+1$,
but this is not sufficient to be able to iterate these estimates
as the scale advances.

To deal with the large-field problem, we do not perform a separate
analysis on regions of space where the field is large and where it is
small, as has been done in other renormalisation group methods, e.g.,
\cite{Bala82,Dimo13,GK85,GK86}.
Instead,
we take advantage of the factor $e^{-g\sum_{x\in B}|\phi_x|^4}$ in $I(B)$ and
exploit it to capture the notion that a typical field should roughly have
size $g^{-1/4}L^{-jd/4}$.
For this, we need information about the size of $g$.

Our ansatz is that at scale $j$,
$g$ is close in size to $\gbar_j$,
which is defined by the recursion
\begin{equation}
\label{e:gbardef}
    \gbar_{j+1} = \gbar_j - \beta_j \gbar_j^2
\end{equation}
of \cite[\eqref{pt-e:newflow-gbar}]{BBS-rg-pt}, with a fixed
initial condition $\gbar_0$,
and with $\beta_j$ given in terms of the covariance $w_j$ of
\refeq{wjdef} by
\begin{equation}
    \beta_j = 8\sum_{x \in \Lambda}(w_{j+1;0,x}^{2}-w_{j;0,x}^{2}).
\end{equation}
The sequence $\beta_j$ is closely related to the \emph{bubble diagram}
$\sum_{x \in \Zd} [(-\Delta_{\Zd}^{-1})_{0x}]^2$, which diverges for
$d=4$ but converges for $d>4$ since the inverse Laplacian is asymptotically
a multiple of $|x|^{-(d-2)}$.  By
\cite[Lemma~\ref{pt-lem:betalim}]{BBS-rg-pt}, $\beta_j \to 0$ for $d>4$
whereas $\beta_j  \to \pi^{-2}\log L$ for $d=4$.  Also, by choosing $\gbar_0$ to be
sufficiently small, it follows that $\gbar_j$ is uniformly small.

In the present paper, our focus is on the advancement of one scale
to the next, rather than on all scales simultaneously.  Because of this,
and to provide flexibility, rather than using $\gbar_j$,
we introduce a small positive $\ggen_j$
and consider $g$ at scale $j$ to be close to $\ggen_j$.
We do not assume that $\ggen_j$ is given by \refeq{gbardef}
(a different but closely related
choice of $\ggen$ is used in \cite[\eqref{log-e:ggendef}]{BBS-saw4-log}),
but we do
assume that $\ggen_j$ is uniformly small for all $j$, and that we are free
to choose how small it is depending on $L$.
Thus we introduce $h_j \propto \ggen_j^{-1/4}L^{-jd/4}$
and seek estimates in terms of the $T_\phi(h_j)$ semi-norm.
Note that for $d=4$, $h_j$ is larger than $\ell_j$ by a factor $\ggen_j^{-1/4}$.

We employ the $T_\phi(h_j)$ semi-norm in conjunction with the large-field regulator
$\tilde G_j$.
An essential property of $\tilde G_j $ (used
in the proofs of
Propositions~\ref{prop:Istab}--\ref{prop:Ianalytic1:5} and
\ref{prop:ip}--\ref{prop:cl} below) is given in the following lemma.
We apply Lemma~\ref{lem:mart} with specific choices of $p$,
and do not thereby lose control of the size of $L$.

\begin{lemma}
\label{lem:mart}
Let $X \subset \Lambda$.
For any fixed $p >0$ (no matter how large),
if $L$ is large enough
depending on $p$, then for all $j+1<N$,
\begin{equation}
\label{e:mart}
    \tilde{G}_{j}(X, \phi)^{p}
    \le
    \tilde{G}_{j+1}(X, \phi).
\end{equation}
\end{lemma}

\begin{proof}
By definition of the regulator in \refeq{9Gdef}, it suffices to prove
that
\begin{equation}
\lbeq{martwant}
    pL^{-dj}\|\phi\|^2_{\tilde\Phi_j(B_{j,x}^\Box,\ell_j)}
    \le L^{-(j+1)d} \|\phi\|^2_{\tilde\Phi_{j+1}(B_{j+1,x}^\Box,\ell_{j+1})}.
\end{equation}
Let $d_{+} = [\phi]+1 = \frac{d-2}{2} + 1 = \frac d2$.  By the
definition of dimension of a polynomial given in
\cite[Section~\ref{loc-sec:oploc}]{BS-rg-loc}, a linear polynomial has
dimension $ [\phi]+1 = d_+$.  It is a consequence of
\cite[Lemma~\ref{loc-lem:phij}]{BS-rg-loc}, with $d_+'= d_{+}+1 =
\frac{d}{2} + 1$, that
\begin{align}
\label{e:phij-pre}
    \|\phi\|_{\tilde{\Phi}_{j} (B_{j,x}^\Box,\ell_j)}
    & \le
    c
    L^{-d/2 - 1}   \|\phi\|_{\tilde{\Phi}_{j+1} (B_{j,x}^\Box,\ell_{j+1})}.
\end{align}
Therefore, since the semi-norm \eqref{e:Phitilnorm} is non-decreasing
in $X$ by definition,
\begin{align}
    \|\phi\|^2_{\tilde\Phi_j(B_{j,x}^\Box,\ell_j)}
    &\le  c^2 L^{-d-2}
    \|\phi\|^2_{\tilde\Phi_{j+1}(B_{j+1,x}^\Box,\ell_{j+1})}
    ,
\label{e:tilnormrescale}
\end{align}
from which \eqref{e:martwant} follows when $L$ is large enough that
$pc^{2} L^{-2} \le 1$.
\end{proof}

The inequality \refeq{mart} does not hold for the regulator $G$: we
have concluded from \refeq{martGfail} that for a constant field we
have $G_j=G_{j+1}^{L^2}$.  In contrast, the norm in $\tilde G$ scales
down, because it does not examine the constant and linear parts of
$\phi$.  By the use of a lattice Sobolev inequality (proved in
Appendix~\ref{sec:Lp}), we take advantage of the decay in
$e^{-g\tau^2}$ to cancel the exponential quadratic $\|\phi\|_\Phi^2$
at the cost of an exponential of $\|\phi\|_{\tilde \Phi}^2$.  By
pursuing this strategy, we prove in Proposition~\ref{prop:Iupper}
below that for $F(B)$ as in \refeq{intro-Ibd},
\begin{equation}
\lbeq{intro-Ibdh}
    \|I_j(V,B)F(B)\|_{T_\phi(h_j)}
    \le 2 \|F(B)\|_{T_0(h_j)}\tilde G_j(B,\phi)
    ,
\end{equation}
and now with \refeq{mart} this leads as above to
\begin{equation}
\lbeq{intro-EIh}
        \|\Ex_{j+1}\theta I_j(V,B)F(B)\|_{T_\phi(h_j)}
        \le
        4 \|F(B)\|_{T_0(h_j)} \tilde G_j(B,\phi)^2
        \le 4 \|F(B)\|_{T_0(h_j)} \tilde G_{j+1}^{\Gtilp}(B,\phi),
\end{equation}
for any fixed choice of $\Gtilp\in (0,1]$, e.g., $\Gtilp = 1/2$, with
$L$ large depending on $\Gtilp$.  Thus the $h$ bound reproduces itself
after expectation and change of scale.  In fact, our ability to choose
$\Gtilp <1$ shows that the $h$ bound \emph{improves}.

On the other hand, the $\ell$ bound degrades after expectation and
change of scale.  However, together the scale-$(j+1)$ $\ell$ and $h$
bounds can be combined using
\cite[Proposition~\ref{norm-prop:KKK}]{BS-rg-norm} to infer a
$G_{j+1}$ bound for all $\phi$ from the $T_{0}(\ell_{j+1})$ and
$\tilde G_{j+1}$ bounds.  In this way it is possible to obtain bounds
at scale $j+1$ of the same form as the bounds at scale $j$.  We
postpone the application of
\cite[Proposition~\ref{norm-prop:KKK}]{BS-rg-norm} to the proof of
\cite[Theorem~\ref{step-thm:mr}]{BS-rg-step}.  With this motivation,
throughout this paper we prove estimates in terms of the two norm
pairs
\begin{equation}
\label{e:np1}
    \|F\|_j = \|F\|_{G_j(\ell_j)}
    \quad \text{and} \quad
    \|F\|_{j+1} = \|F\|_{T_{0,j+1}(\ell_{j+1})},
\end{equation}
and
\begin{equation}
\label{e:np2}
    \|F\|_j = \|F\|_{\tilde{G}_j(h_j)}
    \quad \text{and} \quad
    \|F\|_{j+1} = \|F\|_{\tilde{G}_{j+1}^{\Gtilp}(h_{j+1})},
\end{equation}
i.e., estimates on $\|F\|_{j+1}$ are expressed in terms of $\|F\|_j$
for each of the pairs \refeq{np1} and \refeq{np2}.  We distinguish the
cases \refeq{np1} and \refeq{np2} by writing $\h_j=\ell_j$ to indicate
\refeq{np1}, and $\h_j=h_j$ to indicate \refeq{np2}.  The values of
$\h_\sigma$ in the $T_\phi$ norms, for sheets corresponding to the
observable fields $\sigma,\bar\sigma$, are specified in
\refeq{newhsig} below (see \cite[\eqref{loc-e:Fnormsum}]{BS-rg-loc}
for the $T_\phi$ norm with observables).

Iteration of estimates using \refeq{np2} is possible without the
accompaniment of \refeq{np1}.  However, estimates in terms of the
$\tilde G(h)$ norm are insufficient on their own to make estimates on
remainder terms in the flow of coupling constants, and without such
estimates we are unable to study critical behaviour.  In the flow of
coupling constants determined in \cite{BS-rg-step}, the interaction
polynomial $V_{j+1}$ at scale $j+1$ is expressed in terms of
$V_{\pt,j+1}(V_j)$ plus a non-perturbative remainder $\rho_{j+1}\in
\Qcal$ whose coupling constants must be shown to be third order in
$\ggen_j$.  Our control over these coupling constants is obtained via
the $T_0$ semi-norm.  To illustrate this, consider the case of $d=4$,
and suppose that the $\tau^2$ term in $\rho_{j+1}$ were simply
$\ggen_j^3 \tau^2$.  The calculation of the $T_\phi$ semi-norm of
$\tau^2$ is straightforward, and a small extension of
\cite[Proposition~\ref{norm-prop:taunorm}]{BS-rg-norm} gives
$\|\ggen_j^3\tau^2_x\|_{T_0(\h_j)} \asymp \ggen_j^3 \h_j^4$ (the
symbol $\asymp$ means upper and lower bounds with different
constants).  Focussing only on the power of $\ggen_j$, the choice
$\h_j =h_j$ gives an overall power $\ggen_j^3 ( \ggen_j^{-1/4})^4=
\ggen_j^2$, which is second order rather than the desired third order.
For this reason, estimates in terms of norms with $\h=h$ are
insufficient.  On the other hand, with the $T_0(\ell)$ semi-norm there
is no loss of powers of $\ggen_j$ arising from
$\|\tau^2_x\|_{T_0(\ell_j)} \asymp \ell_j^4$, and the $T_0(\ell)$
semi-norm indeed identifies $\ggen_j^3\tau^2$ as a third-order term.

\begin{rk}\label{rk:h++}
In the $j+1$ members of the norm pairs
\refeq{np1}--\refeq{np2}, the parameter $\h_{j+1}$ may be replaced
by $\h_{++}=c\h_{j+1} >\h_{j+1}$ for any fixed $c>1$.
More precisely, in
Definition~\ref{def:Gnorms}, with $j$ replaced by $j+1$, $\h_{j+1}$ becomes
replaced by $\h_{++}$ in the $T_{\phi,j+1}(\h_{j+1})$ norm.
Our convention is to leave $\ell_{j+1}$ in
the regulator unchanged; it does not become $\ell_{++}$ in the replacement of
$\h_{j+1}$ by $\h_j$.
All our results remain valid with $\h_{++}$ replacing $\h_{j+1}$, with changes in
constants whose precise values are without significance and indeed are
not specified in our results.
To avoid further elaboration of our notation, we do not make the
role of $\h_{++}$ explicit in the rest of the paper, apart from
one additional comment
below \eqref{e:h-assumptions}.
\end{rk}

\begin{rk}
\label{rk:scaleNnorm}
The advancement of estimates to the final scale $N$ is special, since the
$\tilde G$ norm is undefined at that scale.  However, the work of the $\tilde G$
norm is complete
by scale $N$, as there is no further difficulty concerning degradation of
estimates since the scale no longer advances.  Thus, at scale $N$, we can consider the
norm to be the $G$ norm with regulator $G$ replaced by a suitable large power
of $G_{N-1}$, such as $G_{N-1}^{10}$ (using $G_{N-1}^2$ would be sufficient for
\refeq{intro-EI1} but higher powers are required later).
Then a scale-$N$ estimate $\|F\|_N \le C$ is interpreted as stating that
$\|F\|_{T_\phi,N} \le C G_{N-1}(\Lambda,\phi)^{10}$.
In some applications, the $T_0$ estimate obtained by setting $\phi=0$ is sufficient.
More generally, the estimate states that $\|F\|_{T_\phi,N} \le C\exp[O(\|\phi\|^2)]$
(with $L$-dependent constant in the exponent), and this provides additional
information concerning the growth in $\phi$.  We are not always careful to distinguish
the special nature of $\|\cdot\|_N$, but inspection reveals that
our conclusions indeed hold with this choice.
\end{rk}

\subsubsection{Accuracy of perturbative analysis}

One of our main results is a proof of a version of \eqref{e:fps} that
goes beyond formal power series.  The version we prove is a local one,
which permits accurate estimates with errors bounded uniformly in the
volume.  However, the local analysis comes with a cost, which is that
an explicit second-order leading term arises along with the
third-order error.

For simplicity, for the present discussion we set
$\lambdaa=\lambdab=\qa=\qb=0$ so that observables play no role.  In
this setting, a particular case of what we prove is that for $b \in
\Bcal_j$ and $B \in \Bcal_{j+1}$,
\begin{equation}
\lbeq{pt-eg}
    \Itilde_\pt^{B\setminus b}
    \Ex_{j+1} \theta I(V,b)
    \approx
    \Ipttil^B
    \Big(
    1
    -
    \frac 12 \Ex_{j+1} \theta (V_j(b);V_j(\Lambda \setminus b)
    \Big)
    ,
\end{equation}
where the \emph{truncated expectation} (or \emph{covariance}) is defined by
\begin{equation}
\label{e:trun-exp-eg}
    \Ex_C (A; B) = \Ex_C(AB) - (\Ex_CA)(\Ex_CB).
\end{equation}
We prove precise versions of \refeq{pt-eg} with third-order
error estimates, for both norm pairs \refeq{np1}--\refeq{np2}.
For example, in the proof of
Proposition~\ref{prop:h}, for the norm pair \refeq{np1} we show that
\begin{equation}
\label{e:want2-eg}
    \|
    \Itilde_\pt^{B\setminus b}\Ex_{j+1} (\theta I(V,b) - \Ipttil(b))
    +
    \Itilde_\pt^{B}
    \frac 12 \Ex_{j+1} \theta (V_j(b);V_j(\Lambda \setminus b)
    \|_{T_{0,j+1}(\ell_{j+1})}
    \le
    O(\ggen_j^3).
\end{equation}
The bound on the right-hand side is third-order as desired, but there is
a second-order leading term on the left-hand side.
Its origin can be seen from a small
extension of the argument in
 \cite[Section~\ref{pt-sec:WPjobs}]{BBS-rg-pt},
as follows.
Proceeding as in  \cite[Section~\ref{pt-sec:WPjobs}]{BBS-rg-pt},
formally, to a third-order error, we obtain
\begin{align}
    \Ex \theta I(b)
    & \approx
    e^{-\Ex \theta V(b)} \left[
    1 + \Ex \theta W(b) + \frac 12 \Ex \theta (V(b);V(b))
    \right].
\end{align}
The bilinear term $W(b)$ involves $V(b)$ in one argument
and $V(\Lambda)$ in the other, and its partner to make the
argument of \cite[Section~\ref{pt-sec:WPjobs}]{BBS-rg-pt} apply here
has to be $\frac 12 \Ex \theta (V(b);V(\Lambda))$ rather than
$ \frac 12 \Ex \theta (V(b);V(b))$.
Thus we rewrite the right-hand side as
\begin{align}
    \Ex \theta I(b)
    & \approx
    e^{-\Ex \theta V(b)} \left[
    1 + \Ex \theta W(b) + \frac 12 \Ex \theta (V(b);V(\Lambda))
    - \frac 12 \Ex \theta (V(b;V(\Lambda\setminus b))
    \right].
\end{align}
After multiplication by $\Itilde_\pt^{B\setminus b}$, the
extra term produces
$-\Itilde_\pt^{B}\frac 12 \Ex \theta (V(b;V(\Lambda\setminus b))$,
which is what appears in \refeq{want2-eg}.

In Proposition~\ref{prop:hldg}, we prove that the leading term
in the perturbative estimates we require is indeed second order.  This is a
straightforward consequence of the stability bounds.
The fact that the remainder beyond the leading term is third order
is proved in Proposition~\ref{prop:h}, which is more substantial,
and is our full implementation
of the formal arguments of \cite[Section~\ref{pt-sec:WPjobs}]{BBS-rg-pt}.
For the reasons discussed in Section~\ref{sec:lfp}, we need versions of these
two propositions for both norm pairs.

\subsubsection{\texorpdfstring{$\LT$}{Loc} and the crucial contraction}
\label{sec:cc}

The renormalisation group creates an infinite-dimensional dynamical system,
which has a finite number of relevant or marginal directions and
infinitely many irrelevant directions.  A crucial aspect of our
analysis is to employ the operator $\LT$ defined and developed
in \cite{BS-rg-loc}
to extract the relevant and marginal parts of a functional of the
fields, with $(1-\LT)$ projecting onto the irrelevant parts.
The specific result we prove in this respect is Proposition~\ref{prop:cl}.
A special case of Proposition~\ref{prop:cl}  is as follows.

Let $X$ be a small set as defined above \refeq{ssn}.
Let $U$ be the smallest collection of blocks in $\Bcal_{j+1}$ which contains
$X$ ($U$ is the \emph{closure} $U = \overline X$).
Let $F(X) \in \Ncal(X^\Box)$ be such that $\LT_X F =0$; this should be interpreted
as a statement that $F(X)$ is irrelevant for the renormalisation group.
We prove in Proposition~\ref{prop:cl} that, under appropriate assumption on $V$,
\begin{align}
    \label{e:cl-eg}
    \|\tilde I^{U\setminus X} \Ex \theta \left( \Itilde^{X}F (X) \right) \|_{j+1}
    &
    \le {\rm const}\,
    L^{-d-1}
    \|F(X)\|_{j}
    ,
\end{align}
where the pair of norms is given by
either choice of \eqref{e:np1} or \eqref{e:np2}.
The number of distinct $X$ with closure $U$ produces
an entropic factor of order $L^d$, and hence
\begin{align}
    \label{e:cl-sum-eg}
    \sum_{X \in \Scal_j: \overline X =U}\|\tilde I^{U\setminus X} \Ex \theta \left( \Itilde^{X}F (X) \right) \|_{j+1}
    &
    \le {\rm const} \,
    L^{-1}
    \|F(X)\|_{j}
    .
\end{align}
Thus a contractive factor $L^{-1}$ remains also after summation.
This  plays a crucial role in \cite{BS-rg-step} in showing that
the coordinate of the dynamical system that is meant to represent
the irrelevant directions is in actual fact contractive.

\subsection{Parameters and domains}

In this section, we reformulate estimates on the
covariance decomposition that are stated in \cite{BBS-rg-pt},
we specify the parameters that define the $T_\phi$
norms we use,
we define the small parameters $\epV,\epdV$ that permeate
our analysis,
and we discuss the domains for $V$ which ensure stability of $I$.

\subsubsection{Estimate on covariance decomposition}
\label{sec:frp}

We now discuss the size
of the covariances arising in the covariance decomposition, in more detail.
Recall from \refeq{elldef}
the definition $\ell_j = \ell_0 L^{-j[\phi]}$.
We may regard a covariance $C$ as a test function depending on
two arguments $x,y$, and with this identification its $\Phi_j(\ell_j)$
norm is
\begin{align}
&
    \label{e:Phinorm}
    \|C\|_{\Phi_{j}(\ell_j)}  =
    \ell_j^{-2}
    \sup_{x,y\in \Lambda}
    \;
    \sup_{|\alpha_{1}|_1 + |\alpha_{2}|_1 \le p_\Phi}
    L^{(|\alpha_{1}|_1+  |\alpha_{2}|_1)j}
    |\nabla_x^{\alpha_1} \nabla_y^{\alpha_2} C_{x,y}|
,
\end{align}
where $\alpha_i$ is a multi-index.
The norm of the covariance $C_j$ in the covariance decomposition can be estimated using
 an improved version of \refeq{scaling-estimate} from \cite{Baue13a,BBS-rg-pt}.

For this, given $\Omega >1$ we define the $\Omega$-\emph{scale} $\jm$ by
  \begin{equation}
  \lbeq{mass-scale}
    \jm = \inf \{ k \geq 0: |\beta_j| \leq \Omega^{-(j-k)} \|\beta\|_\infty
    \text{ for all $j$} \}
    ,
  \end{equation}
and we set
\begin{equation}
\lbeq{chidef}
    \chi_j = \Omega^{-(j-\jm)_+}.
\end{equation}
The $\Omega$-scale indicates a scale at which the mass term in the covariance
starts to play a dominant role in dramatically reducing the size of the covariance;
further discussion of this point can be found in
\cite[Section~\ref{pt-sec:Greekpfs}]{BBS-rg-pt}.
It is within a constant of the value $j_m$ defined by $j_m=\lfloor \log_{L^2}m^{-2}\rfloor$,
as shown in \cite[Proposition~\ref{pt-prop:rg-pt-flow}]{BBS-rg-pt}, and
$\chi_j$ could alternately be defined in terms of $j_m$.
We always take the infinite volume limit before letting $m^2 \downarrow 0$,
so we may assume that $m^2 \in [\varepsilon L^{-2(N-1)},\delta^2]$ for small fixed $\delta,
\varepsilon$.

It is shown in
\cite[\eqref{pt-e:scaling-estimate-Omega}]{BBS-rg-pt} that
there is an $L$-independent constant $c$
such that for $m^2 \in [0, \delta]$ and $j=1,\ldots,N-1$, or for
$m^2 \in [\varepsilon L^{-2(N-1)},\delta]$ for $N$ large in the
special case $C_j=C_{N,N}$,
      \begin{equation}
      \label{e:scaling-estimate-Omega}
        |\nabla_x^\alpha \nabla_y^\beta C_{j;x,y}|
        \leq c \chi_j
        L^{-(j-1)(2[\phi]+(|\alpha|_1+|\beta|_1))}.
      \end{equation}
Let
\begin{equation}
\lbeq{Ckstardef}
    C_{j*}=
    \begin{cases}
        C_j  & j<N
        \\
        C_{N,N} & j=N.
    \end{cases}
\end{equation}
By \eqref{e:scaling-estimate-Omega},
given $\ellconst \in (0,1]$ we can choose $\ell_0$ large depending on $L$ to obtain,
for $j=1,\ldots,N$,
\begin{equation}
\lbeq{CLbd}
    \|C_{j*}\|_{\Phi_j^+(\ell_j)}
    \le
    \ellconst \chicCov_j
    \le
    \min\{\ellconst, \chicCov_j\},
\end{equation}
where $\Phi^+$ refers to the norm
\refeq{Phinorm} with $p_\Phi$ replaced by $p_\Phi+d$.

Let $c_G=c(\alpha_G)$ be the (small) constant of
\cite[Proposition~\ref{norm-prop:EG2}]{BS-rg-norm}.
We fix the value $\ellconst = \frac{1}{10}c_G$.
Then \cite[Proposition~\ref{norm-prop:EG2}]{BS-rg-norm} ensures that
\begin{align}
    \max_{k=j,j+1}\Ex_{k*} ( G_{j} (X) )^{10}
&
\le
    2^{|X|_j}
\quad\quad
      X \in \Pcal_{j},
\label{e:EG}
\end{align}
where $|X|_j$ denotes the number of scale-$j$ blocks comprising
$X$ (the constants $10$ and $2$ in \refeq{EG}
are convenient but somewhat arbitrary choices).
The use of $\Phi^+$
in \refeq{CLbd} is to satisfy the hypotheses of
\cite[Proposition~\ref{norm-prop:EG2}]{BS-rg-norm}.

\subsubsection{Choice of norm parameters}
\label{sec:hex}

We restrict attention here to $d=4$.

For the $G$ norm,
for the boson and fermion fields we choose $\ell_0$ according to
\refeq{CLbd} and set
\begin{equation}
\label{e:hl}
    \h_j = \ell_j = \ell_0 L^{-j[\phi]}  .
\end{equation}
For the $\tilde G$ norm, we fix a parameter $k_0$
(small,  chosen
as discussed under Proposition~\ref{prop:Iupper}),
we set
\begin{equation}
\label{e:h-def}
    \h_j = h_{j} = k_0 \ggen_j^{-1/4}L^{-jd/4}.
\end{equation}
We assume that $\ggen_j$ can be taken to be as small as desired (uniformly in $j$,
and depending on $L$), and that
\begin{equation}
\label{e:gbarmono}
    \frac 12 \ggen_{j+1} \le
    \ggen_j \le 2 \ggen_{j+1}
\end{equation}
(the above two inequalities hold
for the sequence $\gbar_j$ by \cite[\eqref{pt-e:gbarmono}]{BBS-rg-pt}).
For the observables, we set
\begin{equation}
\lbeq{newhsig}
    \h_{\sigma,j}=
    \begin{cases}
    \ggen_j L^{(j\wedge j_{ab})[\phi]} 2^{(j-j_{ab})_+} & \h=\ell
    \\
    \ggen_j^{1/4} L^{(j\wedge j_{ab})[\phi]} 2^{(j-j_{ab})_+} & \h=h;
    \end{cases}
\end{equation}
see Remark~\ref{rk:hsigmot}
for motivation of this definition.
By \refeq{gbarmono},
the above choices obey:
\begin{align}
\label{e:h-assumptions}
    \h_j \ge \ell_{j},
    \quad\quad
    \frac{\h_{j+1}}{\h_j} L^{[\phi]}
    &\le 2,
    \quad\quad
    \frac{\h_{\sigma,j+1}}{\h_{\sigma,j}}
    \le
    {\rm const}\,
    \begin{cases}
     L^{[\phi]} & j < j_{ab}
     \\
     1 & j \ge j_{ab}.
     \end{cases}
\end{align}
Our results for the norm pairs \eqref{e:np1}--\eqref{e:np2}
require only the bounds on $\h_{j+1}$ in
\eqref{e:h-assumptions}.  However, the choice of $2$ that appears there and in
\eqref{e:gbarmono} is arbitrary, and, e.g., $3$ would do as well.
For this reason,
we can replace $\h_{j+1}$ by a larger $\h_{++}=c\h_{j+1}$, as claimed in
Remark~\ref{rk:h++}.

\subsubsection{Definition of small parameter \texorpdfstring{$\epV$}{epsilonV}}
\label{sec:spdefs}

The stability estimates are expressed in terms of domains defined via
parameters $\epV$ and
$\epVbar$, which we discuss now.  Given $V \in \Qcalnabla$, we write $V_\varnothing =
\sum_{M}M$ for the decomposition of its bulk part
as a sum of individual field
monomials such as $\nu \phi\bar\phi$, $\nu \psi\bar\psi$, $z(\Delta
\phi)\bar\phi$, and so on.  Then, for $0 \le j \le N$, we define
\begin{equation}
\label{e:monobd}
    \epV
    =\epsilon_{V,j}
    =
    L^{dj} \sum_{M : \pi_*M=0} \|M_0\|_{T_{0,j}(\h_j)}
    + (|\lambdaa|+|\lambdab|)\h_j\h_{\sigma,j} + (|\qa|+|\qb|)\h_{\sigma,j}^2
    ,
\end{equation}
where $M_0$ denotes the monomial $M_x$ evaluated at $x=0$.
Thus $\epV$ is a function (in fact, a norm) of the coupling constants in $V$ and
of the parameters $\h_j$ and $\h_{\sigma,j}$ which define
the $T_0$ semi-norm.
The value of $\epV$ depends on the scale $j$, but we
often leave this implicit in the notation.  It measures the size
of $V$ on a block $B \in \Bcal_j$ consisting of $L^{dj}$ points,
and is worst case in the sense that it includes a contribution
from observables whether or not the points $a$ or $b$ lie in $B$.

The term $g\tau^2$ plays a special role in providing the
important factor $e^{-g|\phi|^4}$ in $e^{-V}$, and we define
\begin{equation}
    \label{e:epVbar-def-old}
    \epVbar =
    \epsilon_{g\tau^2,j}
    =
    L^{dj}  \|g\tau^2_0 \|_{T_{0,j}(\h_j)}.
\end{equation}
By definition, $\epVbar \le \epV$.  Also, there is a universal
constant $C_0>0$ such that
\begin{equation}
\label{e:epVbarasymp}
    C_0^{-1} |g|\h_j^4 L^{dj} \le \epVbar \le  C_0 |g|\h_j^4 L^{dj} .
\end{equation}
In fact, the upper bound is proved in
\cite[Proposition~\ref{norm-prop:taunorm}]{BS-rg-norm}, while the
lower bound follows directly from the definition of the $T_{\phi}$
norm (see \cite[Definition~\ref{norm-def:Tphi-norm}]{BS-rg-norm})
since the supremum of the pairing over all unit norm test functions is
larger than the pairing with a constant unit norm test
function.

\subsubsection{Stability domains}

To enable the use of analyticity methods in \cite{BS-rg-step}, we
employ complex coupling constants.  Given a (large) constant
$C_{\DV}$, we define a domain
\begin{align}
\lbeq{DV1-bis}
    \DV_j = \{(g,\nu,z,y,\lambdaa,\lambdab,\qa,\qb)\in \C^{8} :
    C_\DV^{-1} \ggen_j < {\rm Re} g < C_\DV \ggen_j,
    &
    \; |{\rm Im} g| < \textstyle{\frac {1}{10}} {\rm Re} g,
    \nnb & |x| \le r_x
    \; \text{for $x\neq g$}\},
\end{align}
where $r_x$ is defined (with $\lambda$ equal to $\lambdaa$ or
$\lambdab$ and similarly for $q$) by
\begin{gather}
    L^{2j}r_{\nu,j} = r_{z,j}= r_{y,j}= C_{\DV} \ggen_j,
    \quad\quad
    r_{\lambda,j} =
      C_{\DV},
     \nnb
         L^{2j_{ab}[\phi]}2^{2(j-j_{ab})}r_{q,j} = \begin{cases}
        0 & j < j_{ab} \\
        C_{\DV}  & j \ge j_{ab}.
        \end{cases}
\label{e:h-coupling-def-1-bis}
\end{gather}
We also use two additional domains in $\C^{8}$, which depend on the
value of $\h$ (namely $\h=\ell$ or $\h=h$), as well as on parameters
$\alpha,\alpha',\alpha''>0$.  Given these parameters, we define
\begin{align}
\label{e:DVell}
    \bar\DV_j(\ell)
    &=
    \{V \in \Qcal :
    \; |{\rm Im} g| < \textstyle{\frac {1}{5}} {\rm Re} g,
    \;
    \epsilon_{V,j}(\ell_j) \le \alpha''\ggen_j \},
\\
\lbeq{DVh}
    \bar\DV_j(h) &=
    \{V \in \Qcal :
    \; |{\rm Im} g| < \textstyle{\frac {1}{5}} {\rm Re} g,
    \;
    \alpha \le \epsilon_{g\tau^2,j}(h_j), \;
    \epsilon_{V,j}(h_j) \le \alpha'
    \}.
\end{align}
We permit the parameters
$\alpha,\alpha' >0$
to depend on $C_\DV$,
and
$\alpha''=\alpha''_L >0$ to depend on $C_\DV, L$.  Their specific values
are of no importance.  We sometimes need versions with larger $\alpha',\alpha''$ and
smaller $\alpha$, and we denote these by $\bar\DV_j'$.  This is the case in
the following proposition, which is proved in Section~\ref{app:sp}.

\begin{prop}
\label{prop:monobd} Let $d=4$.  If $V \in \DV_j$ then
there is a choice of parameters defining the domains \refeq{DVell}--\refeq{DVh}
such that
\begin{align}
\lbeq{VDbar}
    V &\in \bar\DV_j(\ell) \cap \bar\DV_j(h)
    \quad\quad (j \le N),
\end{align}
and if $V \in \bar\DV_j(\h)$ (for $\h=h$ or $\ell$) then with a new choice of parameters
for $\bar\DV'$,
\begin{align}
\lbeq{Vptbar}
    V_{\pt,j+1}(V) &\in
    \bar\DV_{j}'(\h) \cap \bar\DV_{j+1}'(\h)
    \quad\quad (j < N)
    .
\end{align}
\end{prop}

The domain $\bar\DV_j$ is the principal domain for $V$ throughout the paper.
By Proposition~\ref{prop:monobd}, we know that $\DV_j \subset \bar\DV_j(\h_j)$
for both $\h=\ell$ and $\h=h$, so all
assertions valid for $V \in \bar\DV_j$ are valid for $V\in \DV_j$.
In particular, \refeq{VDbar} asserts that if $V \in \DV_j$, then
\begin{align}
\label{e:epVbarepV}
    \alpha
    &\le
    \epsilon_{g\tau^2,j}(h),
    \quad\quad
    \epsilon_{V,j}
    \le
    \begin{cases}
    \alpha''_L \ggen_j   & \h = \ell
    \\
    \alpha' & \h = h.
    \end{cases}
\end{align}
From \refeq{epVbarasymp} and \refeq{h-def}, we see that
\begin{equation}
\lbeq{epVbark0}
    \epVbar(h) \asymp k_0^4,
\end{equation}
where $k_0$ is the small constant in the definition of $h_j$.
From \refeq{epVbarepV} we see that the
$g\tau^2$ term dominates $V$ in the $h$-norm, in the sense that
\eq
\lbeq{tau2dom}
    \epV(h) \le \alpha'\alpha^{-1} \epVbar(h)
    .
\en
Together with the lower bound on $\epVbar(h)$, this is important in
using the $e^{-g\tau^2}$ factor in $e^{-V}$ to obtain effective
stability bounds.  A bound like \refeq{tau2dom} also holds for the
case $\h=\ell$, but with an $L$-dependent constant; this follows since
$\epV(\ell)$ and $\epVbar(\ell)$ are both of order $\ggen_j$ by
\refeq{epVbarasymp} and \refeq{VDbar}.  However in this case, since we
are interested in situations where $\ggen_j \to 0$ as $j \to \infty$,
we do not have a uniform lower bound on $\epVbar(\ell)$.

\begin{rk}
Our analysis throughout the paper rests on the estimates of
Proposition~\ref{prop:monobd} but does not depend on the particular form of the
observable terms in \refeq{Vx} and their counterparts
on the right-hand side of \refeq{monobd}.  If different observable terms were
used instead then there is no significant change in the analysis as long as the statements
of Proposition~\ref{prop:monobd} remain valid; this observation is useful in
\cite{ST-phi4}.
\end{rk}

\subsubsection{Definition of small parameter \texorpdfstring{$\epdV$}{epsilonbar}}
\label{sec:epdVdef}

An additional small parameter which is important for our analysis
is $\epdV = \epdV(\h)$, which takes on different values for
the two cases $\h=\ell$ and $\h=h$.
Recall that the sequence $\chi_j = \Omega^{-(j-\jm)_+}$ was defined
in \refeq{chidef}.
We define
\eq
\label{e:epdVdef}
    \epdV = \epdV_{j}
    =
    \begin{cases}
    \chicCov_{j}^{1/2} \ggen_{j} & \h_{j}=\ell_{j}
    \\
    \chicCov_{j}^{1/2} \ggen_{j}^{1/4}   & \h_{j}= h_{j}.
    \end{cases}
\en
In view of our assumption throughout the paper that $\ggen_j$ is small (uniformly
in $j$, and small depending on $L$),
we can assume that $\epdV$ is as small as desired
(depending on $L$).
The sequence $\chi_j$ occurring in $\epdV^2$ provides
useful exponential decay beyond the $\Omega$-scale \refeq{mass-scale}.

The small parameter $\epdV$ plays a role in many aspects of the paper.
For example, it arises as an upper bound for $W$ of
\refeq{Wwdef}--\refeq{WLTF} and for $P$ of \refeq{PWdef},
in the sense that there is an $L$-dependent constant $c_L$ such that for
$1 \le j \le N$ and $V \in \bar\DV_j$,
\begin{align}
\lbeq{WBbd}
    \max_{B \in \Bcal_j} \|W_j(V,B)\|_{T_{0,j}(\h_j)}
    &\le
    c_L \epdV^2,
    \\
\lbeq{PBbd}
    \max_{B \in \Bcal_j} \|P_j(V,B)\|_{T_{0,j}(\h_j)}
    &\le
    c_L \epdV^2.
\end{align}
The inequalities \refeq{WBbd}--\refeq{PBbd}
are proved in Proposition~\ref{prop:Wnorms} below.

\section{Main results}
\label{sec:IE}

We now state our main results.  We begin in Section~\ref{sec:stab}
with stability estimates on the interaction $I$ and a statement of the
analyticity of $I$ in the polynomial $V$.  In Section~\ref{sec:pt} we
state our results concerning the accuracy of the perturbative
calculations of \cite{BBS-rg-pt}.  Finally, in
Section~\ref{sec:scale}, we state estimates on Gaussian expectation,
and on the operator $(1-\LT)$ which extracts the irrelevant part of an
element of $\Ncal$; both of these estimates involve advancement of the
scale.  Proofs are deferred to
Sections~\ref{sec:I-estimates}--\ref{sec:ipcl}.

\subsection{Stability estimates}
\label{sec:stab}

In this section, we state stability estimates on $I$, and formulate
the analyticity of $I$ in $V$.  Proofs are given in
Section~\ref{sec:I-estimates}.

Fundamental stability bounds are given in the following proposition,
which is valid for \emph{arbitrary} choice of $\h$ in the definition
of the norms, with corresponding $\epV, \epVbar$ as defined in
Section~\ref{sec:spdefs}.  According to \refeq{WBbd}, if $V \in
\bar\DV$ then $\|W(V,B)\|_{T_0}$ (which occurs in the hypothesis) is
of order $\epdV^2$ so can be made small by the requirement that
$\ggen_j$ be uniformly sufficiently small.  Recall that the norm
$\|\phi\|_{\tilde\Phi(X)}$ was defined in \refeq{Phitilnorm}; it
appears in the last exponent in \eqref{e:Iupper-b}.  All norms in
Proposition~\ref{prop:Iupper} are at scale $j$.  The proof of
\refeq{Iupper-b} makes use of the Sobolev inequality proved in
Appendix~\ref{sec:Lp} to take advantage of the quartic decay in
$e^{-g\tau^2}$.  The restriction to $j<N$ in \refeq{Iupper-b} is
connected with the fact that we do not define the $\tilde G$ norm at
scale $N$.

\begin{prop}
\label{prop:Iupper}
Let $V \in \Qcal$ with
 $0\le |{\rm Im} g| \le \frac 12 {\rm Re}g$.
Let $j \le N$ and $B \in \Bcal_j$.
Let $\omega = \max_{B \in \Bcal_j}\|W(V,B)\|_{T_0}$ and
fix any $u \ge 6(L^{2d}\omega)^{1/3}$.
Let
$F\in \Ncal(B^\Box)$ be a polynomial of degree $r \le p_\Ncal$.
Let $I^*$ denote any one of the following choices:
\\
(a) $I_j(B)$, (b) $\tilde I_j(B)$, (c)  $\Itilde_j(B \setminus X)$ with
$X \in \Scal_{j-1}(B)$,
(d) any of (a-c) with any number of their $1+W$ factors
omitted (thus, in particular, including the case $\Ical(B)$ of \refeq{Icaldef}).
\\
(i) Then
\begin{align}
\label{e:Iupper-a}
&
    \|I^* F\|_{T_{\phi}}
    \le
     \left(\frac{2r}{u} \right)^{r}
     \| F\|_{T_{0}}
    e^{O (\epV +u)(1+ \|\phi\|_{\Phi(B^{\Box})}^2)}
    .
\end{align}
(ii)
Suppose in addition that
there is a constant $C$ such
that $\epV \le C\epVbar$.
Fix any $q \ge 0$, and
let $q_1= q +2u\epVbar^{-1}$.  Then for $j <N$,
\begin{align}
\label{e:Iupper-b}
&
    \|I^* F\|_{T_{\phi}}
    \le
     \left(\frac{2r}{u} \right)^{r}
     \| F\|_{T_{0}}
    e^{O [(1+q_1^{2})\epVbar+u]}
    e^{-q \epVbar \|\phi\|_{\Phi(B^{\Box})}^2}
    e^{O (1+q_1) \epVbar \|\phi\|_{\tilde\Phi(B^{\Box})}^2}.
\end{align}
When $r=0$, \refeq{Iupper-a}--\refeq{Iupper-b} both hold with the prefactor
$\left(\frac{2r}{u} \right)^{r}$ replaced by $1$.
\end{prop}

\smallskip \noindent
\emph{Notation.}
We write $a \prec b$ when there is a constant $c>0$, independent of
$L$ and $j$, such that $a \le c b$.  If there is an $L$-dependent such
constant, we write $a \prec_L b$.  We write $a \asymp b$ when $a \prec
b$ and $b \prec a$.

We now discuss applications of Proposition~\ref{prop:Iupper} under the
assumption that $V \in \bar\DV_j$ of \refeq{DVell}--\refeq{DVh}.

\smallskip \noindent \emph{Application of \refeq{Iupper-a}.}
Let $\h_j=\ell_j$ (defined in \eqref{e:hl}) and let $V \in
\bar\DV_j (\ell)$.  By \eqref{e:epdVdef} and \eqref{e:WBbd}, we obtain
the hypotheses for \eqref{e:Iupper-a} when $\ggen_j$ is small
uniformly in $j$.  Furthermore, $u>0$ can also be chosen small enough, independently
of $j$, so that $\exp [O(\epV +u)] \le 2$.  With these
choices, and with the fluctuation-field regulator defined by
\refeq{GPhidef}, we can restate \refeq{Iupper-a} as
\begin{align}
\label{e:Iupper-a-c}
&
    \|I^* F\|_{T_{\phi}(\ell)}
    \le
     \left(\frac{2r}{u} \right)^{r}
     \| F\|_{T_{0}(\ell)}\,
    2e^{\|\phi\|_{\Phi}^2}
    =
    \left(\frac{2r}{u} \right)^{r}
     \| F\|_{T_{0}(\ell)}\,
    2 G(B,\phi)        ,
\end{align}
again with the convention that
$(\frac{2r}{u} )^{r} =1$
when $r=0$.

\smallskip \noindent
\emph{Application of \refeq{Iupper-b}.}
We apply \refeq{Iupper-a} with the choice
$\h_j=h_j$ of \refeq{h-def}.
By \refeq{epVbark0}  and \refeq{WBbd}, for $V \in \bar\DV_j(h_j)$,
with this choice
\begin{equation}
    \epVbar \asymp k_0^4,
    \quad\quad
    \|W(B)\|_{T_0} \prec_L\; \ggen_j^{1/2}.
\end{equation}
We choose $k_{0}>0$
and take $u=\epVbar$.  Then
\begin{equation}
    (1+q_1^2)\epVbar + u = \left(2+(q+2)^2 \right)\epVbar,
    \quad\quad
    (1+q_1)\epVbar = (3+q) \epVbar.
\end{equation}
We conclude from \refeq{Iupper-b} that there is a constant $a$
such that, for $V \in \bar\DV(h_j)$,
\begin{align}
\label{e:Iupper-b-c}
&
    \|I^* F\|_{T_{\phi}(h)}
    \le
     \left(\frac{2r}{ak_0^4} \right)^{r}
     \| F\|_{T_{0}(h)}\,
    2
    e^{-q\epVbar \|\phi\|_{\Phi(B^{\Box},h)}^2}
    e^{(3+q) \epVbar  \|\phi\|_{\tilde\Phi(B^{\Box},h)}^2}
    ,
\end{align}
with the usual convention when $r=0$.  Since $h \ge \ell$, we have
$\|\phi\|_{\Phi(\h)} \le \|\phi\|_{\Phi(\ell)}$ and hence also
$\|\phi\|_{\tilde\Phi(\h)} \le \|\phi\|_{\tilde\Phi(\ell)}$.  This
allows us to conclude from \refeq{Iupper-b-c} that, for $V \in
\bar\DV(h_j)$, if $q \le \bar q$ for some fixed $\bar q >0$ then we
can choose $k_0$ small depending on $\bar q$ and $\Gtilp$ such that
\begin{align}
\label{e:Iupper-b-d}
&
    \|I^* F\|_{T_{\phi}(h)}
    \le
    \left(\frac{2r}{ak_0^4} \right)^{r}
     \| F\|_{T_{0}(h)}\,
    2
    e^{-q  ak_0^4 \|\phi\|_{\Phi(B^{\Box},h)}^2}
    \tilde{G}^{\Gtilp}(B,\phi).
\end{align}

\smallskip \noindent \emph{Vanishing at weighted infinity.}  In
\refeq{Iupper-a-c}, a stronger bound in which $G(B,\phi)$ is replaced
by a smaller power $G^{\gamma}(B,\phi)$ also holds, by the same proof.
In combination with \refeq{Iupper-b-d}, and with $\Gcal$ denoting $G$
when $\h=\ell$ and $\tilde G$ when $\h=h$, in either case this shows
that if $V \in \bar\DV$ then
\begin{equation}
\lbeq{vai}
    \lim_{\|\phi\|_{\Phi(B^\Box)} \to \infty} \|I^* F\|_{T_\phi}\Gcal(X,\phi)^{-\Gtilp} =0.
\end{equation}
This fact is useful in \cite{BS-rg-step} to establish the property used there
called ``vanishing at weighted infinity.''

\medskip The following proposition extends and reformulates the above
estimates in terms of the four norms $\|\cdot\|_j, \|\cdot\|_{j+1}$
appearing in \emph{either} of \refeq{np1}--\refeq{np2}.  However, here
and throughout the paper, as discussed in Remark~\ref{rk:scaleNnorm},
statements about the scale-$N$ norm are to be interpreted as applying
\emph{only} to the $G_{N-1}^{10}$ norm, and not also to the $\tilde G$
norm: scale-$N$ is always considered to correspond to $j+1$ and never
to $j$ in \refeq{np1}--\refeq{np2}.

\begin{prop}
\label{prop:Istab}
Let $I_*$ denote either of $I_j,\Ipttil$, with $j_*=j$ for $I_j$,
and \emph{either} $j_*=j$ or $j_*=j+1$ for $\Ipttil$.  We assume $j_* \le N$.
Alternately, let $I_*$ denote any of the above with any number of
their $1+W$ factors omitted.
Let $B\in \Bcal_j$.
Let $V \in \bar\DV_j$
and let $F\in \Ncal(B^{\Box})$ be a gauge-invariant
polynomial in the fields of degree at most $p_\Ncal$ with
$\pi_{ab}F=0$ if $j < j_{\pp \qq}$.  Then
\begin{align}
\label{e:IF}
    \|I_*(B) F\|_{j_*}
    &
    \prec
    \|F\|_{T_{0,j}}
,
\\
\label{e:Iass}
    \|I_*(B)\|_{j_*}
    &
    \le 2
,
\\
\label{e:I-b:5}
    \|I_*^{-B}\|_{T_{0,j_*} }
    &
    \le 2
    .
\end{align}
In addition, for $j+1 \le N$
and for a scale-$(j+1)$ block $\hat B \in \Bcal_{j+1}$, and
for $X$ either a small set $X \in \Scal_j$ or the empty set $X=\varnothing$,
\begin{equation}
\label{e:Iptass}
    \|\Ipttil^{\hat B\setminus X}\|_{j+1}
    \le 2
    .
\end{equation}
\end{prop}

The following proposition states our analyticity result for the interaction,
again  in terms of the four norms $\|\cdot\|_j, \|\cdot\|_{j+1}$
appearing
in \refeq{np1}--\refeq{np2}.
We show that $I$ is analytic in $V$ by proving that there is a norm-convergent
expansion of $I$ in powers of $V$.

\begin{prop}
\label{prop:Ianalytic1:5}
Let $I_*$ denote either of $I,\Ipttil$, with $j_*=j$ for $I$,
and \emph{either} $j_*=j$ or $j_*=j+1$ for $\Ipttil$.
 We assume $j_* \le N$.
Alternately, let $I_*$ denote any of the above with any number of
their $1+W$ factors omitted.
Let $B\in \Bcal_j$.
Then $I(B)$ and $\Ipttil(B)$ are analytic functions of $V \in \bar\DV_j$,
taking values in
$\Ncal(B^\Box), \|\cdot\|_{j_*}$.
In addition, $I(B)^{-1}$ is an analytic function of $V \in \bar\DV_j$ taking values in
$\Ncal(B^\Box), \|\cdot\|_{T_{0,j}}$.
\end{prop}

Recall that $\epdV$ was defined in \refeq{epdVdef}, and
that we use $\h=\ell$ for quantities related to the norm pair \refeq{np1},
and $\h=h$ for the norm pair \refeq{np2}.
The following proposition measures the effect of a change in $I$ due to a change in $V$
that is appropriately bounded by $\epdV$.

\begin{prop}
\label{prop:JCK-app-1}
Let $j<N$,
$B \in \Bcal$, $V \in \bar\DV$,  $Q \in \Qcal$ with $\|Q(B)\|_{T_0}
\prec \epdV$, and set $\Ihat = I(V-Q)$ and $I=I(V)$.  Then $V-Q \in \bar\DV'$,
$\Ihat(B)$ obeys the $I_*$ estimates of Proposition~\ref{prop:Istab}, is an
analytic function of $V\in \bar\DV$ taking values in
$\Ncal(B^\Box), \|\cdot\|_j$, and obeys the estimates
\begin{align}
\label{e:JCK1-app}
    \|\Ihat(B)- I(B)\|_j
    &\prec
    \epdV ,
\\
\label{e:JCK2-app}
    \| \Ihat(B)- I(B)(1+ Q(B) ) \|_{T_{0}  }
    &\prec_{L}
    \epdV^{2}
    .
\end{align}
All quantities and norms are at scale $j$, norms are computed with either $\h=\ell$ or
$\h=h$, and \refeq{JCK1-app} holds for
either choice of $\|\cdot \|_j$ in \refeq{np1}--\refeq{np2}.
\end{prop}

\subsection{Perturbative interaction estimates}
\label{sec:pt}

In this section, we formulate two propositions which enable a rigorous
implementation of the formal perturbative calculations
of \cite[Section~\ref{pt-sec:WPjobs}]{BBS-rg-pt}.  The two propositions
are applied in \cite[Section~\ref{step-sec:Map3estimates}]{BS-rg-step}.
Their statements
are in terms of the small parameter $\epdV$ defined in \refeq{epdVdef}.

Recall the map $\theta$ defined below \refeq{LapC}, the polynomial
$\Vpt$ defined in \refeq{Vptdef} (and above \refeq{PNdef} for scale-$N$),
and $\Ipttil$ defined in \refeq{Ipttildef}.
For $B \in \Bcal_j$ and $X \in \Pcal_j$,
we define $\delta I^X \in \Ncal (\Lambdabold \sqcup \Lambdabold')$ by
\begin{equation}
\label{e:dIdef}
    \delta I (B)
    = \theta I_j(B) - \Ipttil(B)
    =
    \theta I_j(V,B)-\Itilde_{j+1}(\Vpt,B) ,
    \quad\quad
    \delta I^X = \prod_{B \in \Bcal_j(X)}\delta I(B).
\end{equation}
For small sets $U\in \Scal_{j+1}$ we define
\begin{equation}
    \hred (U)  =
    \sum_{X \in\overline{\Pcal}_{j}(U): |X|_j \le 2}
    \Ipttil^{-X}\Ex_{j+1} \delta I^X,
\label{e:hred-def}
\end{equation}
where
$|X|_j$ denotes the number of scale-$j$ blocks in $X$, and
$X \in \overline\Pcal_j(U)$ indicates the restriction that $U$ is
the smallest polymer in $\Pcal_{j+1}$ that contains $X$.
The subscript ``red'' indicates that $h$ is ``reduced''
by the restriction $|X|_j\le 2$. (In \cite[\eqref{step-e:hdef}]{BS-rg-step} we
define a version without this restriction.)

For \cite{BS-rg-step}, we need to compute $\hred$
accurately to second order in $\epdV$.
For this, we first recall from \eqref{e:trun-exp-eg} the definition of
the truncated expectation
\begin{equation}
\label{e:trun-exp}
    \Ex_C (A; B) = \Ex_C(AB) - (\Ex_CA)(\Ex_CB).
\end{equation}
We also define (cf.\ \eqref{e:Fpi})
\begin{align}
\label{e:Epi}
    \Ex_{\pi ,C} (A;B)
    &=
    \Ex_C ( A;\pi_{\varnothing}B)
    +
    \Ex_C ( \pi_* A; B)
    .
\end{align}
Then for $(U,B) \in \Scal_{j+1}\times \Bcal_{j+1}$
we define $\hldg (U,B)$
by
\begin{equation}
\label{e:hptdefqq}
    \hldg (U,B)
    =
    \begin{cases}
    -\frac{1}{2}\Ex_{\pi ,j+1} \theta ( V_j(B); V_j(\Lambda \setminus B))
    & U=B
    \\
    \;\;\;
    \frac{1}{2}
    \Ex_{\pi ,j+1}\theta  ( V_j(B);  V_j(U\setminus B)) & U \supset B, |U|_{j+1}=2
    \\
    \;\;\;
    0 &\text{otherwise}
    ,
    \end{cases}
\end{equation}
where we have abbreviated the subscript $C_{j+1}$ to $j+1$ on $\Ex$.
For $U \in \Pcal_{j+1}$ we define
\begin{equation}\label{e:hldgUdef}
    \hldg(U)
    =
    \sum_{B \in \Bcal_{j+1} (U)} \hldg (U,B)
    .
\end{equation}
Due to the finite-range property
\eqref{e:fin-range},
\begin{equation}
    \sum_{U \supset B : U \neq B} \hldg(U,B)  =
    \frac{1}{2} \Ex_{\pi ,j+1}\theta
    (V_{j}(B) ; V_{j}(\Lambda \setminus B)) ,
\end{equation}
and therefore $\hldg$ obeys the identity
\begin{equation}
\label{e:hpt0bis}
    \sum_{U \supset B} \hldg (U,B) = 0.
\end{equation}

The following two propositions, which are proved in
Section~\ref{sec:interaction-estimates444}, show that $\hldg$ is second order in
$\epdV$, and that $\hldg(U)$ is the leading part of $\hred(U)$.  The
latter is a much more substantial result than the former, and is our implementation
of the formal power series statement of
\cite[Proposition~\ref{pt-prop:I-action}]{BBS-rg-pt}.

\begin{prop}
\label{prop:hldg}
There is a positive constant $\cldg = \cldg(L)$
such that for $j<N$, $V \in \bar\DV_j$ and $(U,B) \in \Scal_{j+1}\times \Bcal_{j+1}$,
\begin{align}
\label{e:want1}
    \|\Ipttil (U) \hldg (U,B)\|_{j+1}
    &
    \le \cldg \epdV^2
    ,
\end{align}
where  $\|\cdot \|_{j+1}$ represents either
of the two options \refeq{np1}--\refeq{np2}, with corresponding
$\epdV$ of \refeq{epdVdef}.
\end{prop}

\begin{prop}
\label{prop:h}
There is a positive constant $c_{\pt}=c_{\pt}(L)$
such that for $j<N$, $V \in \bar\DV_j$ and
$U \in\Scal_{j+1}$ with $|U|_{j+1}\in \{1,2\}$,
\begin{align}
\label{e:want2}
    \|\Ipttil (U) [\hred(U)-\hldg (U)]\|_{j+1}
    &
    \le
    c_{\pt} \epdV^{3}
    ,
\end{align}
where $\|\cdot \|_{j+1}$ represents either
of the two options \refeq{np1}--\refeq{np2}, with corresponding
$\epdV$ of \refeq{epdVdef}.
\end{prop}

\subsection{Bound on expectation and crucial contraction}
\label{sec:scale}

The next two propositions
play a key role in our analysis of a single
renormalisation group step in \cite[Section~\ref{step-sec:Map3estimates}]{BS-rg-step}.

\begin{prop}
\label{prop:ip}
There is an $\Econst >0$ (independent of $L$) and a $C_{\delta V}>0$
(depending on $L$) such that
for $j<N$, $V \in \bar\DV_j$, disjoint $X,Y \in \Pcal_j$, and for $F(Y) \in \Ncal(Y^\Box)$,
\begin{equation}
\label{e:integration-property}
    \|\Ex_{j+1} \delta I^X \theta F(Y)  \|_{j+1}
\le
    \Econst^{|X|_j+|Y|_j} (C_{\delta V}\epdV)^{|X|_j}\|F(Y)\|_j,
\end{equation}
where the pair of norms is given by
either of \eqref{e:np1} or \eqref{e:np2}
with corresponding $\epdV$  of \refeq{epdVdef}.
\end{prop}

The proof of Proposition~\ref{prop:ip} is given in
Section~\ref{sec:ippf}.
We refer to the important inequality \eqref{e:integration-property}
as the \emph{integration property}.
It shows that when estimating the
scale-$(j+1)$ norm of an expectation of
a product involving factors of $\delta I(B)$ for scale-$j$ blocks,
each factor gives rise to a small factor $\epdV$.

In the next proposition, the notation $U = \overline X$ again
indicates the restriction that $U$ is the smallest polymer in
$\Pcal_{j+1}$ that contains $X$.  As in
\cite[Definition~\ref{loc-def:LTXYsym}]{BS-rg-loc}, we use the
notation $X(\varnothing)=X$, $X(a) = X \cap \{a\}$, $X(b)=X \cap
\{b\}$, and $X(ab) = X \cap\{a,b\}$.  Given $X \subset \Lambda$, we
define
\begin{equation}
\label{e:gamLdef}
    \gamma(X)
    =
    L^{-d -1} +  L^{-1}\1_{X \cap \{a,b\} \not = \varnothing}.
\end{equation}

\begin{prop}
\label{prop:cl} Let $j<N$ and $V\in \DV_j$.  Let $X \in \Scal_j$ and
$U = \overline X$.  Let $F(X) \in \Ncal(X^\Box)$ be such that
$\pi_\alpha F(X) =0$ when $X(\alpha)=\varnothing$, and such that
$\pi_{ab}F(X)=0$ unless $j \ge j_{ab}$ (recall \eqref{e:Phi-def-jc}).
Then
\begin{align}
    \label{e:contraction3z-new}
    \|\Ipttil^{U\setminus X} \Ex_{C_{j+1}} \theta F (X) \|_{j+1}
    &
    \prec
    \cgam(X)
    \kappa_F
    + \kappa_{\LT F}
    ,
\end{align}
with $\kappa_F=\|F (X)\|_{j}$ and
$\kappa_{\LT F} =\|\Ipttil^X \LT_X \Ipttil^{-X} F(X) \|_j$, and
where the pair of norms is given by
either of \eqref{e:np1} or \eqref{e:np2}.
\end{prop}

The proof of the Proposition~\ref{prop:cl} is given in
Section~\ref{sec:contraction3-proof}.
We refer to the inequality \refeq{contraction3z-new}
as the \emph{crucial contraction}; its importance is discussed  in
Section~\ref{sec:cc} above.

\section{Estimates on small parameters}
\label{app:sp}

In this section, we provide estimates on the small parameters
$\epV,\epdV$ which drive our analysis.  In particular,
we prove Proposition~\ref{prop:monobd}.

\subsection{Preliminaries}

We begin with two general lemmas.  The first relates $\epsilon_{V,j}$ to
$\|V(B)\|_{T_{0,j}}$ for a scale-$j$ block $B \in \Bcal_j$, and the second expresses
an important monotonicity property of the $T_\phi$ semi-norm under change of scale.
Recall from \cite[\eqref{loc-e:Fnormsum}]{BS-rg-loc}
that it follows from the definition of the $T_\phi$
semi-norm that under the direct sum decomposition of $F \in \Ncal$ due to \refeq{Ncaldecomp},
\eqalign
\lbeq{Fnormsum}
    \|F\|_{T_\phi}
    &=
    \sum_{\alpha\in \varnothing,a,b,ab}\|\pi_\alpha F  \|_{T_\phi}
    =
    \|F_\varnothing\|_{T_\phi}
    +
    \h_\sigma\|F_a\|_{T_\phi}
    +
    \h_\sigma\|F_b\|_{T_\phi}
    +
    \h_\sigma^2 \|F_{ab}\|_{T_\phi}
    .
\enalign

\subsubsection{The \texorpdfstring{$T_0$}{T0} semi-norm and \texorpdfstring{$\epV$}{epsilonV}}

\begin{lemma}
\label{lem:T0ep}
For $V \in \Qcal$ and $j<N$, $\epsilon_{V,j} \asymp \max_{B \in \Bcal_j} \|V(B)\|_{T_{0,j}}$.
\end{lemma}

\begin{proof}
Given $V \in \Qcalnabla$, as in \refeq{monobd} we write $V_\varnothing =
\sum_{M}  M$ for the decomposition of its bulk part
as a sum of individual field
monomials such as $\nu \phi\bar\phi$, $\nu \psi\bar\psi$, $z(\Delta
\phi)\bar\phi$, and so on.  For $0 \le j \le N$,
in \refeq{monobd} we defined
\begin{equation}
\label{e:monobd-bis}
    \epV
    =
    L^{dj} \sum_{M : \pi_*M=0} \|M_0\|_{T_{0,j}(\h_j)}
    + (|\lambdaa|+|\lambdab|)\h_j\h_{\sigma,j} + (|\qa|+|\qb|)\h_{\sigma,j}^2
     .
\end{equation}
By direct calculation, $\|\lambda_\pp
\bar\phi_{x}\|_{T_0}=\1_{x=\pp}|\lambda^a|\h$, $\|\lambda_\qq
\phi_{x}\|_{T_0} = \1_{x=\qq}|\lambdab|\h$, and $|q_{\pp\qq}| =
\frac{1}{2}(|\qa|\1_{x=\pp}+|\qb|\1_{x=\qq})$.  Thus the last two
terms on the right-hand side of \refeq{monobd-bis} are
bounded above and below by multiples of
$\max_{B \in \Bcal_j}\|\pi_*V(B)\|_{T_0}$, and it suffices to consider
the case $V=\pi_\varnothing V$, which we assume henceforth.

It follows from the
triangle inequality that
$\|V(B)\|_{T_0} \prec \epV$, and it suffices to prove the reverse inequality.
Let $M$ be a scalar multiple of one
of $g\phi\bar\phi\phi\bar\phi, g\phi\bar\phi\psi\bar\psi, \nu\phi\bar\phi,
\ldots$.
It suffices to prove that
\begin{equation}
\lbeq{M0V}
    \|M_0\|_{T_0}|B| \prec \|V(B)\|_{T_0}.
\end{equation}
For this, we employ the pairing of \cite[Definition~\ref{norm-def:Tphi-norm}]{BS-rg-norm},
and seek dual test functions for the monomials.
In more detail, given a monomial $M$ we seek a unit $\Phi$-norm test function $f_M$ such
that, for all $x\in B$, $\pair{M_x,f_M}=\|M_x\|_{T_0}$
but $\pair{M'_x,f_M} = 0$ if $M' \neq M$.
It then follows
that
\begin{align}
\lbeq{T0epid}
    \|M_{0}\|_{T_{0}} & =
    |\pair{ M_{x},f_M}_{0}|
=
    \frac{1}{|B|}
    |\pair{M (B),f_M}_{0}|
=
    \frac{1}{|B|}
    |\pair{V (B),f_M}_{0}|
\le
    \frac{1}{|B|}
    \|V (B)\|_{T_{0}},
\end{align}
which is equivalent to the desired estimate for this monomial.

For the existence of $f_M$, we proceed as follows
(cf.\ \cite[Lemma~\ref{loc-lem:dualbasis}]{BS-rg-loc} for related ideas).
Consider first the case $M=g\phi\bar\phi\phi\bar\phi$.  We choose $f_M$ to be
zero on all sequences except those
of length four whose components are in the $\phi, \phib, \phi, \phib$
sheets, and choose it to be constant on the set of these sequences,
with the constant such that $f_M$ has unit norm.  This choice can be seen to
have the desired
properties, and it generalises in a straightforward way to all the
monomials arising from $g\tau^2$ and $\nu\tau$.

Next, we consider $M = \frac 12 \sum_{e \in \units}(\nabla^e\phi)(\nabla^e\bar\phi)$
(the coupling constant plays an insignificant role so we omit it for simplicity).
By translation invariance, we may assume that $B$ is centred at $0\in \Lambda$,
and since $j<N$ we can identify $B$ with a subset of $\Zd$.
Let $v_{x_{1},x_{2}}=  x_{1}\cdot
x_{2} + c$ for $x_{1}$ in the $\phi$ sheet and $x_{2}$
in the $\phib$ sheet.
Let $M'=\phi\bar\phi$. Then the pairing of $v$ with any monomial other than $M,M'$
vanishes.
In particular, $\pair{M_x,v}_0=
\frac 12 \sum_{e \in \units}
\nabla_{x_1}^e \nabla_{x_2}^e v_{x,x} = d$.
Also,  $\pair{M_{x}',v}_{0}
\asymp x\cdot x + c$, and by choosing $c\asymp L^{2j}$
such that $\sum_{x \in B}(x\cdot x + c)=0$, we can arrange that $\pair{M'(B),v}_0=0$.
Let $f = v/\|v\|_{\Phi}$.
Then we have $\pair{V(B),f}_0=\pair{M(B),f}_0 = |B|\pair{M_0,f}_0$ and we obtain
\refeq{M0V} in this case, as in \refeq{T0epid}.

The case $M = \bar\phi \Delta\phi$ is similar, with
the test function constructed from $v_{x_{1},x_{2}} = x_{1}\cdot
x_{1} + c$.
This completes the proof.
\end{proof}

\subsubsection{Scale monotonicity}

We now prove a monotonicity property of the $T_\phi$ semi-norm
under change of scale, which is used repeatedly throughout the paper.
The property is more general than our specific application,
and we formulate it under assumptions on $\h=(\h_\phi,\h_\sigma)$ obeyed
by our particular choices.
In our application, \refeq{hprod-bis} with $\h'=\h$
follows from the last two bounds of \refeq{h-assumptions}.

\begin{lemma}
\label{lem:Imono}
Suppose that $F \in \Ncal$ is gauge invariant and such
that $\pi_{ab}F=0$ when $j<j_{ab}$, that
$\h_{\phi,j}'' \le \h_{\phi,j}' \prec L^{-[\phi]}\h_{\phi,j-1}$,
that $\h_{\sigma,j}'' \prec \h_{\sigma,j}'$,
and
that for all $j$,
\begin{equation}
\label{e:hprod-bis}
    \h_{\sigma,j}' \prec
    \begin{cases}
    L^{[\phi]}\h_{\sigma,j-1} & j < j_{ab}
    \\
    \h_{\sigma,j-1} & j \ge j_{ab},
    \end{cases}
    \quad\quad
    \h_{\sigma,j+1}'\h_{\phi,j+1}'
    \prec
    \h_{\sigma,j}\h_{\phi,j}.
\end{equation}
Then, for $L$ large depending on the constant in $\prec$ in the
hypothesis,
\begin{equation}
    \label{e:scale-change}
    \|F\|_{T_{\phi,j}(\h_{j}'')}
    \prec
    \|F\|_{T_{\phi,j}(\h_j') }
    \prec
    \|F\|_{T_{\phi,j-1}(\h_{j-1}) }
.
\end{equation}
\end{lemma}

\begin{proof}
By \refeq{Fnormsum} it suffices to prove that for each $\alpha$,
$\|\pi_\alpha F\|_{T_\phi}$ individually obeys \refeq{scale-change}.

Case $\alpha = \varnothing$.
By definition of the norm on test functions (recall
\cite[Example~\ref{norm-ex:h}]{BS-rg-norm}),
for a test function $g$
with none of its variables corresponding to observable sheets,
\begin{equation}
    \label{e:g-norms}
    \|g\|_{\Phi_{j-1}(\h_{j-1})}
    \le
    \|g\|_{\Phi_{j}(\h_j')}
    \le \|g\|_{\Phi_{j}(\h_j'')} ,
\end{equation}
provided $L$ is chosen large so that the hypothesis
$\h_{\phi,j}' \prec L^{-[\phi]}\h_{\phi,j-1}$ implies that
$\h_{\phi,j}' \le \h_{\phi,j-1}$.  As a direct consequence of the
definition $\|F\|_{T_\phi} = \sup_{\|g\|_\Phi \le
1}|\pair{F,g}_\phi|$ of the $T_\phi$ semi-norm in
\cite[Definition~\ref{norm-def:Tphi-norm}]{BS-rg-norm}, we obtain
\eqref{e:scale-change} with $F$ replaced by $\pi_{\varnothing}F$ as
was to be proved.  In fact for this case we obtain the stronger
inequality with $\prec$ replaced by $\le$.

Case $\alpha = \pp\qq$. By \refeq{Fnormsum} $\|\pi_{ab}
F\|_{T_\phi} = \h_\sigma^2 \|F_{ab}\|_{T_\phi}$, so the first
inequality of \refeq{scale-change} follows immediately from the
hypothesis $\h_{\sigma,j}'' \prec \h_{\sigma,j}'$ and case $\alpha
=\varnothing$.  Likewise the second inequality for $j \ge j_{ab}$
follows from the hypothesis $\h_{\sigma,j}' \prec \h_{\sigma,j-1}$.
The second inequality for $j<j_{ab}$ is vacuous because, by
hypothesis, $\pi_{ab}F=0$ for $j<j_{ab}$.

Cases $\alpha=\pp$ or $\alpha=\qq$.
These are similar, and we consider only $\alpha =\pp$.  The fact that
$\pi_\pp F = \sigma F_{\pp}$ is gauge invariant implies that
its pairing with a test function $g$ is zero unless exactly one
argument of $g$ has species $\sigma$ and at least one other argument
has species $\phi$ or $\psi$.  Therefore, for gauge invariant $F$, we
can refine \cite[Definition~\ref{norm-def:Tphi-norm}]{BS-rg-norm} by
restricting the supremum to unit norm test functions with this support
property.  By the second inequality of \refeq{hprod-bis} test
functions with this support property satisfy \eqref{e:g-norms} with
$\prec$ in place of $\le$. The constants in $\prec$ must be
independent of $j$, and they are because there is only one $\sigma$ and
$L$ is large. This implies \eqref{e:scale-change} for
case $\alpha =\pp$ and completes the proof.
\end{proof}

\subsection{The small parameter \texorpdfstring{$\epV$}{epsilonV}: Proof of Proposition~\ref{prop:monobd}}
\label{sec:epVW}

\begin{proof}[Proof of Proposition~\ref{prop:monobd}.]
It suffices to prove that:
\\
(i)
For $j \le N$ and $V \in \DV_j$, there exist $a,A >0$
(depending on $C_\DV$)
and
$A_L >0$
(depending on $C_\DV, L$) such that
\begin{align}
\label{e:epVbardefz-app}
    |{\rm Im} g| < \textstyle{\frac {1}{5}} {\rm Re} g,
\quad\quad
    a k_0^4
    &\le
    \epsilon_{g\tau^2,j}(h_j),
\quad\quad
    \epsilon_{V,j}
    \le
    \begin{cases}
    A_L \ggen_j    & \h = \ell
    \\
    A k_0 & \h = h.
    \end{cases}
\end{align}
(ii)
For $j < N$ and $V \in \bar\DV_j$,
the bounds \refeq{epVbardefz-app}
hold (with different constants) when $V$ is replaced by
$V_{\pt,j+1}$ (and $g$ by $\gpt$), and also when $j$ is replaced by $j+1$.

\smallskip\noindent
We prove the above two statements in sequence.
\\
(i)
For $j \le N$ and $V \in \DV_j$, the coupling constants obey
\begin{equation}
    C_{\DV}^{-1} \ggen_j < {\rm Re} g < C_{\DV} \ggen_j,
    \quad |{\rm Im} g| <
    \textstyle{\frac {1}{10}}{\rm Re} g
    <
    \textstyle{\frac {1}{5}} \ggen_j, \quad
\end{equation}
\begin{gather}
\label{e:ccbds}
    L^{2j}|\nu|,|z| , |y| \le C_{\DV} \ggen_j,
    \quad
    |\lambda| \le
    C_{\DV}
    ,
    \quad
    L^{2j_{ab}[\phi]}2^{2(j-j_{ab})}|q|\le \begin{cases}
        0 & j < j_{ab} \\
        C_{\DV}   & j \ge j_{ab},
        \end{cases}
\end{gather}
for $\lambda$ equal to $\lambdaa$ or $\lambdab$ and similarly for $q$.
The first inequality of \eqref{e:epVbardefz-app} holds by definition.
As noted in \refeq{epVbarasymp},
\begin{equation}
\label{e:epVbardef-h-i}
    \epsilon_{g\tau^2,j}(\h_j) \asymp L^{dj} |g| \h_j^4 .
\end{equation}
In particular, since $|g|\asymp \ggen_j $ by hypothesis,
\begin{equation}
\label{e:epVbardef-h}
    \epsilon_{g\tau^2,j}(h_j) \asymp L^{dj} |g| h_j^4  \asymp k_0^4  ,
\end{equation}
which proves the second bound of \refeq{epVbardefz-app}.
The last bound of \refeq{epVbardefz-app} for the bulk part
$V_\varnothing$ of $V$
similarly
follows from direct calculation as in
\cite[Proposition~\ref{norm-prop:taunorm}]{BS-rg-norm};
e.g., $\|\phi_x\bar\phi_x\|_{T_{0,i}} = \h_i^2$,
$\|\phi_x\bar\phi_x\phi_x\bar\phi_x\|_{T_{0,i}} = \h_i^4$,
$\|\phi_x\Delta \bar\phi_x\|_{T_{0,i}} = L^{-2i} \h_i^2$,
while the observables contribute
\begin{align}
\lbeq{hsigh}
    |\lambda| \h_i \h_{\sigma,i}
    &=
    |\lambda| \times
    \begin{cases}
    \ggen_i \ell_0
    (2/L)^{(i-j_{ab})_+ }
    & \h =\ell
    \\
    k_0
    (2/L)^{(i-j_{ab})_+ }
    & \h = h
    ,
    \end{cases}
\\
\lbeq{qhsig}
    |q|\h_{\sigma,i}^2 & \prec
    \begin{cases}
     \ggen_i^2 & \h=\ell
    \\
    \ggen_i^{1/2} & \h=h
    \end{cases}
\end{align}
(for \refeq{qhsig} we can restrict to $i \ge j_{ab}$ since otherwise
$q=0$).  The combination of these bounds completes the proof of \refeq{epVbardefz-app},
after taking into account that $\ell_0$ depends on $L$ and $k_0^4 \le k_0$.

\smallskip \noindent
(ii)
Let $V \in \bar\DV_j$.
We first consider the case $j+1<N$, $n=j$ of \refeq{Vptbar}.
By \refeq{VptE},
$\Vpt =V +  2gC_{0,0}\tau - P$, with $C=C_{j+1}$ and $P=P_j$.
By \refeq{scaling-estimate-Omega},
\begin{equation}
    \|2gC_{0,0}\tau_x\|_{T_{0,j}} \prec  |g|L^{-2j}\h_j^2   .
\end{equation}
By \refeq{monobd}, Lemma~\ref{lem:T0ep}, and \refeq{PBbd},
\begin{align}
\lbeq{epVptbd}
    \epsilon_{\Vpt}
    &
    \prec \epV + |g|L^{2j} \h_j^2 + \epsilon_P
    \prec
    \epV + |g|L^{2j}\h_j^2 + \max_{B \in \Bcal_j}\|P(B)\|_{T_{0,j}}
    \nnb &
    \prec
    \epV + |g|L^{2j}\h_j^2 + O_L(\epdV^2).
\end{align}
With the definition of $\h_j$ in \refeq{hl}--\refeq{h-def},
this shows that $\epsilon_{\Vpt}$ obeys the last
bound of \refeq{epVbardefz-app}.
For the second bound of \refeq{epVbardefz-app},
we restrict to $\h=h$, and note that the lower bound
follows from the lower bound on the $\tau^2$ term of $V$,
together with the fact that the contribution to $\tau^2$ from $P$ is
bounded above by $\epsilon_P \prec_L \ggen_j^{1/2}$.
Finally, for the bound on the imaginary part of $\gpt$ we use the
fact that it changes insignificantly from the imaginary part of $g$,
since the coupling constant $g_P$ of $P$ obeys $|g_P| \prec
\epsilon_{g_p\tau^2,j}(\ell_j) \le \epsilon_P(\ell_j) \prec_L \ggen_j^2$
(the first of these inequalities follows from \refeq{epVbardef-h-i}).

For the case $j+1=N$, $n=j$ of \refeq{Vptbar},
we simply observe that our definition of $V_{\pt,N}$
is identical to what it would be on a torus of scale larger than $N$, so the
bound in this case follows from the above argument applied to the torus
of scale $N+1$.

For the $n=j+1$ case of \refeq{Vptbar}, note that
the computations in the proof of (i) lead to the same conclusion when $\h_j$ is
replaced by $\h_{j+1}$ and $|B|=L^{dj}$ is replaced by $L^{d(j+1)}$, and since
$\ggen_{j+1}\asymp \ggen_j$ by \refeq{gbarmono},
we conclude that
$V \in \bar\DV_{j+1}(\ell_{j+1}) \cap \bar\DV_{j+1}(h_j)$
(with adjusted constants).
The desired result then follows exactly as in the proof of (ii),
now with
\refeq{PBbd} applied at scale-$(j+1)$.
This completes the proof.
\end{proof}

\begin{rk}
\label{rk:hsigmot} The choice of $\h_\sigma$ in \refeq{newhsig} can be
motivated as follows; we discuss this for the case $\h=\ell$.  Since
the norm gives a better bound on the observables when $\h_\sigma$ is
chosen large, as a first attempt it would be natural to choose
$\ell_\sigma$ as large as possible to make the norm of
$\lambdaa\sigma\bar\phi$ agree with (or be bounded by) that of
$g\tau^{2}$ on a block, namely $\ggen_j\ell_j^4
L^{dj}=\ggen_j\ell_0^4$.  The coupling constant $\lambdaa$ is $O(1)$.
The $T_0$ norm of $\sigma\bar\phi$ is $\ell_\sigma \ell$, and to make
this no larger than the norm of $g\tau^{2}$ a block, we could choose
$\ell_{\sigma,j} = \ggen_j L^{[\phi]j}$.  In addition, our choice of
$\ell_\sigma$ must also be appropriate for the $\sigma\bar\sigma$ term
which arises in $\Vpt$.  Our procedure is to take $\qa=\qb=0$ in $V$.
Thus, according to the flow of $q$ given in
\cite[\eqref{pt-e:qpt2}]{BBS-rg-pt}, the $\sigma\bar\sigma$ term in
$\Vpt$ is given by the increment $\lambdaa\lambdab
C_{j+1;a,b}\sigma\bar\sigma$ (which is only nonzero above the
coalescence scale $j_{ab}$).  According to \refeq{scaling-estimate},
with the above choice of $\ell_\sigma$ the norm of this term is of
order $L^{-2[\phi]j} \ell_\sigma^2 = \ggen_j^2$, and this is
significantly smaller than the norm of the $\lambdaa \sigma \bar\phi$
term (which is good).  However, a disadvantage of the choice
$\ell_{\sigma,j} = \ggen_j L^{[\phi]j}$ is that it would make the
monomial $\sigma\bar\sigma\phi\bar\phi$ be marginal (scale invariant),
hence in the range of $\LT$ and thus in $\Vpt$.  This monomial only
appears after the coalescence scale, and we would prefer it to be
irrelevant.  To achieve this, we decrease the size of $\ell_\sigma$ to
the choice $\ell_\sigma = \ggen_j 2^{(j-j_{ab})_+}L^{[\phi](j \wedge
j_{ab})}$ made in \refeq{newhsig}.  Then $\ell_\sigma$ grows as a
power of $L$ below the coalescence scale, but only by a power of $2$
above the coalescence scale.  This power of 2 plays a role in the
proof of \cite[Theorem~\ref{saw4-thm:wsaw4}]{BBS-saw4}.
\end{rk}

\subsection{The small parameter \texorpdfstring{$\epdV$}{epsilonbar}}
\label{sec:epdV-app}

For $j <N$, we define
\begin{equation}
\lbeq{ellhatdef}
    \hat \ell_j^2 = \hat\ell_0^2 \ell_j^2 \|C_{(j+1)*}\|_{\Phi_{j}^+(\ell_{j})},
\end{equation}
with $C_{k*}$ defined by \refeq{Ckstardef}.  We choose $\hat\ell_0^2 =
100/ c_G$, where $c_G=c(\alpha_G)$ is the constant of
\cite[Proposition~\ref{norm-prop:EG2}]{BS-rg-norm} (this choice is
useful in the proof of Lemma~\ref{lem:dIipV} below), so that
\begin{equation}
\lbeq{ellhatdef-1}
    \|C_{(j+1)*}\|_{\Phi_{j}^+(\hat\ell_{j})}
    =
    \|C_{(j+1)*}\|_{\Phi_{j}^+(\ell_{j})}
    \frac{\ell_j^2}{\hat\ell_j^2}
    =
    \hat\ell_0^{-2}
    =
    \frac{1}{100} c_G.
\end{equation}
Below the $\Omega$-scale defined by \refeq{mass-scale}, $\hat\ell_j$ and $\ell_j$
are of the same order of magnitude, but
well above the $\Omega$-scale $\hat\ell_j \ll \ell_j$.
We use $\hat\ell_j$ in estimates involving integration, as a parameter which captures
the size of the covariance effectively.

Let
\begin{equation}
\label{e:dVdef}
    \delta V = \theta V - \Vpt
    =
    \theta V - V_{\pt,j+1}(V).
\end{equation}
Recall the definition of
$\epdV$ from \refeq{epdVdef}.
The following lemma justifies the notation used for $\epdV$, by showing
that it provides an upper bound for $\delta V$.
Its restriction to $j<N$ is to keep $\delta V$ defined in \refeq{dVdef}.

\begin{lemma}
\label{lem:epdV}
Let $j<N$.
There is an $L$-dependent constant $C_{\delta V}$ such that
for all $V \in \bar\DV_j$, and for $j_*=j$ or $j_*=j+1$,
\begin{equation}
\label{e:dVbd}
    \max_{b \in \Bcal_j}
    \|\delta V(b)\|_{T_{0,j_*}(\h_{j_*} \sqcup \hat\ell_{j_*})}
    \le
    C_{\delta V} \epdV.
\end{equation}
\end{lemma}

\begin{proof}
We fix $j<N$, concentrate first on the case $j_*=j$, and
drop subscripts $j$.
We show that for $V \in \Qcalnabla$ and $b \in \Bcal_j$,
\begin{equation}
\label{e:dVbdz-old}
    \|\delta V(b)\|_{T_{0}(\h \sqcup \hat\ell)}
    \prec_L
    \frac{\hat\ell}{\h} \epsilon_{V}
    + \epdV^2  .
\end{equation}
This suffices, since (using \refeq{CLbd}) the first term on the
right-hand side of \refeq{dVbdz-old} obeys
\begin{equation}
    \frac{\hat\ell}{\h} \epsilon_{V}
    =
    \|C\|_{\Phi^+(\ell)}^{1/2}
    \frac{\ell}{\h} \epsilon_{V}
    \prec
    \chicCov_{j}^{1/2}
    \frac{\ell}{\h} \epsilon_{V}
    \prec_L
    \begin{cases}
    \chicCov_{j}^{1/2}\ggen = \epdV(\ell) & \h=\ell
\\
    \chicCov_{j}^{1/2}\ggen^{1/4} = \epdV(h)
     & \h=h.
    \end{cases}
\end{equation}
This gives \refeq{dVbd} and reduces the proof to showing \refeq{dVbdz-old}.

We now prove \refeq{dVbdz-old}.
By \refeq{VptE} and \refeq{PNdef}, with $C=C_{j+1}$,
\begin{equation}
    \Vpt -V=   2gC_{0,0}\tau - P.
\end{equation}
Therefore, by definition of $\delta V$ in \refeq{dVdef} and by the triangle
inequality,
\begin{align}
    \|\delta V(b)\|_{T_{0}(\h \sqcup \hat\ell)}
    & \le \|\theta V(b) - V(b)\|_{T_{0}(\h\sqcup \hat\ell)}
    +
    \|V(b)-\Vpt (b)\|_{T_{0}(\h)}
    \nnb & \le
    \|\theta V(b) - V(b)\|_{T_{0}(\h \sqcup \hat\ell)}
    +
    \|C\|_{\Phi(\h)} \h^2 \|2g \tau(b)\|_{T_0(\h)}
    + \|P(b)\|_{T_{0}(\h)}
    .
\label{e:dVbd1}
\end{align}
For the first term on the right-hand side of \refeq{dVbd1}, we use the
triangle inequality to work term by term in the monomials in $V$.  For
example, the $\tau$ term makes a contribution
\begin{equation}
     \|\nu(\theta \tau(b) - \tau(b))\|_{T_{0}(\h \sqcup \hat\ell)}.
\end{equation}
After expansion in the fluctuation fields $\xi,\bar\xi,\eta,\bar\eta$,
the difference $\theta \tau(b) - \tau(b)$ is given by a sum of
products of fluctuation fields and $\phi,\bar\phi,\psi,\bar\psi$
fields, with each term containing two fields of which at least one is
a fluctuation field.  Thus it is bounded by $O(\hat\ell \h)$.  The end
result is a bound on $\|\theta V(b) - V(b)\|_{T_{0}(\h \sqcup \hat\ell)}$
equal to $\hat\ell/\h$ times the $T_0(\h)$ semi-norm of the worst monomial in
$V$ (but without the $\sigma\bar\sigma$ term which cancels).  This
gives
\begin{equation}
    \label{e:epsthetaV}
    \|\theta V(b) - V(b)\|_{T_{0}(\h \sqcup \hat\ell)}
    \prec
    \frac{\hat\ell}{\h} \gh.
\end{equation}
For the second term on the right-hand side of \refeq{dVbd1},
\begin{equation}
    \|C\|_\Phi \h^2 \|2g \tau(b)\|_{T_0}
    \prec   \|C\|_{\Phi(\h)}  \epV
    =
     \frac{\hat\ell^2}{\h^2}  \|C\|_{\Phi(\hat\ell)} \gh
     .
\end{equation}
For the last term, we use \refeq{PBbd} to obtain
\begin{equation}
    \|P(b)\|_{T_0} \prec_L
     \epdV^2
    .
\end{equation}
The combination of the last three inequalities gives \refeq{dVbdz-old}
and the proof for the case $j_*=j$ is complete.

Finally, for the case $j_*=j+1$, we start with the first line of \refeq{dVbd1}
with norms at scale $j+1$.  The norm of $V-\Vpt$ is bounded by its scale-$j$
counterpart, by Lemma~\ref{lem:Imono}.  In addition, \refeq{epsthetaV} applies also at
scale $j+1$, and this give the desired conclusion and completes the proof.
\end{proof}

\section{Estimates on field polynomials}
\label{sec:W}

In this section, we prove
the following proposition, which gives our
main estimates on the field polynomials $F,W,P$.
As usual, $\epdV$ depends on whether $\h=\ell$ or $\h=h$,
as indicated in \refeq{epdVdef}.
Recall that $P_j$ is defined for $0 \le j <N$, so there is no bound
missing in \refeq{epP}.

\begin{prop}
\label{prop:Wnorms}
For $L$ sufficiently large and
$V \in \bar\DV_j$,
\begin{align}
\label{e:Fepbd}
    \max_{B \in \Bcal_{j}}
    \sum_{x \in B}
    \sum_{B' \in \Bcal_j(\Lambda)}
    \|F_{\pi,C_{j*}}(V_x,V(B'))\|_{T_{0,j}(\h_{j})}
    &\prec_L \epdV_{j}^2
    \quad (j \le N)
    ,
\\
\label{e:Wbomega}
    \max_{B \in \Bcal_{j}}
    \sum_{x\in B}
    \|W_j(V,x)\|_{T_{0,j}(\h_j)}
    &\prec_L \epdV_{j}^2
    \quad (j \le N)
    ,
\\
\label{e:epP}
    \max_{B \in \Bcal_{j}}
    \sum_{x\in B}
    \|P_j(V,x)\|_{T_{0,j}(\h_{j})}
    &\prec_L \epdV_{j}^2
    \quad (j < N)
    .
\end{align}
\end{prop}

\begin{rk}
\label{rk:sm}
\emph{Scale mismatch.}
The bounds of Proposition~\ref{prop:Wnorms} continue to hold if
$T_{0,j\pm 1}(\h_{j\pm 1})$ would be used on the left-hand sides
instead of $T_{0,j}(\h_{j})$ (for indices that do not exceed the final
scale).  In fact, $F$ and $W$ are (non-local) polynomials of
degree at most six, and $P$ is a (local) polynomial of degree at most four.
A change of $\pm 1$ in $j$ in the evaluation of on of these $T_0$ semi-norms
can therefore only give rise to a bounded power of $L$, and constants
in \refeq{Fepbd}--\refeq{epP} are permitted to depend on $L$.
\end{rk}

We prepare for the proof
in Section~\ref{sec:Pformula} with useful identities for
$P$ and $W$, and the proof is concluded in Section~\ref{sec:examples4}.
The proof is based on a crucial contraction estimate
from \cite{BS-rg-loc} for the operator $\LT$, which we recall
below as Proposition~\ref{prop:1-LTdefXY-loc}.

\subsection{Preliminary identities}
\label{sec:Pformula}

The first lemma provides a formula for the expectation of $F$.

\begin{lemma}
\label{lem:EthF}
For polynomials $A,B$ in the fields, and for covariances $C,w$,
\begin{equation}
\label{e:EthF}
    e^{\Lcal_C} F_{\pi,w} (A,B)
    =
    F_{\pi,w+C}(e^{\Lcal_C} A, e^{\Lcal_C} B)
    -
    F_{\pi,C}(e^{\Lcal_C} A, e^{\Lcal_C} B).
\end{equation}
\end{lemma}

\begin{proof}
By the definition of $F$ in \eqref{e:FCAB},
\begin{align}
    F_{w+C}(e^{\Lcal_C}A,e^{\Lcal_C}B)
    &=
    e^{\Lcal_C}
    e^{\Lcal_w}
    \big(e^{-\Lcal_{w}}A\big)
    \big(e^{-\Lcal_{w}} B\big)
    -(e^{\Lcal_C}A)(e^{\Lcal_C}B)
    \nnb
    &=
    e^{\Lcal_C}
    F_{w}( A, B) +
    e^{\Lcal_C} \left(AB \right)
    - (e^{\Lcal_C}A)(e^{\Lcal_C}B)
    \nnb
    &=
    e^{\Lcal_C}
    F_{w}( A, B) + F_{C}(e^{\Lcal_C}A,e^{\Lcal_C}B)
    .
\label{e:EFAB}
\end{align}
Rearrangement gives
\begin{equation}
\label{e:EthFnopi}
    e^{\Lcal_C} F_{w} (A,B)
    =
    F_{w+C}(e^{\Lcal_C} A, e^{\Lcal_C} B)
    -
    F_{C}(e^{\Lcal_C} A, e^{\Lcal_C} B),
\end{equation}
and, by \eqref{e:Fpi} and the fact that the projection operators commute with
$e^{\Lcal_C}$,
\eqref{e:EthFnopi} extends to the same equation with $F$ replaced by
$F_\pi$.
\end{proof}

For the next lemma, we define
\begin{align}
\label{e:Pxydef}
    P_j(V'_x,V''_y)
    &=
    \frac{1}{2} \LT_{x}
    F_{\pi,w_{j+1}} (e^{\Lcal_{j+1}} V'_x,e^{\Lcal_{j+1}} V''_y)
    - \frac{1}{2} e^{\Lcal_{j+1}}\LT_{x} F_{\pi,w_{j}} ( V'_x, V''_y)
    \nnb
    & \hspace{60mm}
    (0 \le j < N-1)
    ,
    \\
\label{e:Wpoint}
    W_j(V'_x,V''_y)
    &=
    \frac 12
    (1-\LT_x) F_{\pi,w_j}(V'_x,V''_y)
    \quad
    (1 \le j < N)
    .
\end{align}
Both definitions will be extended to the final scale in Section~\ref{sec:Pbd},
but this extension is not yet needed here.
By definition, for $j<N$,
\begin{equation}
\lbeq{WVVsum}
    W_j(V,x) = \sum_{y \in \Lambda}W_j(V_x,V_y).
\end{equation}
With the definition of $P_j(V)$ in \refeq{PWdef}, the next lemma shows that,
for $j<N-1$,
\begin{equation}
\lbeq{PVVsum}
    P_j(V,x) = \sum_{y \in \Lambda}P_j(V_x,V_y).
\end{equation}
For its proof, we observe that
since $e^{\Lcal_C}$ reduces the dimension of a monomial in the fields,
$e^{\Lcal_C} : \Vcal \to \Vcal$, and since $\LT_X$ acts as the identity
on $\Vcal$, it follows that
\begin{equation}
\label{e:LTELT}
    \LT_X e^{\Lcal_C} \LT_X = e^{\Lcal_C} \LT_X.
\end{equation}

\begin{lemma}
\label{lem:Palt}
For $x,y\in \Lambda$, $0\le j < N-1$, and for $V',V'' \in \Vcal$,
\begin{align}
\label{e:Palt0}
    P_j(V'_x,V''_y)
    &=
    \LT_{x}\left(
    e^{\Lcal_{j+1}} W_{j}(V_x',V_y'')
    + \frac{1}{2}
    F_{\pi,C_{j+1}}(e^{\Lcal_{j+1}}V_x',e^{\Lcal_{j+1}} V_y'')
    \right)
    .
\end{align}
\end{lemma}

\begin{proof}
Consider first the case $j<N-1$.
By definition of $W_j$ in \eqref{e:WLTF},
the right-hand side of \refeq{Palt0} can be rewritten as
\begin{equation}
\label{e:Pxformula}
    \frac 12 \LT_x \big(
    e^{\Lcal_{j+1}}
    (1-\LT_x)F_{\pi,w_{j}}( V_x', V_y'')
    + F_{\pi,C_{j+1}} (e^{\Lcal_{j+1}} V_x',e^{\Lcal_{j+1}} V_y'')
    \big).
\end{equation}
Application of \eqref{e:LTELT} shows that \eqref{e:Pxformula} is equal to
\begin{align}
    &
    \frac 12 \LT_x \big(
    e^{\Lcal_{j+1}}
    F_{\pi,w_{j}}( V_x', V_y'')
    + F_{\pi,C_{j+1}} (e^{\Lcal_{j+1}} V_x',e^{\Lcal_{j+1}} V_y'')
    \big)
    -
    \frac{1}{2}
    e^{\Lcal_{j+1}}
    \LT_x F_{\pi,w_{j}}( V_x', V_y'')
    .
\label{e:Pxformula1}
\end{align}
By \eqref{e:EthF}, \refeq{Pxformula1} is equal to
\begin{align}
    \frac 12
    \LT_{x} F_{\pi,w_{j+1}} (e^{\Lcal_{j+1}} V_x',e^{\Lcal_{j+1}} V_y'')
    -
    \frac 12
    e^{\Lcal_{j+1}}\LT_{x} F_{\pi,w_{j}} ( V_x', V_y'')
    ,
\end{align}
which is \eqref{e:Pxydef}.
\end{proof}

The following lemma computes the expectation of $W$.

\begin{lemma}
\label{lem:EW}
For $x,y \in \Lambda$, $j < N$,
and for $V',V'' \in \Vcal$,
\begin{align}
\label{e:Palt1b}
    e^{\Lcal_j} W_{j-1} (V'_x,V''_y)
    &=
   W_{j} (e^{\Lcal_j} V'_x, e^{\Lcal_j} V''_y)
   + P_{j-1}(V'_x,V''_y)
   -
   \frac{1}{2} F_{\pi,C_j }(e^{\Lcal_j} V_x,e^{\Lcal_j} V''_y)
   .
\end{align}
\end{lemma}

\begin{proof}
By \eqref{e:Wpoint} and  the formula \eqref{e:Pxydef} for $P$,
\begin{align}
\label{e:Palt5}
    e^{\Lcal_j} W_{j-1} (V'_x,V''_y)
    &=
   \frac{1}{2} e^{\Lcal_j} F_{\pi,w_{j-1}}( V'_x,V''_y)
   -
   \frac{1}{2} e^{\Lcal_j} \LT_x F_{\pi,w_{j-1} }(V'_x,V''_y)
   \nnb
   & =
   \frac{1}{2} e^{\Lcal_j} F_{\pi,w_{j-1}}( V'_x,V''_y)
   +
       P_{j-1}(V'_x,V''_y)
   \nnb & \qquad
    -
    \frac{1}{2} \LT_{x}
    F_{\pi,w_j} (e^{\Lcal_j} V'_x,e^{\Lcal_j} V''_y)
   .
\end{align}
Substitution of \eqref{e:EthF} into
\eqref{e:Palt5} gives
\begin{align}
    e^{\Lcal_j} W_{j-1} (V'_x,V''_y)
    &
    =
   F_{\pi,w_j}(e^{\Lcal_j}V_x',e^{\Lcal_j}V_y'')
   -F_{\pi,C_j}(e^{\Lcal_j}V_x',e^{\Lcal_j}V_y'')
   \nnb & \quad
   +
       P_{j-1}(V'_x,V''_y)
    -
    \frac{1}{2} \LT_{x}
    F_{\pi,w_j} (e^{\Lcal_j} V'_x,e^{\Lcal_j} V''_y)
   ,
\end{align}
which is the same as \eqref{e:Palt1b}.
\end{proof}

The next lemma applies Lemma~\ref{lem:EW} to obtain a formula that
enables us to bound $W$ recursively, in Proposition~\ref{prop:Wbounds} below.

\begin{lemma}
\label{lem:W-explicit}
For $x,y \in \Lambda$, $j<N$, and $V',V'' \in \Vcal$,
\begin{equation}
\label{e:WWF}
    W_{j} (V'_x,V''_y)
    =
    (1-\LT_{x})
    \Big(
    e^{\Lcal_j}
    W_{j-1} (e^{-\Lcal_j} V'_x , e^{-\Lcal_j} V''_y)
    +
    \frac{1}{2} F_{\pi ,C_j} ( V'_x,V''_y )
    \Big)
    .
\end{equation}
\end{lemma}

\begin{proof}
The equalities
\begin{align}
    &W_{j}(e^{\Lcal_j} V'_x,e^{\Lcal_j} V''_y)
    = e^{\Lcal_j} W_{j-1} (V'_x,V''_y)
    - P_{j-1}(V'_x,V''_y)
   +
   \frac{1}{2} F_{\pi,C_j }(e^{\Lcal_j} V_x,e^{\Lcal_j} V''_y)
    \nnb
    &\quad =
    e^{\Lcal_j} W_{j-1} (V'_x,V''_y)
    -\frac{1}{2} \LT_{x}
    F_{\pi,w_j} (e^{\Lcal_j} V'_x,e^{\Lcal_j} V''_y)
    + \frac{1}{2} e^{\Lcal_j}\LT_{x} F_{\pi,w_{j-1}} ( V'_x, V''_y)
    \nnb & \quad
    +
   \frac{1}{2} F_{\pi,C_j }(e^{\Lcal_j} V_x,e^{\Lcal_j} V''_y)
   \nnb &
   \quad
   =
       e^{\Lcal_j} W_{j-1} (V'_x,V''_y)
    + \frac{1}{2} e^{\Lcal_j}\LT_{x} F_{\pi,w_{j-1}} ( V'_x, V''_y)
    +\frac{1}{2}
    F_{\pi,C_j} (e^{\Lcal_j} V'_x,e^{\Lcal_j} V''_y)
    \nnb & \quad
    - \frac{1}{2}\LT_x F_{\pi,C_j} (e^{\Lcal_j}V'_x,V''_y)
    - \frac{1}{2}\LT_x e^{\Lcal_j} F_{\pi,w_{j-1}}(V'_x,V''_y)
    \nnb & \quad
    =
       e^{\Lcal_j} W_{j-1} (V'_x,V''_y)
    +\frac{1}{2}(1-\LT_x)
    F_{\pi,C_j} (e^{\Lcal_j} V'_x,e^{\Lcal_j} V''_y)
    \nnb & \quad
    + \frac{1}{2} \LT_x e^{\Lcal_j}\LT_{x} F_{\pi,w_{j-1}} ( V'_x, V''_y)
    - \frac{1}{2}\LT_x e^{\Lcal_j} F_{\pi,w_{j-1}}(V'_x,V''_y)
    \end{align}
give the desired result.  The first equality is \refeq{Palt1b},
the second follows from the formula for $P$ in \refeq{Pxydef},
the third uses \refeq{EthF}, and for the last we used
\eqref{e:LTELT} to insert an operator
$\LT_x$ acting on the second term of the third right-hand side.
\end{proof}

\subsection{Proof of Proposition~\ref{prop:Wnorms}}
\label{sec:examples4}

We now prove the estimates on $F,W,P$ stated in
Proposition~\ref{prop:Wnorms}.
We first consider $F$, then
recall the crucial contraction estimate from
\cite{BS-rg-loc} concerning the operator $\LT$, then apply
the contraction estimate to obtain bounds on $W$ and $P$.

\subsubsection{Bound on \texorpdfstring{$F$}{F}}
\label{sec:Fbds}

We now prove the bound \refeq{Fepbd} on $F$.

Operator bounds on the Laplacian as a map on $T_\phi$
are given in
\cite[Proposition~\ref{norm-prop:Etau-bound}]{BS-rg-norm}, which
asserts that the
operators $\Lcal_C$ and $e^{\pm \Lcal_C}$, restricted to the subspace of
$\Ncal$ consisting of polynomials of degree $A$ with semi-norm
$\|\cdot \|_{T_{\phi}}$, are bounded operators whose norms obey
\begin{equation}
\label{e:eDC}
    \|\Lcal_C\| \leq A^2 \|C\|_\Phi,
    \quad
    \quad
    \|e^{\pm \Lcal_C }\|
    \le
    e^{A^2  \|C\|_{\Phi}}.
\end{equation}
The above operator norms are for operators acting
on $T_{\phi}$, with the scale fixed.

Let $Y(C,x)= \{y : C_{x,y} \neq 0\}$.
Recall \eqref{e:fin-range}, which implies
that the diameter and volume of $Y(C_k,x)$ obey
\begin{equation}
\label{e:YCbd}
    {\rm diam}\left(Y(C_k,x)\right) \le L^k,
    \quad\quad
    |Y (C_{k},x)| \le L^{dk}.
\end{equation}

We recall
the definition of $\Lcallr_w$ from
\cite[\eqref{pt-e:Lcallrdef}]{BBS-rg-pt}, and also recall
\cite[Lemma~\ref{pt-lem:Fexpand}]{BBS-rg-pt},
which asserts that for
$V',V''$ of degree at most $A$,
\begin{equation}
\label{e:FCsum}
    F_{w} (V'_{x} , V''_{y} )
    =
    \sum_{n=1}^A \frac{1}{n!} V_x' (\Lcallr_w)^n V_y''.
\end{equation}

\begin{lemma}
\label{lem:Fpibd-bis}
Suppose that $\|C\|_\Phi \le 1$.
Then
for $x,y \in \Lambda$ and $V' ,V'' \in \Vcal$,
\begin{equation}
    \label{e:FCABX}
    \|  F_{\pi ,C} (V'_{x} , V''_{y} ) \|_{T_{0}}
    \prec
    \|C\|_{\Phi}
    \|V'_{x}\|_{T_{0}}
    \|V''_y\|_{T_{0}}\1_{y \in Y(C,x)}
.
\end{equation}
Also, $ F_{\pi ,C} (V'_{x} , V''_{y} ) \in
\Ncal(Y(C,x))$ and $ \sum_{y \in \Lambda}W_{w} (V_x' , V_y'' ) \in \Ncal(Y(w,x))$.
\end{lemma}

\begin{proof}
By \eqref{e:Fpi}, it follows from \refeq{FCsum} that $ F_{\pi ,C} (V'_{x} ,
V''_{y} ) \in \Ncal(Y(C,x))$.
It then follows from \eqref{e:WLTF} that  $ W_{w} (V' , V'',\{x\} ) \in
\Ncal(Y(w,x))$.

Now we prove \eqref{e:FCABX}.  We have already shown that
the left-hand side is zero for $y \not \in Y(C,x)$, so it suffices to
prove \eqref{e:FCABX} without the factor $\1_{y \in Y(C,x)}$.
Furthermore, by \eqref{e:Fpi}, it is enough to prove \eqref{e:FCABX}
with $F_{\pi ,C}$ replaced by $F_{C}$.  For $t\ge 0$, let
\begin{equation}
    \label{e:F1}
    F (t)
    =
    e^{\Lcal_{tC}}
    \left(
    (e^{-\Lcal_{tC}}V_x')
    (e^{-\Lcal_{tC}}V_y'')
    \right)
.
\end{equation}
Since $V',V''$ are
polynomials in fields, by expanding each of the exponentials we find that $F
(t)$ is a polynomial $\sum_{n= 0}^m F_{n}t^{n}$, for some finite $m$.
According to the second inequality of
\eqref{e:eDC}, there is a $k>0$ determined by
$\|tC \|_{\Phi}$ such that
\begin{equation}
    \sum_{n= 0}^m \|F_{n}\|_{T_{0}}|t|^{n}
    \le
    k
    \|V'_{x}\|_{T_{0}}
    \|V''_y\|_{T_{0}}
.
\end{equation}
Although $k$ depends on $\|tC \|_{\Phi}$, it is uniform for $\|tC
\|_{\Phi} \le 1$. By \eqref{e:FCAB},
\begin{equation}
    F_{C} (V'_{x} , V''_{y} )
    =
    F (1) - F (0)
    =
    \sum_{n= 1}^m F_{n}
.
\end{equation}
Therefore, taking $t = \|C\|_{\Phi}^{-1} \ge 1$, we obtain
\begin{equation}
    \|F_{C} (V'_{x} , V''_{y} )\|_{T_{0}}
    \le
    \sum_{n= 1}^m \|F_{n}\|_{T_{0}}
    \le
    \frac{1}{t}
    \sum_{n= 1}^m \|F_{n}\|_{T_{0}}t^{n}
    \le
    k \|C\|_{\Phi}
    \|V'_{x}\|_{T_{0}}
    \|V''_y\|_{T_{0}}
,
\end{equation}
which completes the proof.
\end{proof}

To estimate the covariance of $C_j$, we use \refeq{CLbd} to conclude
that for $j \le N$,
\begin{align}
\lbeq{Chbd}
        \|C_{j*}\|_{\Phi_j(\h_k)}
        &\prec_L
        \begin{cases}
        \chi_j & \h_j=\ell_j
        \\
        \chi_j \ggen_j^{1/2} & \h_j=h_j .
        \end{cases}
\end{align}
The case $\h_j=\ell_j$ follows immediately from \refeq{CLbd},
and the case $\h_j=h_j$ follows from
\begin{equation}
\label{e:ellh}
    \|C_{j*}\|_{\Phi_j(h_j)}
    =
    \left( \frac{\ell_{j}}{h_j}\right)^2
    \|C_{j*}\|_{\Phi_j(\ell_j)}
    =
    \left( \frac{\ell_0}{k_0} \right)^2
    \ggen_j^{1/2}
     \|C_{j*}\|_{\Phi_j(\ell_j)}
    \prec_L \;
    \ggen_j^{1/2}
    \chi_j.
\end{equation}

\begin{proof}[Proof of \eqref{e:Fepbd}]
Let $1 \le j \le N$.
Summation of \eqref{e:FCABX}
gives, for any $B \in \Bcal_{j}$, the upper bound
\begin{equation}
\label{e:FepWbd}
    \sum_{x \in B}
    \sum_{y \in \Lambda } \|F_{\pi,C_{j*}}(V'_x,V'_y)\|_{T_{0,j}(\h_j)}
    \prec
    \|C_j \|_{\Phi_j(\h_j)} \epsilon_{V',j} \epsilon_{V'',j}.
\end{equation}
We set $V'=V''=V$ in
\eqref{e:FepWbd}.  Since $V \in \bar\DV_j$, $\epsilon_{V,j}$ is bounded by
a multiple of $\ggen_j$ for $\h=\ell$, and of $1$ for $\h=h$.
With \refeq{Chbd}, this gives
\begin{equation}
\label{e:Fepbd-bis}
    \max_{B \in \Bcal_{j}}
    \sum_{x \in B}
    \sum_{y \in \Lambda } \|F_{\pi,C_{j*}}(V_x,V_y)\|_{T_{0,j}}
    \prec_L \epdV^2
    ,
\end{equation}
which is the desired estimate \eqref{e:Fepbd}.
\end{proof}

\subsubsection{\texorpdfstring{$\LT$}{Loc} and the crucial contraction}
\label{sec:cl}

It is shown in \cite[Proposition~\ref{loc-prop:opLTdefXY}]{BS-rg-loc}
(with $R=L^{-j}$)
that $\LT_X$ is a bounded operator on $T_0$ in the sense
that if $F \in \Ncal(X)$ then
\begin{equation}
\label{e:LTXY5}
    \|\LT_{X}F\|_{T_0} \le \bar{C}' \|F\|_{T_0},
\end{equation}
where $\bar{C}'$ depends on $L^{-j}{\rm diam}( X)$.  We also recall
\cite[Proposition~\ref{loc-prop:1-LTdefXY}]{BS-rg-loc}, which is the
crucial contraction estimate which we state here as follows.  As in
\cite[Definition~\ref{loc-def:LTXYsym}]{BS-rg-loc}, we use the
notation $X(\varnothing)=X$, $X(a) = X \cap \{a\}$, $X(b)=X \cap
\{b\}$, and $X(ab) = X \cap\{a,b\}$.  As discussed in
Section~\ref{sec:formint}, $d_+ = d$ on $\Ncal^\varnothing$, $d_+=0$
on $\Ncal^{ab}$, whereas when $\LT$ acts at scale $k$ on $\Ncal^a$ and
$\Ncal^b$, $d_+=[\phi]=\frac{d-2}{2}=1$ if $k<j_{\pp\qq}$ and $d_+=0$
for $k \ge j_{\pp\qq}$.  For $\alpha,\beta \in \{ \varnothing,
a,b,ab\}$, we define $d_{\alpha}'=d_\alpha +1$, and
\begin{equation}
\label{e:cgamobs-loc}
    \cgam_{\alpha,\beta}
        =
    (L^{-d_{\alpha}'} +  L^{-(A+1)[\phi]})
    \left( \frac{\h'_\sigma}{\h_\sigma} \right)^{|\alpha \cup \beta|}
    .
\end{equation}

\begin{prop}
\label{prop:1-LTdefXY-loc} Let $A < p_\Ncal$ be a positive integer,
and let $\varnothing \not = Y \subset X \in \Pcal_j$.  Let $F_{1} \in
\Ncal (X)$, and let $F_2 \in \Ncal(Y)$ with
$\pi_{\alpha}F_{2}=0$ when $Y(\alpha)=\varnothing$.
Let $F =
F_1(1-\LT_{Y})F_2$.  Let $T_\phi'$ denote the
$T_{\phi,j+1}(c\h_{j+1})$ semi-norm for any fixed $c \ge 1$, and let
$T_\phi$ denote the $T_{\phi,j}(\h_{j})$ semi-norm.  Then
\begin{align}
\label{e:LTXY5a-loc}
    \|F\|_{T_{\phi}'}
&\le
    \bar{C}
    \sum_{\alpha,\beta=\varnothing ,\pp ,\qq,\pp\qq}
    \cgam_{\alpha,\beta}
    \left(1 + \|\phi\|_{\Phi'}\right)^{A'}
    \nnb & \quad \quad \times
    \sup_{0\le t \le 1}
    \big(
    \|\pi_\beta F_{1} \pi_\alpha F_{2}\|_{T_{t\phi}}
    +
    \|\pi_\beta F_{1}\|_{T_{t\phi}}\|\pi_\alpha F_{2}\|_{T_{0}}\big)
    \|\sigma^{\alpha\cup\beta}\|_{T_0},
\end{align}
where
$A'=A+d_{\alpha}/[\phi] +1$, and
$\bar C$ depends on $c$ and $L^{-j}{\rm diam}( X)$.
\end{prop}

As a corollary, we specialise to our particular setting
to obtain the following proposition.  We
state Proposition~\ref{prop:1-LTdefXY} in a more general form than is
needed to bound $W$, but
the additional generality is used in the proof of Proposition~\ref{prop:cl}.

\begin{prop}
\label{prop:1-LTdefXY} Let $d\ge 4$, $A=\lceil 2(d+1)/(d-2)\rceil$,
and assume that $p_\Ncal >A$.  Let $\varnothing \not = Y \subset X \in
\Pcal_j$.  Let $F_{1} \in \Ncal (X)$, let $F_2 \in \Ncal(Y)$
with $\pi_{\alpha}F_{2} =0$ when $Y(\alpha)=\varnothing$,
and let $F = F_1(1-\LT_{Y})F_2$.  Suppose
that each of $F_1,F_2, F_1F_2$ has no component in $\Ncal_{ab}$ unless
$j \ge j_{ab}$
(recall \eqref{e:Phi-def-jc}).  Let $T_\phi'$ denote the
$T_{\phi,j+1}(c\h_{j+1})$ semi-norm for some fixed $c \ge 1$, and let
$T_\phi$ denote the $T_{\phi,j}(\h_{j})$ semi-norm.  There is a
constant $\bar C$ depending on $c$ and $L^{-j}{\rm diam}( X)$ such
that
\begin{align}
\label{e:LTXY5a}
    \|F\|_{T_{\phi}'}
&\le
    \bar{C}
    \cgam
    \left(1 + \|\phi\|_{\Phi'}\right)^{A+d+1}
    \sup_{0\le t \le 1}
    \big(
    \|F_{1}F_{2}\|_{T_{t\phi}}
    +
    \|F_{1}\|_{T_{t\phi}}\|F_{2}\|_{T_{0}}\big)
    ,
\end{align}
with
\begin{equation}
\label{e:cgamobs}
    \gamma
    =
    \gamma(Y)
    =
    L^{-d -1} +
    L^{-1}\1_{Y \cap \{a,b\} \not = \varnothing}.
\end{equation}
Moreover, if $\pi_*F_2=0$ then we can replace \eqref{e:cgamobs}
by $\gamma =
L^{-d-1}$.
\end{prop}

\begin{proof}
In our setting, $d_\varnothing =d$, $d_{ab}=0$, and $d_a=d_b=1$ if
$j<j_{ab}$ whereas $d_a=d_b=0$ if $j\ge j_{ab}$.  Also,
$[\phi]=\frac{d-2}{2}\ge 1$ for all $\alpha$.  In particular,
$A+d_{\alpha}/[\phi] +1 \le A+d+1$.  Our choice of $A$ ensures that
$(A+1) [\phi] \ge d+1 \ge d_\alpha + 1$ for all $\alpha$.  By
\eqref{e:h-assumptions},
\begin{equation}
    \frac{\h_{\sigma,j+1}}{\h_{\sigma,j}}
    \le
    {\rm const}\,
    \begin{cases}
     L^{[\phi]}
     & j < j_{ab}
     \\
     1 & j \ge j_{ab}.
     \end{cases}
\end{equation}
By assumption, when $|\alpha \cup \beta|=2$ we can use the $j \ge
j_{ab}$ version of the above bound.  Also by assumption, for $\alpha
=a,b,ab$ we have $\pi_\alpha F_{2} =0$ when $Y \cap \{a,b\} =
\varnothing$.  Taking these points into account, from
\eqref{e:cgamobs-loc} we obtain
\begin{equation}
    \gamma_{\alpha,\beta}
    \le
    2\,
    \begin{cases}
    L^{-d-1} & |\alpha \cup \beta| =0
    \\
    L^{-1}\1_{Y \cap \{a,b\} \not = \varnothing} & |\alpha \cup \beta| =1,2.
    \end{cases}
\end{equation}
This shows that $\gamma_{\alpha,\beta} \le 2 \gamma$ uniformly in
$\alpha,\beta$.  It follows from \refeq{Fnormsum} that
\begin{equation}
\big(
    \|\pi_\beta F_{1} \pi_\alpha F_{2}\|_{T_{t\phi}}
    +
    \|\pi_\beta F_{1}\|_{T_{t\phi}}\|\pi_\alpha F_{2}\|_{T_{0}}\big)
    \|\sigma^{\alpha\cup\beta}\|_{T_0}
    \le
    \|F_{1}F_{2}\|_{T_{t\phi}}
    +
    \|F_{1}\|_{T_{t\phi}}\|F_{2}\|_{T_{0}}.
\end{equation}
Together with Proposition~\ref{prop:1-LTdefXY-loc}, these facts
give the desired estimate and the proof is complete.
\end{proof}

\subsubsection{Bound on \texorpdfstring{$W$}{W}}
\label{sec:Wbds}

We now prove \refeq{Wbomega} for $j<N$, beginning with the following proposition,
whose proof requires our assumption that $L$ is large.
We defer the case $j=N$ of \refeq{Wbomega}
(and also of \refeq{Wbound2az}) to Sections~\ref{sec:Pbd}--\ref{sec:Waux}.

\begin{prop}
\label{prop:Wbounds}
Let $j<N$.
In general, $\pi_{ab}W_j =0$.  Let $V',V'' \in \Vcal$.
Suppose there is a sequence $v_k'$ with $v_{k-1}' \prec v_k'$ for all $k \le j$,
such that $\max_{B \in \Bcal_k} \sum_{x \in B}\|V'_x\|_{T_{0,k}} \le v_k'$,
and similarly for $V''$.  Then there is a constant $c$ such that
\begin{equation}
\label{e:Wbound2az}
     \max_{B \in \Bcal_j} \sum_{x \in B} \sum_{y \in \Lambda}
     \|W_{j}(V'_x,V''_y)\|_{T_{0,j}(\h_j)}
\le
    c
    \chi_j \left( \frac{\ell_j}{\h_j} \right)^2 v_j' v_j''
    .
\end{equation}
\end{prop}

\begin{proof}
In $W$, we can exclude the $\sigma\bar\sigma$ terms in each of
$V',V''$ since these contribute zero to $F$.  Thus the only possible
$\sigma\bar\sigma$ contribution to $W$ can be due to the contribution to $F$
due to a contraction of $\sigma\bar\phi_a$ with $\bar\sigma \phi_b$.
Such a contraction contains no boson or fermion fields, so is
annihilated by $1-\LT_{\{x\}}$.  This proves that $\pi_{ab}W=0$,
and it remains to prove \eqref{e:Wbound2az}.

We prove \eqref{e:Wbound2az},
by induction on $j$.
Our induction
hypothesis is that \eqref{e:Wbound2az} holds for $j-1$,
and we use this
to prove that it also holds for $j$.
Initially
$W_0=0$, so it is trivial to begin the induction.
The starting point is Lemma~\ref{lem:W-explicit},
which implies that
\begin{equation}
\label{e:WWFj}
    W_{j} (V'_x,V''_y)
    =
    (1-\LT_{x})
    \Big(
    e^{\Lcal_{j}}
    W_{j-1} (e^{-\Lcal_{j}} V'_x , e^{-\Lcal_{j}} V''_y)
    +
    \frac{1}{2} F_{\pi ,C_{j}} ( V'_x,V''_y )
    \Big)
    .
\end{equation}
We estimate \eqref{e:WWFj}
using the triangle inequality on the right-hand side, retaining
the cancellation in
$1-\LT_{\{x\}}$ for the first term but not for the second.
With \eqref{e:LTXY5},
this gives
\begin{align}
\label{e:WWFj2}
    \|W_{j} (V'_x,V''_y)\|_{T_{0,j}} & \le
    \|(1-\LT_{x})
    e^{\Lcal_{j}}
    W_{j-1} (e^{-\Lcal_{j}} V'_x , e^{-\Lcal_{j}} V''_y)\|_{T_{0,j}}
    \nnb & \quad
    + \frac{1}{2}(1+\bar C)
    \| F_{\pi ,C_{j}} ( V'_x,V''_y )\|_{T_{0,j}}.
\end{align}
The constant $\bar C$ is independent of $j$, as a consequence
of \eqref{e:YCbd} together
with the fact that
$ F_{\pi ,C_j} (V'_{x} , V''_{y} ) \in \Ncal(Y(C_j,x))$
by Lemma~\ref{lem:Fpibd-bis}.

We begin with the second term on the right-hand side of \refeq{WWFj2}.
After application of Lemma~\ref{lem:Fpibd-bis} and
\refeq{Chbd}, and summation over $x,y$, we find that there is a constant $\Fconst$ such that
\begin{equation}
\label{e:ind2nd}
    \frac{1}{2}(1+\bar C)
     \max_{B \in \Bcal_j} \sum_{x \in B} \sum_{y \in \Lambda}
     \| F_{\pi ,C_{j}} ( V'_x,V''_y )\|_{T_{0,j}}
     \le
    \bar\Fconst \chi_j
     \left( \frac{\ell_j}{\h_j} \right)^2 v_j' v_j''.
\end{equation}

For the first term on the right-hand side, we apply
Proposition~\ref{prop:1-LTdefXY} with $F_1=1$ and $F_2 = e^{\Lcal_{j}}
W_{j-1} (e^{-\Lcal_{j}} V'_x , e^{-\Lcal_{j}} V''_y)$.  Note that, as
required by the hypotheses of Proposition~\ref{prop:1-LTdefXY},
$\pi_*F_2 =0$ unless $x \in \{a,b\}$; this is a consequence of the
careful definition of $F_{\pi,C}$ in \eqref{e:Fpi}, which ensures that
if one of $\pi_*V'$ or $\pi_*V''$ is nonzero then $\pi_*V'$ must be
nonzero.  The application of Proposition~\ref{prop:1-LTdefXY} gives
the estimate
\begin{equation}
\lbeq{ind-1Loc}
    \|(1-\LT_{x})F_2\|_{T_{0,j}}
    \le
    \bar{C} \gamma_x
    \|F_2\|_{T_{0,j-1}}
    ,
\end{equation}
with
\begin{equation}
    \gamma_x = L^{-d-1}+L^{-1}\1_{x \in \{a,b\}},
\end{equation}
and with a scale-independent constant $\bar C$ since $F_2 \in \Ncal(Y
(C_{j-1},x))$ by Lemma~\ref{lem:Fpibd-bis}.
The operators $e^{\pm \Lcal_j}$ are
bounded on $T_{0,j-1}$, by \eqref{e:eDC} and the fact that
\begin{equation}
\lbeq{Cnormcomp}
    \|C_{j}\|_{\Phi_{j-1}(\h_{j-1})} \le
    \|C_{j}\|_{\Phi_{j}(\h_{j-1})}
    = (\ell_{j}/\h_{j-1})^2 \|C_{j}\|_{\Phi_{j}(\ell_{j})}
   \le (\ell_{j}/\h_{j-1})^2
    \ellconst \le 1
\end{equation}
using \eqref{e:h-assumptions} and \eqref{e:CLbd}.  Thus, by the
induction hypothesis, there is a constant $\bar A$ such that
\begin{equation}
\label{e:W22}
     \max_{B \in \Bcal_j} \sum_{x \in B} \sum_{y \in \Lambda}
     \gamma_x
     \|F_2\|_{T_{0,j}}
     \le
     \bar{A}   c  L^{-1}
    \chi_j
     \left( \frac{\ell_j}{\h_j} \right)^2 v_j' v_j''
    ,
\end{equation}
where we have used the fact that $B$ contains $L^d$ blocks of scale $j-1$,
our assumption on the sequences $v_k'$ and $v_k''$,
and that the factors involving $\chi$ and $\ell/\h$ change only by a constant
factor under a single advance of scale.

The combination of \refeq{ind2nd}, \refeq{ind-1Loc} and \refeq{W22} gives
\begin{equation}
\label{e:Wind}
     \max_{B \in \Bcal_j} \sum_{x \in B} \sum_{y \in \Lambda}
     \|W_{j}(V'_x,V''_y)\|_{T_{0,j}(\h_j)}
\le
    (\bar C \bar A c L^{-1} + \Fconst)
    \chi_j \left( \frac{\ell_j}{\h_j} \right)^2 v_j' v_j''
    \1_{y \in Y(C_k,x)}
    .
\end{equation}
We require that $L > \bar C \bar A$ (which we can do in view of our general
hypothesis that $L$ is large enough).  Then \refeq{Wind} advances the induction
with the choice $c= \Fconst/(1 - \bar C \bar A  L^{-1})$, since this choice
gives $\bar C \bar A c L^{-1} + \Fconst =c$.
This completes the proof.
\end{proof}

\begin{proof}[Proof of \eqref{e:Wbomega} for $j<N$.]
Let $j<N$.
For $V \in \bar\DV_j$, by direct computation
as in the proof of Proposition~\ref{prop:monobd}, we find that, for any $k \le j$ and
$b \in \Bcal_k$, $\sum_{x \in B}\|V_x\|_{T_{0,k}}$ is bounded above by a multiple
of $\ggen_j$ for $\h=\ell$, and of $\ggen_j/\ggen_k$ for $\h=h$.
We apply Proposition~\ref{prop:Wbounds} with these two
choices for $v_k$, which do obey its hypothesis by \refeq{gbarmono}.  This gives
\begin{align}
    \sum_{x \in B} \|W_{j}(V,x)\|_{T_{0,j}}
    &
    \prec_L\;
    \begin{cases}
    \chi_j \ggen_j^2 & \h=\ell
    \\
    \chi_j \ggen_j^{1/2} & \h=h.
    \end{cases}
\label{e:Wepbd-bis}
\end{align}
The right-hand side is $\epdV_j^2$ and this completes the proof.
\end{proof}

\subsubsection{Bound on \texorpdfstring{$P$}{P}}
\label{sec:Pbd}

We now prove \refeq{epP}, and also prove the case $j=N$ of \refeq{Wbomega}.

\begin{proof}[Proof of \eqref{e:epP}.]
We first consider $j<N-1$, and recall from Lemma~\ref{lem:Palt} that
\begin{align}
\lbeq{PVV}
    P_j(V_x',V_y'')
    & =
    \LT_{x}\left(
    e^{\Lcal_{j+1}} W_{j}(V_x',V_y'')
    + \frac{1}{2} F_{\pi,C_{j+1}}(e^{\Lcal_{j+1}} V_x',e^{\Lcal_{j+1}} V_y'')
    \right)
     .
\end{align}
We bound the operator norms of $\LT_x$ and
$e^{\Lcal_{j+1}}$ as discussed previously (using \refeq{Cnormcomp}),
and apply \eqref{e:FepWbd}
and \eqref{e:Wbound2az},
to conclude that under the same hypothesis on $V',V''$ as in Proposition~\ref{prop:Wbounds},
\begin{align}
\lbeq{PVVsumbd}
    \max_{B \in \Bcal_j}
    \sum_{x\in B}
    \sum_{y \in \Lambda}
    \|P_j(V_x',V_y'')\|_{T_{0,j}(\h_{j})}
    &
    \prec
    \chi_j
    \left(
    \frac{\ell_j}{\h_j}
    \right)^2
    v_j'v_j''
   .
\end{align}
Then we set $V'=V''=V \in \bar\DV_j$ and as in the proof of
\refeq{Wepbd-bis} we obtain
\begin{equation}
    \max_{B \in \Bcal_{j}}
    \sum_{x\in B}
    \|P_j(V,x)\|_{T_{0,j}(\h_{j})}
    \prec_L\;
    \epdV_{j}^2
\end{equation}
as desired.
This completes the proof of \refeq{epP}  for $j<N-1$.

As discussed in Section~\ref{sec:finalscale},
our definition of $P_{N-1}$ is designed so that $P_{N-1}$ for
the torus of scale $N$ is the same local polynomial as $P_{N-1,N+1}$ on the torus of scale $N+1$.
Consequently we can apply \refeq{epP} on the torus of scale $N+1$
to obtain the desired estimate \refeq{epP} on $P_{N-1}$.
\end{proof}

\begin{proof}[Proof of \eqref{e:Wbomega} for $j=N$.]
According to \refeq{WNdef},
\begin{align}
\label{e:WNdef-bis}
    W_{N}(V,x) & =
    e^{\Lcal_{N,N}} W_{N-1}(e^{-\Lcal_{N,N}}V,x)
    -P_{N-1}(e^{-\Lcal_{N,N}}V,x)
    + \frac 12  F_{\pi,C_{N,N}}(V_x,V(\Lambda))
     .
\end{align}
This obeys \eqref{e:Wbomega} by using \refeq{CLbd} and \refeq{eDC}
together with the estimates on $W_{N-1}$, $P_{N-1}$, $F_{\pi,C_{N,N}}$ obtained
above.
\end{proof}

\subsubsection{Auxiliary estimates on \texorpdfstring{$W$}{W}}
\label{sec:Waux}

In \refeq{PNdef}, we defined $P_{N-1}(V)$ to be equal to the common
value that \refeq{PWdef}
would give on any torus of scale larger than $N$.  Similarly, we extend
the definition of $P_j(V'_x,V''_y)$ to $j=N-1$ by defining it to be the
common value of the right-hand side of \refeq{Pxydef}, with $j=N-1$, on any torus of
scale larger than $N$.
In addition, we adapt the identity \refeq{Palt1b} to
define $W_N(V_x',V''_y)$ (which has not yet been defined for distinct
$V',V''$) as
\begin{align}
\label{e:WNVV}
    W_{N} ( V'_x,  V''_y)
    &=
   e^{\Lcal_{N,N}} W_{N-1} (e^{-\Lcal_{N,N}}V'_x,e^{-\Lcal_{N,N}}V''_y)
   - P_{N-1}(e^{-\Lcal_{N,N}}V'_x,e^{-\Lcal_{N,N}}V''_y)
   \nnb & \quad
   +
   \frac{1}{2} F_{\pi,C_{N,N} }(V_x,V''_y)
   .
\end{align}
Then from \refeq{WNdef-bis} we see that the identity \refeq{WVVsum} extends
to scale $j=N$:
\begin{equation}
    W_N(V,x) = \sum_{y \in \Lambda} W_N(V_x,V_y).
\end{equation}
Also, the estimate \refeq{Wbound2az} of Proposition~\ref{prop:Wbounds} now extends
to scale $N$.  To see this, we use the definition \refeq{WNVV},
the fact that $e^{\pm\Lcal_{N,N}}$ is a bounded operator,
the bounds on $W_{N-1}$ and $F$ obtained previously, and finally the fact that
\refeq{PVVsumbd} extends to its final scale $N-1$ by application of
\refeq{PVVsumbd} on a larger torus.

The next lemma provides a concrete upper bound on $W_j(V',V'')$ when
observables are absent.

\begin{lemma}
\label{lem:W-logwish}
Suppose that $V',V'' \in \pi_\varnothing \Vcal$, and let
$|V|_j = \max \{ |g|,L^{2j}|\nu|, |z|, |y|\}$.
For $j \le N$,
\begin{align}
    \max_{B \in \Bcal_j}
    \sum_{x \in B}\sum_{y \in \Lambda} \|W_{j}(V'_x,V_y'')\|_{T_{0,j}(\ell_j)}
    &
    \prec_L\;
    \chi_j |V'|_j |V''|_j.
\label{e:W-logwish}
\end{align}
\end{lemma}

\begin{proof}
Let $v_k=L^{dk}\|V_x\|_{T_{0,k}(\ell_k)}$.
Direct computation of the $T_{0,k}(\ell_k)$ norm shows that
$v_k \prec |V|_k \le |V|_j$ for $k \le j$.
Then Proposition~\ref{prop:Wbounds} (extended as noted above to include $j=N$)
gives a bound on the left-hand side of \refeq{W-logwish} of order $v_{j}^2$,
as desired.
\end{proof}

Finally, the next lemma provides estimates for later use.

\begin{lemma}
\label{lem:Wbil}
For $j+1\le N$, $B \in \Bcal_j$, and $V\in \bar\DV_j$,
\begin{align}
\label{e:Wbil}
    \|W_{j+1}(e^{\Lcal_{j+1}} V,B) -W_{j+1}(\Vpt,B)\|_{T_{0,j+1}}
    &\prec_L
    \epdV_j^3.
\end{align}
For $j \le N$, $B \in \Bcal_j$, $V\in \bar\DV_j$,
and for $Q \in \Qcal$ with $\|Q(B)\|_{T_{0,j}} \prec \epdV_{j}$,
\begin{align}
\lbeq{Wprimebd-app}
    \|W_j(Q(B),V(\Lambda))\|_{T_0,j}
    &\prec_L
    \begin{cases}
    \epdV_j^2   & \h=\ell
    \\
    \epdV_j^2 \ggen_j^{1/4} & \h=h,
    \end{cases}
\\
\lbeq{Wprimebd-app2}
    \|W_j(Q(B),Q(\Lambda))\|_{T_0,j}
    &\prec_L
    \begin{cases}
    \epdV_j^2   & \h=\ell
    \\
    \epdV_j^2 \ggen_j^{1/2}  & \h=h.
    \end{cases}
\end{align}
\end{lemma}

\begin{proof}
By linearity and the triangle inequality,
\eqalign
    \|W_{j+1}(e^{\Lcal_{j+1}} V) -W_{j+1}(\Vpt)\|_{T_{0,j+1}}
    &\le
    \|W_{j+1}(P,e^{\Lcal_{j+1}} V)\|_{T_{0,j+1}}
    +
    \|W_{j+1}(\Vpt,P)\|_{T_{0,j+1}}.
\enalign
We apply Proposition~\ref{prop:monobd}, use
Proposition~\ref{prop:Wnorms} to see that for $B_k \in \Bcal_k$
it is the case that $\sum_{x \in B_k}\|P_x\|_{T_{0,k}} \prec_L \epdV_k^2$,
and then apply Proposition~\ref{prop:Wbounds} (including its extension to scale $N$),
to see that
\eqalign
    \|W_{j+1}(e^{\Lcal_{j+1}} V) -W_{j+1}(\Vpt)\|_{T_{0,j+1}}
    &
    \prec_L
    \chi_j ( \ell_j/\h_j)^2
    \epdV_j^2 \times \begin{cases}\ggen_j & \h=\ell \\ 1& \h=h \end{cases}
    \;\;\;\prec
    \epdV_j^3
    ,
\label{e:Wbilbdell-app}
\enalign
as required.
For \refeq{Wprimebd-app}--\refeq{Wprimebd-app2}, a similar calculation, using
$\epsilon_Q \prec \epdV$ (by Lemma~\ref{lem:T0ep} and assumption)
gives the desired result.
This completes the proof.
\end{proof}

\section{Proof of Propositions~\ref{prop:Iupper}--\ref{prop:JCK-app-1}}
\label{sec:I-estimates}

In this section, we prove
Propositions~\ref{prop:Iupper}--\ref{prop:JCK-app-1}.  Attention is
restricted here to $d=4$.

We begin by proving estimates on $\Ical=e^{-V}$ of \eqref{e:Icaldef}.
Since norms in the global space $\Phi=\Phi(\Lambda)$ can be replaced
in upper bounds by the local space $\Phi (X)$ whenever an element of
$\Ncal (X)$ is being estimated (as discussed around
\cite[\eqref{norm-e:NXdef}]{BS-rg-norm}), we sometimes write simply
$\Phi$ rather than $\Phi(X)$.  However, decay estimates (such as
\refeq{IcalB} below) must always be stated in localised form.

Temporarily, we write $a_0,b_0$ (rather than the usual $a,b$) for the
points where observables are located in $V$, and instead we use $b$
for a block in $\Bcal_{j-1}$.  Also, we write
\begin{equation}
    \epV(b) =
    \begin{cases}
    L^{-d} \epsilon_{V_\varnothing} & \{a_0,b_0\} \cap b = \varnothing
    \\
    L^{-d} \epsilon_{V_\varnothing} +
    (|\lambda^{a_0}|+|\lambda^{b_0}|)\h\h_\sigma
    + \textstyle{\frac 12}(|\q^{a_0}|+|\q^{b_0}|)\h_\sigma^2
    & \{a_0,b_0\} \cap b \neq \varnothing,
    \end{cases}
\end{equation}
as opposed to $\epV$ which always includes the contribution from the
observables.

\begin{prop}
\label{prop:Iupperzz}
Let $j \le N$.  Let $V \in \Qcal$ with
$0 \le |{\rm Im}g | \le \frac 12 {\rm Re}g$.
\\
(i) For  $b \in \Bcal_{j-1}$,
\begin{equation}
\label{e:Iupper0zz}
    \|\Ical(b)\|_{T_{\phi}}
    \le
    e^{O (\epV(b)) (1+ \|\phi\|_\Phi^2)}
.
\end{equation}
(ii) Fix any $q \ge 0$.  Suppose that
 $\epV \le C \epVbar$ for some
$C>0$.
For
$B \in \Bcal_j$, and $X \in \Scal_{j-1}(B)$ or $X=\varnothing$,
\begin{align}
&
    \|\Ical(B\setminus X) \|_{T_{\phi}}
    \le
    e^{O (1+q^{2})\epV}
    e^{-q \epVbar \|\phi\|_{\Phi(B^{\Box})}^2}
    e^{O (1+q) \epV \|\phi\|_{\tilde\Phi(B^{\Box})}^2}
    .
    \label{e:IcalB}
\end{align}
\end{prop}

\begin{proof}
(i)
We write $V=g\tau^2 +Q$.
By \cite[Proposition~\ref{norm-prop:taunorm}]{BS-rg-norm}
(with $q_2 =0$) and \refeq{epVbar-def-old},
\begin{equation}
    \|e^{-g\tau_x^2}\|_{T_{\phi}}
\le
    e^{O (|g|\h^4)} = e^{O (L^{-dj}\epVbar)}
.
\end{equation}
By the product property,
\begin{gather}
\label{e:egtbd}
    \|e^{-g\tau^2(b)}\|_{T_{\phi}}
\le
    \prod_{x\in b} \|e^{-g\tau_{x}^2}\|_{T_{\phi}}
\le
    e^{O (L^{-d}\epVbar)}
.
\end{gather}
Also, since $Q$ is quadratic, from
\cite[Proposition~\ref{norm-prop:T0K}]{BS-rg-norm} and \refeq{monobd}
we obtain
\begin{equation}
    \|Q (b)\|_{T_{\phi}}
\le
    \|Q (b)\|_{T_{0}} (1+\|\phi\|_\Phi)^2
\le
    2
    \epV(b)
    (1+\|\phi\|_\Phi^2)
.
\end{equation}
Therefore, by the power series expansion of the exponential and the product
property,
\begin{equation}
\label{e:apos-e}
    \|e^{-Q (b)}\|_{T_{\phi}}
\le
    e^{\|Q (b)\|_{T_{\phi}}}
\le
    e^{2\epV(b) (1+\|\phi\|_\Phi^2)}
.
\end{equation}
With the product property, \refeq{Iupper0zz} then follows
from \refeq{egtbd}, \refeq{apos-e},  and the fact that
$\epVbar \le \epsilon_{V_\varnothing}$.

\smallskip \noindent (ii)
Fix any $q'\ge 0$.
Since ${\rm Re} g \le |g| \le \frac 32 {\rm Re}g$ by hypothesis,
we can conclude from \cite[Proposition~\ref{norm-prop:taunorm}]{BS-rg-norm} that
\begin{equation}
    \|e^{-g\tau_x^2}\|_{T_{\phi}}
\le
    e^{O (1+q'^{2})| g|\h^4}
    e^{-q' |g|\h^4 |\phi_x/\h|^2}
.
\end{equation}
By the product property and \refeq{epVbardef-h-i}, this gives
\begin{gather}
\label{e:egtau2}
    \|e^{-g\tau^2(B \setminus X)}\|_{T_{\phi}}
\le
    e^{O (1+q'^{2}) \epVbar}
    e^{-q' |g|\h^4 \sum_{x\in B \setminus X}|\phi_x/\h|^2}
.
\end{gather}
For $Y \subset \Lambda$, we define the $L^2(Y)$ norm by
\begin{equation}
    \|\phi \|^{2}_{L^2(Y)}
    =
    \frac{1}{|Y|} \sum_{x\in Y}\frac{|\phi_{x} |^{2}}{\h^{2}}
.
\end{equation}
Then, again writing $V=g\tau^2 +Q$, we combine \refeq{egtau2} with
\eqref{e:apos-e}, using the product property, \refeq{epVbarasymp}, and
$|B\setminus X| \ge \frac 12 |B|$, to obtain
\begin{align}
    \label{e:Ical-1}
    \|\Ical(B\setminus X) \|_{T_{\phi}}
&\le
    e^{O (1+q'^{2}) \epVbar}
    e^{- q' |g|\h^4 |B \setminus X| \, \|\phi\|_{L^{2} (B\setminus X)}^2}
    e^{2\epV(1+\|\phi\|_\Phi^2)}
\nnb
&\le
    e^{O (1+q'^{2}) \epVbar}
    e^{- \frac 12 q' C_0^{-1}\epVbar \, \|\phi\|_{L^{2} (B\setminus X)}^2}
    e^{2\epV(1+\|\phi\|_\Phi^2)}
\end{align}
(no $L^d$ factor is produced for the observables).
By our hypothesis on $X$ and
Proposition~\ref{prop:equivalent-norms},
\begin{equation}
\label{e:Sob2}
    \|\phi\|_{L^2(B\setminus X)}^2
    \geq
    \frac{1}{2c_2^2} \|\phi\|_{\Phi(B^{\Box})}^2
    -
    \|\phi\|_{\tilde \Phi(B^{\Box}) }^2
    .
\end{equation}
We insert this into \eqref{e:Ical-1} and localise the $\Phi $ norm to
$\Phi (B^{\Box})$
to obtain
\begin{align}
    \|\Ical(B\setminus X) \|_{T_{\phi}}
&\le
    e^{O (1+q'^{2}) \epV}
    e^{- (\frac {1}{4}C_0^{-1} c_2^{-2} q'\epVbar -2\epV)  \, \|\phi\|_{\Phi(B^\Box)}^2}
    e^{\frac 12 q' \epV \|\phi\|_{\tilde\Phi(B^\Box)}^2}
.
\end{align}
Then \refeq{IcalB} follows by choosing $q'=4 C_0 c_2^{2}
(q+2C)$, which is $O(q)$.
\end{proof}

We prove Proposition~\ref{prop:Iupper} by combining
Proposition~\ref{prop:Iupperzz} with the following
elementary lemma.

\begin{lemma}
\label{lem:exp-bounds}
For $x, u>0$ and any integer $r \ge \max\{1,u\}$,
\begin{align}
\label{e:exp1}
    (1 + x)^{2r}
&
    \le
    (2r/ u)^{r}   e^{u x^{2}}
\\
\label{e:exp2}
    1 + u^{r} (1+x)^{2r}
&
    \le
    e^{2ru (1+ x^{2})}
.
\end{align}
\end{lemma}

\begin{proof}
For the first bound, we
combine $(1+x)^{2r} \le 2^r (1+x^2)^r$ with the inequality
$1+x^2 \le (r/u) e^{u x^2/r}$
(since $r \ge u$).
The second bound follows from
\begin{align}
    1 + u^{r} (1+x)^{2r}
    &\le
    1 + ( 2u)^{r} (1+x^{2})^{r}
    \le
    (1+2u + 2u x^{2})^{r}
    \le
    (e^{2u + 2 u x^{2}})^{r},
\end{align}
where we used $r \ge 1$ in the second inequality.
\end{proof}

\begin{proof}[Proof of Proposition~\ref{prop:Iupper}]
We first consider the choice $I^*=I(B)$.
By the product property and
\cite[Proposition~\ref{norm-prop:T0K}]{BS-rg-norm},
\begin{align}
    \|I (B) F\|_{T_{\phi}}
    &\le
    \|\Ical (B)\|_{T_{\phi}}
    \|1+W(B)\|_{T_{\phi}}
    \| F\|_{T_{\phi}}
    \nnb &
    \le
    \|\Ical (B)\|_{T_{\phi}}
    \|1+W(B)\|_{T_{\phi}}
     \| F\|_{T_{0}}
    \left( 1 + \|\phi\|_{\Phi}\right)^r,
\end{align}
where $r$ denotes the degree of $F$.
By \refeq{FCsum}, $W$ is a degree-six  polynomial in the boson and fermion fields.
By \refeq{exp2} and \cite[Proposition~\ref{norm-prop:T0K}]{BS-rg-norm},
\begin{align}
\label{e:1Wbd}
    \|1+W(B)\|_{T_{\phi}}
    &\le
    1+\|W(B)\|_{T_{\phi}}
    \le
    1+\|W(B)\|_{T_{0}}\left( 1 + \|\phi\|_{\Phi}\right)^6
    \le
    e^{6\omega^{1/3} (1+\|\phi\|_{\Phi}^2)}.
\end{align}
where $\omega = \max_{B \in \Bcal_j}\|W(B)\|_{T_0}$.
Then, since $6(L^{2d}\omega)^{1/3} \le u$ by hypothesis, \refeq{exp1} gives
\begin{equation}
    \|I(B) F\|_{T_{\phi}}
    \le
    \|\Ical (B)\|_{T_{\phi}} \| F\|_{T_{0}}
    \left(\frac{2r}{u} \right)^{r} e^{u + 2u \|\phi\|_{\Phi}^{2}}
    .
\end{equation}
Then \refeq{Iupper-a} with $I^*=I(B)$
follows from \refeq{Iupper0zz}.
For \refeq{Iupper-b}, fix $q \ge 0$ to be the desired parameter in
\refeq{Iupper-b},
and choose the variable
called $q$ in \refeq{IcalB} to
be $q_1$ defined by $q_1= q+2u\epVbar^{-1}$.
This gives \refeq{Iupper-b} for the choice $I^*=I(B)$.

For the case $\tilde{I}(B \setminus X)$ with $X=\varnothing$
or $X\in \Scal_{j-1}$, we replace \refeq{1Wbd} by
\begin{equation}
    \|\prod_{b \in \Bcal_{j-1}(B\setminus X)}(1+W(b))\|_{T_{\phi}}
    \le
    e^{6L^d (L^{-d } \omega)^{1/3} (1+\|\phi\|_{\Phi}^2)}
    \le
    e^{u (1+\|\phi\|_{\Phi}^2)},
\end{equation}
and proceed similarly.

Omitting factors $1+W$ in the above bounds
only makes it easier, so we also have the
bounds if we choose $I^*$ with factors of
$1+W$ missing, and the proof is complete.
\end{proof}

\begin{proof}[Proof of Proposition~\ref{prop:Istab}]
Let $V \in \bar\DV$.
We first consider the case $I_*=I$ (possibly with some $1+W$ factors omitted) and $j_*=j$.
The bound \refeq{IF} follows from
\refeq{Iupper-a-c} and \refeq{Iupper-b-d} (with $q=0$),
and \refeq{Iass} follows similarly from the case $r=0$.
Also, for $B
\in \Bcal_j$, it follows from the definition of $I$, the product property, \refeq{monobd}
and \refeq{Wbomega}, that
\begin{equation}
    \label{e:I-b}
    \|I(B)^{-1}\|_{T_0}
    \le
    e^{\|V(B)\|_{T_0}} \frac{1}{1-\|W(V,B)\|_{T_0}}
    \le
    (1 +O (\epV+\epW))
    \le 2
,
\end{equation}
which gives \refeq{I-b:5}.  This completes the proof for the case $I_*=I$.

Next, we consider the case $I_*=\Ipttil$.
It follows from Proposition~\ref{prop:monobd} that
$\Vpt \in \bar\DV'$, and the above result for $I_*=I$ then gives
\refeq{IF}--\refeq{I-b:5} also for $\Ipttil$ when $j_*=j$.

This leaves \refeq{IF}--\refeq{I-b:5} for
the case $I_*=\Ipttil$ with %
$j_*=j+1$, as well as %
\refeq{Iptass}.
For \refeq{IF}, we apply Lemma~\ref{lem:Imono}
and the scale-$j$ case of \refeq{IF} (now $W_{j+1}$ occurs rather than $W_j$
but it is bounded by Remark~\ref{rk:sm}) to obtain
\begin{align}
    \|\Ipttil (B)F\|_{T_{\phi,j+1} (\h_{j+1})}
&\le
    \|\Ipttil (B)F\|_{T_{\phi,j} (\h_{j})}
    \prec
    \|F\|_{T_{0,j}}
    \Gcal_{j}  (B,\phi)
,
\end{align}
where $\Gcal_j = G_j$ for $\h_j=\ell_j$, and $\Gcal_j=\tilde G_j$ for
$\h_j=h_j$.  For $\h=\ell$ we set $\phi=0$ and \refeq{IF} immediately
follows for $j+1$.  For $\h=h$ we use the fact that $\tilde
G_{j}(X,\phi) \le \tilde G_{j+1}^{\Gtilp}(X,\phi)$ by
Lemma~\ref{lem:mart}, and \refeq{IF} also follows in this case.  Note
that $\|F\|_{T_0,j}$ occurs in \refeq{IF} both for $j_*=j$ and
$j_*=j+1$.  The estimate \refeq{Iass} follows similarly, and
\refeq{I-b:5} for $j+1$ follows from \refeq{I-b:5} for $j$ by
Lemma~\ref{lem:Imono}, which implies that the $T_{\phi,j+1}$ norm is
bounded above by the $T_{\phi,j}$ norm.

Finally, to prove \refeq{Iptass},
we recall from Proposition~\ref{prop:monobd} that
$\Vpt\in \bar\DV_{j+1}'$, and then \refeq{Iptass} follows exactly as
the scale-$j$ case of \refeq{Iass} for $\Ipttil$.
This completes the proof.
\end{proof}

\begin{proof}[Proof of Proposition~\ref{prop:Ianalytic1:5}]
We first prove the analyticity of $V \mapsto \Ical = e^{-V}$ for $V$ in
$\bar\DV_j$;
in this case $j_*=j$.
We fix $B$ and drop it from the notation.

Fix $V \in \bar\DV_j$ and let $\dot{V} \in \Qcal$.
We prove analyticity by showing that $I(V+\dot V)$ has a norm
convergent power series expansion in $\dot V$,
if $|\dot g| \le \frac{1}{8} {\rm Re}g$ and $\epsilon_{\dot{V}}$
is sufficiently small.
By the integral form of the remainder in Taylor's theorem,
together with the product property of the $T_\phi$ semi-norm,
\begin{align}
    \big\|e^{-(V+\dot V)} - \sum_{n=0}^N e^{-V} \frac{(-\dot V)^n}{n!}\big\|_{j}
    &=
    \big\| \int_0^1 \frac{1}{N!} e^{-(V+s\dot V)}\dot V^{N+1}(1-s)^{N} ds\big\|_{j}
    \nnb
    &
    \le
    \sup_\phi \Gcal(\phi)^{-1}
    \frac{1}{(N+1)!}\|e^{-V}\dot V^{N+1}\|_{T_\phi} e^{\|\dot V\|_{T_\phi}},
\end{align}
where $\Gcal$ denotes the regulator, either $G_j$ or $\tilde G_j$.
It suffices to show that the above right-hand side goes to zero as $N \to
\infty$, and for this it suffices to show that insertion of summation
over $N$ under the supremum leads to a convergent result.  Since
\eq
    \sum_{N=0}^\infty
    \frac{1}{(N+1)!}\|e^{-V}\dot V^{N+1}\|_{T_\phi} e^{\|\dot V\|_{T_\phi}}
    \le
    \|e^{-V} \|_{T_\phi} e^{2\|\dot V\|_{T_\phi}},
\en
it suffices to show that
\begin{equation}
\lbeq{eVanal}
    \sup_\phi \Gcal(\phi)^{-1} \|e^{-V} \|_{T_\phi} e^{2\|\dot V\|_{T_\phi}}
    < \infty.
\end{equation}

We isolate the $\tau^2$
terms by writing $V=g\tau^2 + Q$ and $\dot V =\dot g \tau^2 + \dot Q$.
By \cite[Proposition~\ref{norm-prop:taunorm}]{BS-rg-norm},
$\|\tau_x\|_{T_\phi}=\h^2P(t)$,
where $P(t)=t^2+2t+2$ and $t=|\phi_x|/\h$.
Let $\epsilon = \epV + 2\epsilon_{\dot{V}}$.
We use the product property of the $T_\phi$ norm,
as well as \cite[Proposition~\ref{norm-prop:T0K}]{BS-rg-norm}, to obtain
\begin{align}
    \|e^{-V_x} \|_{T_\phi} e^{2\|\dot V_x\|_{T_\phi}}
    &\le
    \|e^{-g\tau_x^2 }\|_{T_\phi}
    e^{2|\dot g| \,\|\tau_x^2\|_{T_\phi}+
    \|Q_x\|_{T_\phi} + 2\|\dot Q_x\|_{T_\phi}}
    \nnb &
    \le
    \|e^{-g\tau_x^2 }\|_{T_\phi}
    e^{2|\dot g |\h^4 P(t)^2 + \epsilon L^{-dj}(1+\|\phi\|_\Phi^2)}.
\end{align}
By \cite[Proposition~\ref{norm-prop:taunorm}]{BS-rg-norm},
together with the assumption in the definition of $\bar\DV$ that
$|{\rm Im}g| < \frac 15 {\rm Re}g$,
\eq
    \|e^{-g\tau_x^2}\|_{T_\phi}
    \le
    e^{({\rm Re}g)\h^4[-2t^2 + \frac 32 P(t)^2]}.
\en
Since $|\dot g| \le \frac18 {\rm Re}g$, this gives
\eq
    \|e^{-g\tau_x^2 }\|_{T_\phi}
    e^{2|\dot g |\h^4 P(t)^2 }
    \le
    e^{({\rm Re}g)\h^4[-2t^4 + \frac 74 P(t)^2]}
    \le e^{({\rm Re}g)\h^4[q_1-q_2t^2]},
\en
where $q_2 \ge 0$ can be chosen arbitrarily with a corresponding
choice of $q_1$.  Therefore,
\begin{equation}
    \|e^{-V_x} \|_{T_\phi} e^{2\|\dot V_x\|_{T_\phi}}
    \le
    e^{({\rm Re}g)\h^4[q_1-q_2t^2] + \epsilon L^{-dj}(1+\|\phi\|_\Phi^2)}
    .
\end{equation}

To conclude \refeq{eVanal} for the $G$ norm, we take $q_2=0$ and
$\epsilon_{\dot V} =\epV$, and the desired estimate follows for uniformly small $\ggen_j$.
The proof of \refeq{eVanal} for the $\tilde G$ norm can be
completed by applying the Sobolev inequality exactly as in the proof of
Proposition~\ref{prop:Iupperzz}, using the fact that we do have
$\epV \le C\epVbar$ in this case by \refeq{tau2dom}.

It remains to consider the effect of $1+W$ on the above argument.
Since $1+W$ is a degree-6 polynomial in the fields, it is analytic for the case
of the $G$ norm, and its effect is therefore unimportant.  For the case of the
$\tilde G$ norm, $1+W$ is not analytic because polynomial growth in the absolute
value of $\phi$
is not cancelled by the regulator in this case (since the regulator has linear functions
factored out).  However, it is an exercise to include the factor $1+W$ alongside
the $e^{-V}$ factor in the above argument and thereby conclude analyticity also in this
case.

To prove the analyticity of $\Ipttil$ in $V \in \bar\DV_j$,
it again suffices to consider $e^{-\Vpt}$.
Let  $V \in \bar\DV_j$ and consider first the case $j_*=j$.
We can regard
$e^{-\Vpt}$  as the composition of $V \mapsto \Vpt$ and $\Vpt \mapsto e^{-\Vpt}$.
The first of these maps is polynomial in $V$.
Thus, for the case of the $G$ norm, $V \mapsto \Vpt$ is analytic, while the
second map is analytic by the previous argument together with the fact that
$\Vpt \in \bar\DV'$ when $V \in \bar\DV$ by Proposition~\ref{prop:monobd}.
This proves the desired analyticity when $j_*=j$ for the $G$ norm.
The analyticity for the case of the $\tilde G$ norm can be established with
small additional effort.

Next, we consider the case $j_*=j+1$.  As above, the main work
lies in showing that
$e^{-\Vpt}$ is an analytic function of $\Vpt \in \bar\DV$ when measured in the
$\|\cdot\|_{j+1}$ norm.  But it follows from Lemmas~\ref{lem:Imono}
and \ref{lem:mart} that
for either of the choices \refeq{np1}--\refeq{np2} for the norm pairs,
$\|F\|_{j+1} \le C\|F\|_j$ for some $C>0$ and for all $F$.  Thus convergence
of a power series in a neighbourhood in the $j$-norm implies convergence
in a neighbourhood in the $j+1$-norm, and the analyticity for $j_*=j+1$ follows
from the analyticity for $j_*=j$.

Finally, it follows similarly that $I(B)^{-1}$ is
analytic in $V$, as a map into the space with norm $\|\cdot\|_{T_{0,j}}$.
For example, the factor $e^{g\tau^{2} (Y)}$ in
$I(B)^{-1}$ is analytic in $g$ because it has an absolutely convergent
power series,
\begin{gather}
    \|e^{g \tau^{2} (B)}\|_{T_{0} (\ell)}
    \le
    \sum_{n\ge 0} \frac{1}{n!} \|g\tau^{2} (B)\|_{T_{0} (\ell)}^{n}
    \le
    \sum_{n\ge 0}  \frac{1}{n!} \epsilon_{g\tau^{2}}^{n}
.
\end{gather}
A similar argument applies to the inverse of $1-W$.
This completes the proof.
\end{proof}

\begin{proof}[Proof of Proposition~\ref{prop:JCK-app-1}.]
Let $j<N$, $V \in \bar\DV$, and $Q\in\Qcal$ with $\|Q(B)\|_{T_0} \prec \epdV$.
We first show that $V-Q \in \bar\DV'$.  This implies that
the estimates of Proposition~\ref{prop:Istab}
apply to $\Ihat$, and that the desired analyticity follows
from Proposition~\ref{prop:Ianalytic1:5}, so then it will remain
only to prove the estimates \refeq{JCK1-app}--\refeq{JCK2-app}.

By Lemma~\ref{lem:T0ep},
$\epsilon_{V-Q} \le \epV + \epsilon_{Q} \prec \epV + \max_B \|Q(B)\|_{T_0}$.
The last bound of \refeq{epVbardefz-app} (with worse constants)
then follows from the assumption on $Q$.
For the middle bound of \refeq{epVbardefz-app}, let
$g_Q$ denote the coefficient of $\tau^2$ in $Q$.  By hypothesis,
$L^{dj} |g_Q| \|\tau^2_0\|_{T_0(h)} \prec \ggen_j^{1/4}$, and hence
\begin{equation}
    L^{dj} |g- g_Q| \|\tau^2_0\|_{T_0(h)}
    \ge   L^{dj} |g| \|\tau^2_0\|_{T_0(h)}-c_L\ggen_j^\eta
    \ge ak_0^4 - c \ggen_j^{1/4} \ge \frac 12 ak_0^4,
\end{equation}
by taking $\ggen_j$ sufficiently small.
Finally, for the first inequality of \refeq{epVbardefz-app},
we apply \refeq{epVbardef-h-i} to see that
\begin{equation}
    |{\rm Im}g_Q| \le |g_Q| \prec \frac{\epsilon_{Q,j}(\h_j)}{L^{dj}\h_j^4}
    .
\end{equation}
By the hypothesis on $Q$, for $\h=\ell$ the right-hand side is at most
$c\ell_0^{-4} \ggen_j$, which is at most
$\frac{1}{10}C_{\DV}^{-1}\ggen_j < \frac{1}{10}{\rm Re}g$ for $L$ sufficiently
large (hence $\ell_0$ large).  Similarly, for $\h=h$ the right-hand side is
$\prec \ggen_j^{5/4}$,
and hence the effect of $Q$ on the imaginary part of $g$ is negligible.
This completes the proof that $V-Q\in \bar\DV'$.

It remains to prove \refeq{JCK1-app}--\refeq{JCK2-app}.
For $s \in [0,1]$, we write $V_s=V-sQ$, $I_s=I(V_s)$,
$\Ical_s = e^{-V_{s}}$, and $W_s=W(V_{s})$, and omit the $B$ arguments.
Direct calculation gives
\begin{align}
\label{e:Isprime}
    I_s' & = I_s Q + \Ical_s W_s',
    \\
\label{e:Isprime2}
    I_s'' &= I_s Q^2 + 2\Ical_s Q W_s' + \Ical_s W_s'',
    \\
\lbeq{Wprime}
    W_s' &= -W(Q,V_s) - W(V_s,Q),
    \\
\lbeq{Wprime2}
    W_s'' & = -2W(Q,Q).
\end{align}
By Lemma~\ref{lem:Wbil},
\eq
\lbeq{Wprimebd}
    \|W_s'\|_{T_0}
    \prec_L
    \begin{cases}
    \chi_j \ggen_j^2   & \h=\ell
    \\
    \chi_j \ggen_j^{3/4} & \h=h
    \end{cases}
    \qquad
    \qquad
    \|W_s''\|_{T_0}
    \prec_L
    \begin{cases}
    \chi_j \ggen_j^2   & \h=\ell
    \\
    \chi_j \ggen_j  & \h=h.
    \end{cases}
\en

Let $\Ihat(B) = I(V-Q,B)$.
By the Fundamental Theorem of Calculus, $\Ihat-I=\int_0^1 I_s' ds$,
and hence by \eqref{e:Isprime}
\begin{equation}
\lbeq{IIhatdif}
    \|\hat I - I\|_{j}
    \le
    \sup_{s\in [0,1]}
    \left( \|I_s Q\|_{j} + \|\Ical_s W_s'\|_j \right).
\end{equation}
We have shown above that $V-sQ \in \bar\DV'$ (in fact this holds
uniformly in $s$),
and consequently \refeq{IF} holds with $V$ replaced by $V-sQ$.
By \eqref{e:IF}, the first term on the
right-hand side of \refeq{IIhatdif} is of order $\|Q\|_{T_0} \prec \epdV$.
By \refeq{IF} and \refeq{Wprimebd},
the second term of \refeq{IIhatdif} is negligible compared to the first.
This proves \eqref{e:JCK1-app}.

For \eqref{e:JCK2-app}, we first note that $I_1 -I_0-I_0' = \hat I - I - IQ -
\Ical_0 W_0'$.  Using this, with a second-order Taylor remainder estimate
followed
by \eqref{e:IF}, gives
\begin{align}
    \|\hat I - I - IQ\|_{T_0}
    &\le
    \|\Ical_0 W_0'\|_{T_0}
    +
    \sup_{s\in [0,1]}
    \| I_s''\|_{T_0}
    \nnb
    & \prec
    \|W_0'\|_{T_0}
    +
    \|Q\|_{T_0}^2
    +
    \sup_{s\in [0,1]}
    \left( \| Q \|_{T_0} \| W_s'\|_{T_0} + \|  W_s''\|_{T_0} \right)
    \prec_{L}
    \epdV^2
    ,
\end{align}
where for the last step we used \eqref{e:Wprimebd} together with the fact that
its right-hand sides are at most $\epdV^2$.
This proves \refeq{JCK2-app}.
\end{proof}

\section{Proof of Propositions~\ref{prop:hldg}--\ref{prop:h}}
\label{sec:interaction-estimates444}

In this section, we prove Propositions~\ref{prop:hldg}--\ref{prop:h}.
The proof of Proposition~\ref{prop:hldg} is short, whereas
the proof of Proposition~\ref{prop:h} is substantial.
In the proof of Proposition~\ref{prop:h} it
is important that $W$ and $\Vpt$ be defined as they are,
and it is here that we implement the ideas in
\cite[Section~\ref{pt-sec:WPjobs}]{BBS-rg-pt}.

\subsection{Proof of Proposition~\ref{prop:hldg}}

\begin{proof}[Proof of Proposition~\ref{prop:hldg}.]
Let $j<N$ and $V \in \bar\DV_j$.
Recall from \eqref{e:hptdefqq} that $\hldg (U,B)$ is defined
for $(U,B) \in \Scal_{j+1}\times \Bcal_{j+1}$ by
\begin{equation}
\label{e:hptdefqqz}
    \hldg (U,B)
    =
    \begin{cases}
    -\frac{1}{2}
    \Ex_{\pi ,j+1} \theta ( V_j(B); V_j(\Lambda \setminus B))
    & U=B
    \\
    \;\;\;
    \frac{1}{2}
    \Ex_{\pi ,j+1}\theta  ( V_j(B);  V_j(U\setminus B)) & U \supset B, |U|_{j+1}=2
    \\
    \;\;\;
    0 &\text{otherwise}
    .
    \end{cases}
\end{equation}
By \eqref{e:trun-exp}, \eqref{e:FCAB}
and \eqref{e:EWick},
\begin{equation}
\label{e:EABbis}
   \Ex_{\pi,C} (\theta A; \theta B)
   =
   F_{\pi,C}( e^{ \Lcal_{C}}  A , e^{ \Lcal_{C}}  B).
\end{equation}
By Proposition~\ref{prop:Wnorms},
\begin{align}
\label{e:Fepbd-xxx}
    \max_{B \in \Bcal_{j+1}}
    \sum_{x \in B}
    \sum_{B' \in \Bcal_{j+1}(\Lambda)}
    \|F_{\pi,C_{j+1}}(V_x,V(B'))\|_{T_{0,j+1}}
    &
    \prec_L\;
    \epdV^2
.
\end{align}
As an operator on the subspace of $\Ncal$
consisting of bounded-degree polynomials in the fields,
$e^{\pm \Lcal_{C_k}}$
is bounded (uniformly in $k$), due to
\eqref{e:eDC} and \refeq{CLbd}.
With \eqref{e:Fepbd} and \eqref{e:EABbis}, this gives
\begin{equation}
\label{e:EtruncT0}
    \max_{B \in \Bcal_{j+1}}
    \sum_{x \in B}
    \sum_{B' \in \Bcal_{j+1}(\Lambda)}
    \|\Ex_{\pi,C_{j+1}}(\theta V_x;\theta V(B'))\|_{T_{0,j+1}}
    \prec_{L} \epdV^2,
\end{equation}
from which we conclude that
\begin{align}
\label{e:Etruncbd}
    \|\hldg(U,B)\|_{T_0,j+1}
&\prec_{L}
    \epdV^2
    .
\end{align}
By Proposition~\ref{prop:Istab}, this implies that
\begin{align}
\label{e:2Lprimeh1}
    \|\Ipttil(U)\hldg(U,B) \|_{j+1}
& \prec_L
    \epdV^2
    .
\end{align}
This gives \eqref{e:want1} and completes the
proof of Proposition~\ref{prop:hldg}.
\end{proof}

\subsection{Proof of Proposition~\ref{prop:h}}

We require some preparation for the
proof of Proposition~\ref{prop:h}.
By \eqref{e:hptdefqq}--\eqref{e:hldgUdef},
\begin{align}
    \hldg (B)
    &=
    -\sum_{b \in \Bcal_j(B)} \frac{1}{2} \Ex_{\pi} (\theta V(b);\theta V(\Lambda \setminus B))
    \nnb
    &=
    -\sum_{b \in \Bcal_j(B)} \frac{1}{2} \Ex_{\pi} ( \theta V(b);\theta V(\Lambda \setminus b)) +
    \sum_{b\not =b' \in \Bcal_j(B)} \frac{1}{2} \Ex_{\pi} (\theta V(b);\theta V(b'))
    .
\end{align}
It follows from \eqref{e:Epi} that
\begin{equation}
\label{e:EpiE}
    \frac 12 \Ex_{\pi} (V';V'') + \frac 12 \Ex_{\pi}(V'';V') =  \Ex (V';V''),
\end{equation}
from which we conclude that
\begin{align}
    \hldg (B)
    &=
    -\sum_{b \in \Bcal_j(B)}
    \frac{1}{2} \Ex_{\pi} (\theta V(b);\theta V(\Lambda \setminus b)) +
    \sum_{b\not =b' \in \Bcal_j(B)} \frac{1}{2} \Ex (\theta V(b);\theta V(b')).
\label{e:hleadnew}
\end{align}

For distinct $b,b' \in \Bcal_j$, $B \in \Bcal_{j+1}$,
and for $U \in \Scal_{j+1}$
with $|U|_{j+1}\in \{1,2\}$, we define
\begin{align}
\label{e:R1def}
    R_1(b;B) & =
    \Ipttil^{B\setminus b} \Ex \delta I^b
    +
    \Ipttil^{B}
    \frac{1}{2} \Ex_{\pi} (\theta V_j(b);\theta V_j(\Lambda \setminus b)) ,
    \\
\label{e:R2def}
    R_2(b,b';U)
    & =
    \frac{1}{2}
    \left[
    \Ipttil^{U\setminus (b\cup b')}\Ex \delta I^{b\cup b'}
    -
    \Ipttil^{U} \Ex (\theta V_j(b);\theta V_j(b'))
    \right]
    ;
\end{align}
note that $\Ex_{\pi}$ appears in $R_1$ but not in $R_2$.
Then, by \eqref{e:hred-def},
\eqref{e:hptdefqq}--\eqref{e:hldgUdef}, and \eqref{e:hleadnew},
\begin{align}
\label{e:hhptB}
    \Ipttil^B [\hred (B) - \hldg (B)]
    & =
    \sum_{b \in \Bcal_j(B)}
    R_1(b;B)
    + \sum_{b\neq b' \in \Bcal_j}
    R_2(b,b';B)
    ,
    \\
\label{e:hhptU}
    \Ipttil^U [\hred (U) - \hldg (U)]
    & =
    \sum_{b\neq b' : \overline{b\cup b'}=U}
    R_2(b,b';U)
    \quad \quad \quad\quad |U|_{j+1}= 2.
\end{align}
By the triangle inequality and \refeq{hhptB}--\refeq{hhptU},
to prove Proposition~\ref{prop:h}
it suffices to show that
\begin{align}
\label{e:R12suff}
    \|
    R_1(b;B)
    \|_{j+1}
    &
    \prec_L \, \epdV^3
    ,
\quad\quad
    \|
    R_2(b,b';U)
    \|_{j+1}
    \prec_L \, \epdV^3
    ,
\end{align}
where the constants in the upper bounds depend on $L$,
and $\epdV$ is given by \refeq{epdVdef}.

The appearance of $\delta I$ leads naturally to the study of
$\delta V$, which was defined in \refeq{dVdef} as $\delta V = \theta V - \Vpt$.
As a first step in the proof of \refeq{R12suff}, we prove the
following lemma which relies heavily on results from \cite{BS-rg-norm}.
The ``5'' appearing in its statement
has been chosen as a convenient positive constant and is not significant.
The parameter $\hat\ell_j >0$ is defined in \refeq{ellhatdef}.
The constant $C_{\delta V}$ is the $L$-dependent constant of
Lemma~\ref{lem:epdV}.

\begin{lemma}
\label{lem:dIipV}
Let $j<N$, $b,b'  \in \Bcal_j$, and $n,n' \ge 0$.
Suppose
that $F \in \Ncal((b\cup b')^\Box)$ obeys
$\|F\|_{T_\phi(\h+\hat\ell)} \le c_F e^{\alpha \|\phi\|_{\Phi(\h)}^{2}}$
for some $c_F,\alpha >0$.  If $u \in (0,2]$ obeys $\alpha + \frac{1}{20}(n+n')u \le 5$, then
 \begin{align}
   \label{e:ip-V1}
     \|
     \Ex_{j+1}
     \left[(\delta V(b))^n (\delta V(b'))^{n'}
     \theta F \right]
     \|_{T_{\phi}(\h)}
     &\prec_L
     c_F (C_{\delta V}\epdV)^{n+n'}
     e^{(2\alpha + (n+n')u) \|\phi\|_{\Phi(\h)}^2},
 \end{align}
where the constant in the upper bound depends on $u,n,n'$,
and where $\h$, $\hat\ell$ and all norms are at scale $j+1$.
\end{lemma}

\begin{proof}
By \cite[Proposition~\ref{norm-prop:EK}]{BS-rg-norm}
(with \refeq{ellhatdef-1} to provide its hypothesis on the covariance),
and by the product property of the $\Ttimes_{\phi\sqcup\xi}$ semi-norm,
\begin{align}
    &\|\Ex
    \left[
    (\delta V(b))^{n} (\delta V(b'))^{n'} \theta F
    \right]
    \|_{T_{\phi}}
    \leq
    \Ex
    \|
    (\delta V(b))^{n} (\delta V(b'))^{n'} \theta F
    \|_{\Ttimes_{\phi\sqcup\xi}(\h\sqcup\hat\ell)}
    \nnb
    & \hspace{30mm}
    \leq
    \Ex \left[\| \delta V(b)\|_{\Ttimes_{\phi\sqcup\xi}(\h\sqcup\hat\ell)}^{n}
    \| \delta V(b')\|_{\Ttimes_{\phi\sqcup\xi}(\h\sqcup\hat\ell)}^{n'}
    \|\theta  F \|_{\Ttimes_{\phi\sqcup\xi}(\h\sqcup\hat\ell)}
    \right].
\end{align}
We apply
\cite[Proposition~\ref{norm-prop:T0K}]{BS-rg-norm}
to the $\Ttimes_{\phi\sqcup\xi}(\h\sqcup\hat\ell)$ semi-norm of $\delta V$,
with
a multi-component field with $\h=\hat\ell$ for $\xi$.
With \eqref{e:dVbd}, this gives
\begin{equation}
    \| \delta V(b)\|_{\Ttimes_{\phi\sqcup\xi}(\h\sqcup\hat\ell)}
    \le
    \epdV
    (1+\|\phi\|_{\Phi(\h)})^4 (1 + \|\xi\|_{\Phi(\hat\ell)})^4.
\end{equation}
For any $u \in (0,2]$, \refeq{exp1} then gives (with a $u$-dependent constant
and with $\hat{u}=u(\ell/\hat\ell)^2$)
\begin{equation}
\label{e:dVbdpf}
    \| \delta V(b)\|_{\Ttimes_{\phi\sqcup\xi}(\h\sqcup\hat\ell)}
    \prec
    C_{\delta V} \epdV
    e^{u(\|\phi\|_{\Phi(\h)}^2 + \|\xi\|_{\Phi(\hat\ell)}^2)}
    =
    C_{\delta V} \epdV
    e^{u\|\phi\|_{\Phi(\h )}^2 } G(b,\xi)^{\hat{u}}
    .
\end{equation}
Similarly,
by \cite[Proposition~\ref{norm-prop:derivs-of-tau-bis}]{BS-rg-norm},
by hypothesis, by
$\|\phi+\xi\|^2 \le 2(\|\phi\|^2+\|\xi\|^2)$, and by $\h \ge \ell$,
\begin{align}
    \|\theta  F \|_{\Ttimes_{\phi\sqcup\xi}(\h\sqcup\hat\ell)}
    & \le
    \| F \|_{T_{\phi+\xi}(\h+\hat\ell)}
    \le
    c_F e^{2\alpha(\|\phi\|_{\Phi(\h)}^2 + \|\xi\|_{\Phi(\h)}^2)}
    \nnb &
    \le
    c_F e^{2\alpha\|\phi\|_{\Phi(\h)}^2 }G(b\cup b', \xi)^{2\alpha}.
\end{align}
Therefore, with $s=n+n'$, since $G(b\cup b')=G(b)G( b')$
by \cite[\eqref{norm-e:GXYfluct}]{BS-rg-norm},
\begin{align}
\label{e:EGpapp}
    \|\Ex
    [
    (\delta V(b))^{n} (\delta V(b'))^{n'} \theta F
    ]
    \|_{T_{\phi}}
    & \prec
    (C_{\delta V}\epdV)^{s}
    c_F
    e^{(2\alpha +su)\|\phi\|_{\Phi(\h )}^2 }
    \Ex
    \left[
    G(b \cup b',\xi)^{2\alpha +s\hat{u}}
    \right]
    .
\end{align}
It suffices now to bound the expectation on the right-hand side by a constant.
By \refeq{ellhatdef-1}, by our choice
$\ellconst = \frac{1}{10}c_G$ above \refeq{EG}, and by \refeq{CLbd} and \refeq{ellhatdef-1},
\begin{align}
    (2\alpha +s\hat{u}) \|C\|_{\Phi^+(\ell)}
    & =
    2\alpha \|C\|_{\Phi^+(\ell)}  +su \|C\|_{\Phi^+(\hat\ell)}
    \nnb &
    \le
    2\alpha \ellconst + su \frac{c_G}{100}
    \le
    \left(
    \frac{\alpha}{5} + \frac{su}{100}
    \right)
    c_G
    \le c_G
    ,
\end{align}
with the last inequality true by hypothesis.  Then
\cite[Proposition~\ref{norm-prop:EG2}]{BS-rg-norm} yields the desired bound
on the expectation, and the proof is complete.
\end{proof}

For $j \ge 1$, we define $A_j$ by
\begin{equation}
\label{e:Ajdef}
    A_j = e^{-\delta V} - \sum_{i=0}^{j-1} \frac{(-\delta V)^i}{i!}.
\end{equation}
By Taylor's theorem with integral form of the remainder,
\begin{equation}
\label{e:AjTaylor}
    A_j = \frac{1}{(j-1)!} \int_0^1 (1-t)^{j-1} (\delta V)^j e^{-t\delta V} dt.
\end{equation}
It follows from the definitions that
$e^{-\theta V}=e^{-\Vpt}e^{-\delta V}$, and that for $b \in \Bcal_j$,
\begin{align}
    \delta I(b) & =
    e^{-\Vpt(b)}\left( A_1(b) + Z(b) \right),
\label{e:dIexpansion}
\end{align}
with
\begin{equation}
\label{e:ZdefW}
    Z = e^{-\delta V}\theta W - W_{j+1}.
\end{equation}

It is in the following proof that it
is important that $W$ and $\Vpt$ be defined as they are,
and our implementation of the ideas laid out in
\cite[Section~\ref{pt-sec:WPjobs}]{BBS-rg-pt} occurs here.  In
particular, the identity
\begin{equation}
\label{e:EW}
    \Ex_{j+1} \theta W_{j} (V, X)
    =
    W_{j+1} (\Ex_{j+1}\theta V ,X)
    + P (X)
    -
    \frac{1}{2} \Ex_{\pi ,j+1}\big(\theta V(X);\theta V(\Lambda)\big)
\end{equation}
of Lemma~\ref{lem:EW} enters the proof
of \refeq{ExIb} in a crucial manner, as does the definition
$\Vpt = \Ex \theta V - P$
of \refeq{Vptdef} (recall \refeq{EWick}).

\begin{proof}[Proof of Proposition~\ref{prop:h}.]
All norms in this proof are at scale $j+1$.
Fix $B\in \Bcal_{j+1}$ and $b \in \Bcal_j(B)$ for $R_1$,
and fix $U \in \Scal_{j+1}$ with $|U|\in \{1,2\}$ and $b\neq b' \in \Bcal_j$
with $\overline{b \cup b'}=U$ for $R_2$.
To prove \refeq{R12suff},
it suffices to prove that
\begin{align}
\label{e:ExIb}
    \|
    R_1(b;B)
    \|_{T_\phi}
    &\prec_L
    \epdV^3
    \Gcal(B,\phi)
    ,
\\
\label{e:ExIbb}
    \|
    R_2(b,b';U)
    \|_{T_\phi}
    &\prec_L \epdV^3
    \Gcal(U,\phi)
    ,
\end{align}
where $\Gcal$ represents $G$ or $\tilde G^{\Gtilp}$ according to the
choice $\h=\ell$ or $\h=h$.  We first prove the bound \refeq{ExIb} for
$R_1$, and then the bound \refeq{ExIbb} for $R_2$.

\smallskip\noindent \emph{Identity for $R_1$.}
We apply \refeq{dIexpansion}, \refeq{Ajdef},
and the definition $\Vpt=\Ex\theta V -P$, to obtain
\begin{align}
    \delta I & =
    e^{-\Vpt}\left( -\delta V + A_2  + Z \right)
    \nnb & =
    e^{-\Vpt}\left( -\delta V + \frac 12 (\delta V)^2
    + A_3 + (1+A_1)\theta W  -W_{j+1} \right)
    \nnb
    & =
    e^{-\Vpt}\left( \Ex\theta V -\theta V - P
    + \frac 12 (\theta V - \Ex\theta V)^2 + \theta W -W_{j+1}(\Ex\theta V)
    + \Ecal_1 \right),
\label{e:dIexpansion2}
\end{align}
where
\begin{equation}
    \Ecal_1 = (\theta V - \Ex\theta V)P + \frac 12 P^2 +
    A_3 + A_1 \theta  W
    + W_{j+1}(\Ex\theta V) -W_{j+1}(\Vpt)
    .
\end{equation}
Then, taking the expectation,  we obtain
\begin{align}
    \Ex \delta I(b)
    & =
    e^{-\Vpt}\left( - P
    + \frac 12 \Ex(\theta V(b) ; \theta V(b)) + \Ex\theta W -W_{j+1}(\Ex\theta V)
    + \Ex\Ecal_1  \right)
    ,
\end{align}
with
\begin{equation}
\label{e:ExEcal1}
    \Ex \Ecal_1
    =
    \frac 12 P^2 +
    \Ex A_3 + \Ex (A_1 \theta  W)
    +
    \big[
    W_{j+1}(\Ex\theta V) -W_{j+1}(\Vpt)
    \big]
    .
\end{equation}
It follows from \refeq{EpiE} that
$\Ex(\theta V(b) ; \theta V(b)) = \Ex_\pi(\theta V(b) ; \theta V(b))$.
Thus, after application of
\eqref{e:EW},
together with use of the identity
\begin{equation}
    \Ex_\pi(\theta V(b); \theta V(b)) -
    \Ex_\pi(\theta V(b); \theta V(\Lambda))
    =
    -
    \Ex_\pi(\theta V(b); \theta V(\Lambda \setminus b)),
\end{equation}
we obtain
\begin{align}
    \Ipttil^{B\setminus b}\Ex \delta I(b)
    & =
    \Ipttil^{B \setminus b} e^{-\Vpt(b)}\left(
    - \frac{1}{2} \Ex_{\pi}(\theta V  (b); \theta V(\Lambda \setminus b))
    + \Ex\Ecal_1 (b) \right)
    \nnb & =
    -\Ipttil^B \frac{1}{2} \Ex_{\pi}(\theta V  (b); \theta V(\Lambda \setminus b))
    \nnb & \quad
    + \Ipttil^{B \setminus b} e^{-\Vpt(b)}W_{j+1}
    \frac{1}{2} \Ex_{\pi}(\theta V  (b); \theta V(\Lambda \setminus b))
    +
    \Ipttil^{B \setminus b} e^{-\Vpt(b)}
    \Ex\Ecal_1 (b)
    .
\label{e:EdIid}
\end{align}
By definition of $R_1$, this gives
\begin{equation}
\lbeq{R1identity}
    R_1(b;B)
    =
    \Ipttil^{B \setminus b} e^{-\Vpt(b)}W_{j+1}
    \frac{1}{2} \Ex_{\pi}(\theta V  (b); \theta V(\Lambda \setminus b))
    +
    \Ipttil^{B \setminus b} e^{-\Vpt(b)}
    \Ex\Ecal_1 (b)
    .
\end{equation}
The use of \eqref{e:EW} has led to an important cancellation
which the definitions of $W$ and $\Vpt$ were engineered
to create.

\smallskip\noindent \emph{Bound on $R_1$.}
It suffices to obtain a bound of the form $\epdV^3
\Gcal(B,\phi)$ for the $T_\phi$ semi-norms of each of the two terms on
the right-hand side of \refeq{R1identity}, with the last of these
terms given by
\refeq{ExEcal1}.  The resulting five terms are of two types: one type
involves $\Ipttil^{B \setminus b} e^{-\Vpt}$ multiplied by the
polynomials $W_{j+1}\Ex_{\pi}(\theta V (b); \theta V(\Lambda \setminus
b))$, $P^2$, $[ W_{j+1}(\Ex\theta V) -W_{j+1}(\Vpt)]$, and the second
type involves two terms with expectations of the non-polynomial
quantities $A_1$ and $A_3$.

For the first type of term, we apply
\refeq{IF} (the version with factor $(1+W(b))$ omitted)
to conclude that, for a polynomial $Q$,
\begin{equation}
    \|\Ipttil^{B \setminus b} e^{-\Vpt(b)} Q(b)\|_{j+1}
    \prec
    \|Q(b)\|_{T_{0,j}}.
\end{equation}
Bounds on the $T_0$ semi-norm of $W_{j+1}$,
$\Ex_{\pi}(\theta V  (b);\theta V(\Lambda \setminus b))$ and $P$
follow from \refeq{Wbomega}, \refeq{EtruncT0}, and \refeq{epP}.
Also, the norm of $W_{j+1}(\Ex\theta V) -W_{j+1}(\Vpt)$ is bounded
in \refeq{Wbil}.  With these bounds,
we obtain an upper bound of order $\epdV^3$ for
the $(j+1)$-norm of the three terms with polynomials.

For the second type of term, we apply
Lemma~\ref{lem:dIipV}.  For the
$A_3$ term, it follows from \refeq{AjTaylor} and the product property that
\begin{align}
\label{e:A3term}
    \|\Ipttil^{B \setminus b} e^{-\Vpt(b)}\Ex  A_3(b)\|_{T_{\phi}}
    & \le
    \sup_{t\in [0,1]}
    \|\Ipttil^{B \setminus b} e^{-(1-t)\Vpt(b)}\|_{T_{\phi}}
    \| \Ex \delta V(b)^3 e^{-t\theta V(b)}\|_{T_{\phi}}
    .
\end{align}
By \refeq{scale-change} and \refeq{Iupper-a}
(for its hypothesis on $\omega$ we see from \refeq{Wbomega} that $\omega \prec_L \epdV^2$),
given any small $u_1>0$,
\begin{equation}
    \|e^{-t  V(b)}\|_{T_{\phi}(\h+\hat\ell)}
    \le \|e^{-t  V(b)}\|_{T_{\phi}(2\h)}
    \le e^{O(\epV(2\h) +u_1) \|\phi\|_{\Phi(2\h)}^2}
    \le e^{O(\epV(\h) +u_1) \|\phi\|_{\Phi(\h)}^2}
    .
\end{equation}
It therefore follows from Lemma~\ref{lem:dIipV}
that given any small $u>0$, with a constant depending on $u$ we have
\begin{equation}
\label{e:A3term-a}
    \| \Ex \delta V(b)^3 e^{-t\theta V(b)}\|_{T_{\phi}}
    \prec_L
    \epdV^3 e^{O(\epV  + u) \|\phi\|_\Phi^2}.
\end{equation}

For the case of the regulator
$G$, we bound the first factor on the right-hand side
of \refeq{A3term} as follows.  By \refeq{Iupper-a},
the product property, and \refeq{scale-change},
\begin{equation}
    \|\Ipttil^{B \setminus b} e^{-(1-t)\Vpt(b)}\|_{T_{\phi,j+1}}
    \le
    \|\Ipttil^{B \setminus b}\|_{T_{\phi,j+1}}
    \| e^{-(1-t)\Vpt(b)}\|_{T_{\phi,j}}
    \le
    e^{O(\epV  + u) \|\phi\|_\Phi^2}.
\end{equation}
Thus we obtain
\begin{align}
    \|\Ipttil^{B \setminus b} e^{-\Vpt(b)}\Ex  A_3(b)\|_{T_{\phi}}
    & \prec_L
    \epdV^3 G(B,\phi)
    ,
\end{align}
as required.

For the case of the regulator
$\tilde G$, we take $u=u_1=\epVbar$ and recall
from \refeq{tau2dom} and \refeq{epVbardef-h} that $\epV  \prec \epVbar  \asymp k_0^4$,
with $k_0$ chosen small (recall the discussion above \refeq{Iupper-b-d}).
Then \refeq{A3term-a} gives, for some $c_0>0$,
\begin{equation}
\label{e:A3term-b}
    \| \Ex \delta V(b)^3 e^{-t\theta V(b)}\|_{T_{\phi}}
    \prec_L
    \epdV^3 e^{c_0\epVbar  \|\phi\|_\Phi^2}.
\end{equation}
We apply \refeq{Iupper-b}, with $q = c_0$, to
see that
\begin{align}
    \|\Ipttil^{B \setminus b} e^{-\Vpt(b)}\Ex  A_3(b)\|_{T_{\phi}}
    & \prec_L
    \epdV^3 \tilde G^{\Gtilp} (B,\phi)
    ,
\end{align}
as required.

The $A_1\theta W_j$ term can be treated similarly, using
Lemma~\ref{lem:dIipV} with $F=e^{-tV}W_j$.
This completes the discussion of the bound on $R_1$.

\smallskip\noindent \emph{Bound on $R_2$.}
Starting from the first line of \refeq{dIexpansion2},
and recalling that $Z$ is defined in \refeq{ZdefW},
a little algebra leads to
\begin{equation}
    \Ex \delta I^{b \cup b'}
    =
    e^{-\Vpt(b\cup b')}
    \big(
    \Ex (\theta V(b) ; \theta V(b'))
    + \Ecal_2(b,b')
    \big),
\label{e:EdIbb}
\end{equation}
where
\begin{align}
    \Ecal_2(b,b') & =
    P(b)P(b') - \Ex \big(\delta V(b) A_2(b')\big)
    - \Ex \big(A_2(b) \delta V(b')\big)
    + \Ex \big(A_2(b)A_2(b')\big)
    \nnb & \quad\quad
    + \Ex \big(A_1(b) Z(b')\big) + \Ex \big(Z(b)A_1(b')\big)
    + \Ex \big(Z(b)Z(b')\big).
\end{align}
Therefore,
\begin{align}
    &
    2R_2(b,b';U)
    =
    \Ipttil^{U\setminus (b\cup b')}e^{-\Vpt(b \cup b')} \Ecal_2(b,b')
    \\ \nonumber & \quad
    +
    \Ipttil^{U\setminus (b\cup b')}e^{-\Vpt(b\cup b')}[(1+W_{j+1}(b))(1+W_{j+1}(b'))-1]
    \Ex (\theta V(b) ; \theta V(b')) .
\end{align}
By \refeq{IF} (with two missing $1+W$ factors),
the $(j+1)$-norm of the second term
on the right-hand side is bounded by a multiple of
the $T_0$ semi-norm of the polynomial factor, which
by \refeq{Wbomega} and \refeq{EtruncT0} is of order $\epdV^4$.
The contribution due to the $PP$ term in $\Ecal_2$ can be bounded in
the same way, using \refeq{epP}.
The six remaining terms in $\Ecal_2$ can be handled in the same way
as the $A_3$ and $A_1$ terms in $\Ecal_1$, and we omit the details.
Using Lemma~\ref{lem:dIipV},
the $\delta V A_2$ and $A_1Z$ terms are seen to be order
$\epdV^3$, while the $A_2A_2$ and $ZZ$ terms are order $\epdV^4$.
In particular, it is not necessary
to make use of any cancellation within $Z$.
Together, these estimates produce an overall bound of
order $\epdV^3$,
and the proof is complete.
\end{proof}

\section{Proof of Propositions~\ref{prop:ip}--\ref{prop:cl}}
\label{sec:ipcl}

In this section, we prove Propositions~\ref{prop:ip}--\ref{prop:cl}.

\subsection{Proof of Proposition~\ref{prop:ip}}
\label{sec:ippf}

The main step in the
proof of Proposition~\ref{prop:ip} is provided by the following lemma.
The constant $C_{\delta L}$ is the $L$-dependent constant of Lemma~\ref{lem:epdV}.

\begin{lemma}
\label{lem:dIip}
Let $X,Y \in \Pcal_j$ be disjoint.
Let $F(Y) \in \Ncal (Y^{\Box})$.
There is an $\Econst>0$ (independent of $L$) and a $C_{\delta V}>0$
(depending on $L$) such that
 \begin{align}
  \label{e:integration-property-a}
     \|\Ex \delta I^X \theta F(Y) \|_{T_{\phi}(\h/2)}
     &\leq
     \Econst^{|X|_j+|Y|_j}
     (C_{\delta V} \epdV)^{|X|_j}
     \| F(Y) \|_{\Gcal(\h)}
    \Gcal(X\cup Y,\phi)^5,
 \end{align}
where $\Gcal$ denotes $G$ or $\tilde G$ when $\h=\ell$ or $\h=h$, respectively.
Norms and regulators are at
scale $j$, the expectation
represents $\Ex_{C_{j+1}}$, and $\delta I$ is given by
\refeq{dIdef}.
\end{lemma}

\begin{proof}
We write $\h'=\h/2$ and $\hat\ell' = \hat\ell/2$.
By \cite[Proposition~\ref{norm-prop:EK}]{BS-rg-norm} (with \refeq{ellhatdef-1}
to provide its hypothesis),
and by the product property of the $\Ttimes_{\phi\sqcup\xi}$ semi-norm,
\begin{align}
    \|\Ex \delta I^X \theta F(Y) \|_{T_{\phi}(\h')}
    & \leq
    \Ex \left[\| \delta I^X
    \|_{\Ttimes_{\phi\sqcup\xi}(\h'\sqcup \hat\ell')}
    \|\theta  F(Y) \|_{\Ttimes_{\phi\sqcup\xi}(\h'\sqcup \hat\ell')}
    \right].
\label{e:ip1}
\end{align}
By \cite[Proposition~\ref{norm-prop:derivs-of-tau-bis}]{BS-rg-norm}
and \refeq{scale-change}
(with the fact that $\h \ge \hat\ell$ for uniformly small $\ggen_j$),
\begin{align}
\lbeq{ip1.1}
    \|\theta  F(Y) \|_{\Ttimes_{\phi\sqcup\xi}(\h'\sqcup\hat\ell')}
    & \!\!
    \le
    \| F(Y) \|_{T_{\phi+\xi}(\h'+\hat\ell')}
    \leq
    \| F(Y) \|_{T_{\phi+\xi}(\h)}
    \leq
    \| F(Y) \|_{\Gcal(\h)}
    \Gcal(Y,\phi+\xi).
\end{align}
Since
$\|\phi+\xi\|^2 \le 2\|\phi \|^2 + 2\|\xi\|^2$,
and since $\Gcal \le G$ because
$\tilde G \le G$, this gives
\begin{align}
    \|\theta  F(Y) \|_{\Ttimes_{\phi\sqcup\xi}(\h'\sqcup\hat\ell')}
    & \!\!
    \le
    \| F(Y) \|_{\Gcal(\h)}
    \Gcal(Y,\phi)^2 \Gcal(Y,\xi)^2
    \le
    \| F(Y) \|_{\Gcal(\h)}
    \Gcal(Y,\phi)^2 G(Y,\xi)^2.
\label{e:ip1.5}
\end{align}
By \eqref{e:AjTaylor}--\eqref{e:ZdefW}, for $b \in \Bcal_j$,
\begin{align}
    \|\delta I(b)\|_{T_{\phi\sqcup\xi}(\h'\sqcup\hat\ell')}
    & \!\!
    \le
    \!\!
    \|\delta V(b)\|_{T_{\phi\sqcup\xi}(\h'\sqcup\hat\ell')}
    \!\!
    \sup_{t\in [0,1]} \|e^{-(1-t)\Vpt(b)}\|_{T_{\phi}(\h')}
    \|\theta e^{-t V(b)}\|_{T_{\phi\sqcup\xi}(\h'\sqcup\hat\ell')}
    \nnb & \quad \quad
    +
    \|\theta(e^{-V(b)}W(b)) \|_{T_{\phi\sqcup\xi}(\h'\sqcup\hat\ell')}
    +
     \|e^{-\Vpt}W_{j+1}(b)\|_{T_{\phi}(\h')}.
\label{e:delItimes}
\end{align}
By \refeq{dVbdpf} (now interpreted at scale $j$ rather than $j+1$;
recall that the bound of Lemma~\ref{lem:epdV} applies to either scale),
for any choice of small positive $u$,
and with $\hat{u}=u(\ell/\hat\ell)^2$,
\begin{equation}
    \| \delta V(b)\|_{\Ttimes_{\phi\sqcup\xi}(\h'\sqcup\hat\ell')}
    \prec
    C_{\delta V}
    \epdV
    e^{u\|\phi\|_{\Phi(\h' )}^2 } G(b,\xi)^{\hat{u}}
    .
\end{equation}

We now consider the supremum on the right-hand side of \refeq{delItimes}.
Either $t \ge \frac 12$ or $1-t
\ge \frac 12$.  Suppose that $t \ge \frac 12$; the other
case is simpler and we omit its details.  By \refeq{Iass}
and \refeq{scale-change},
$\|e^{-(1-t)\Vpt(b)}\|_{T_{\phi}(\h')} \le 2\Gcal(b,\phi)$.
By \cite[Proposition~\ref{norm-prop:derivs-of-tau-bis}]{BS-rg-norm},
\refeq{scale-change},
the inequality $\|\phi\|^2 \le 2\|\phi+\xi\|^2 + 2\|\xi\|^2$,
and the identity $\|\phi\|_{\Phi(\h')}=2\|\phi\|_{\Phi(\h)}$,
\begin{align}
    \|\theta e^{-t V(b)}\|_{T_{\phi\sqcup\xi}(\h'\sqcup\hat\ell')}
    e^{u\|\phi\|_{\Phi(\h' )}^2 }
    &\le
    \|e^{-t V(b)}\|_{T_{\phi+\xi}(\h'+\hat\ell')}e^{u\|\phi\|_{\Phi(\h' )}^2 }
    \nnb &
    \le
    \|e^{-t V(b)}\|_{T_{\phi+\xi}(\h)}
    e^{8u\|\phi+\xi\|_{\Phi(\h )}^2 }
    e^{8u\|\xi\|_{\Phi(\h )}^2 }
    \nnb &
    \le
    \|e^{-t V(b)}\|_{T_{\phi+\xi}(\h)}
    e^{8u\|\phi+\xi\|_{\Phi(\h )}^2 }
     G(b,\xi)^{1/2},
\label{e:thetaeV}
\end{align}
where we used
$8u\|\xi\|_{\Phi(\h)}^2 \le \frac 12 \|\xi\|_{\Phi(\ell)}^2$ in the last step
(we can take $u \le \frac {1}{16}$).
Next, we apply \refeq{Iupper-a} when $\Gcal =G$,
and \refeq{Iupper-b} with $u=\epVbar$ and $q= 8$ when $\Gcal =\tilde G$ ,
to obtain
\begin{equation}
    \|e^{-t V(b)}\|_{T_{\phi+\xi}(\h)}
    e^{8u\|\phi+\xi\|_{\Phi(\h )}^2 }
    \prec
    \Gcal(b,\phi+\xi),
\end{equation}
and hence
\begin{align}
    \|\theta e^{-t V(b)}\|_{T_{\phi\sqcup\xi}(\h'\sqcup\hat\ell')}
    e^{u\|\phi\|_{\Phi(\h' )}^2 }
    &
    \prec
    \Gcal(b,\phi+\xi)G(b,\xi)^{1/2}
    \nnb &
    \le
    \Gcal(b,\phi)^2 \Gcal(b, \xi)^2 G(b,\xi)^{1/2}.
\label{e:thetaeV-bis}
\end{align}
Since $\Gcal \le G$, we conclude from the above estimates that
\begin{align}
\label{e:supdV}
    &\|\delta V(b)\|_{T_{\phi\sqcup\xi}}
    \sup_{t\in [0,1]} \|e^{-(1-t)\Vpt(b)}\|_{T_{\phi}}
    \|\theta e^{-t V(b)}\|_{T_{\phi\sqcup\xi}}
    \nnb & \quad\quad \prec
    C_{\delta V}\epdV
    \Gcal (b,\phi)^3 G(b,\xi)^{\hat u + 5/2}
    \prec
    C_{\delta V}\epdV
    \Gcal (b,\phi)^3 G(b,\xi)^{3(\ell/\hat\ell)^2}
    ,
\end{align}
using the fact that $u$ is small and that $\hat\ell \le \ell$ by definition.

To complete the estimate on $\delta I(b)$, we now consider the two $W$ terms
in \refeq{delItimes}.
By \cite[Proposition~\ref{norm-prop:derivs-of-tau-bis}]{BS-rg-norm},
\refeq{scale-change}
and the fact that $\h \ge \ell$, \refeq{IF}, and \refeq{Wbomega},
\begin{align}
    \|\theta(e^{-V(b)}W(b)) \|_{T_{\phi\sqcup\xi}(\h'\sqcup\hat\ell')}
    &\le
    \|e^{-V(b)}W(b) \|_{T_{\phi+\xi}(\h'+\hat\ell')}
    \prec
    \|e^{-V(b)}W(b) \|_{T_{\phi+\xi}(\h)}
    \nnb & \prec
    \|W(b)\|_{T_0} \Gcal(b,\phi+\xi)
    \prec_L
    \epdV^2 \Gcal(b,\phi)^2 \Gcal(b,\xi)^2.
\label{e:eVW}
\end{align}
Similarly (recall Remark~\ref{rk:sm}),
\begin{align}
    \|e^{-\Vpt}W_{j+1}(b)\|_{T_{\phi}(\h')}
    & \!\!
    \prec
    \|e^{-\Vpt}W_{j+1}(b)\|_{T_{\phi}(\h)}
    \!\!
    \prec
    \|W_{j+1}(b)\|_{T_{0}(\h)}\Gcal(b,\phi)
    \prec_L
    \epdV^2 \Gcal(b,\phi).
\label{e:eVW+}
\end{align}
We are free to take $\epdV$ small depending on $L$,
so that in the
above two bounds $\prec_L \epdV^2$ can be replaced
by a bound $\prec  \epdV$.

The combination of \refeq{delItimes} with \refeq{supdV}--\refeq{eVW+}
gives
\begin{align}
\label{e:dIass}
    \|\delta I (b) \|_{\Ttimes_{\phi\sqcup\xi} (\h\sqcup\ell)}
    &\prec
    C_{\delta V}\epdV
    \Gcal (b,\phi)^3 G(b,\xi)^{3(\ell/\hat\ell)^2}.
\end{align}
As noted below Definition~\ref{def:Gnorms},
$\Gcal(X)\Gcal(Y)=\Gcal(X\cup Y)$.
Thus there is a constant $c$ (independent of $L$) such that
\begin{equation}
\label{e:ip2}
    \| \delta I^X \|_{\Ttimes_{\phi\sqcup\xi}}
    \leq
    \prod_{b \in \Bcal_j(X)} \| \delta I(b) \|_{\Ttimes_{\phi\sqcup\xi}}
    \leq
    (cC_{\delta V}\epdV)^{|X|_j}
    \Gcal(X,\phi)^3 G(X,\xi)^{3(\ell/\hat\ell)^2}.
\end{equation}
The proof is completed by inserting \refeq{ip1.5} and
\refeq{ip2} into \refeq{ip1}, also noting that
\begin{equation}
    \Ex G(X \cup Y,\xi)^{3(\ell/\hat\ell)^2}
    \le
     2^{|X|_j+|Y|_j}.
\end{equation}
This last inequality is a consequence of
\cite[Proposition~\ref{norm-prop:EG2}]{BS-rg-norm}, whose hypothesis
is supplied by the fact that $3(\ell/\hat\ell)^2
\|C\|_{\Phi^+(\hat\ell)} = 3\|C\|_{\Phi^+(\ell)} \le 3 \ellconst \le
c_G$ by our choice of $\ellconst$.
\end{proof}

\begin{proof}[Proof of Proposition~\ref{prop:ip}.]  We apply
Lemma~\ref{lem:dIip} with scale-$j$ norms and $\h=\h_{j}$.  Since
$\h_{j+1} \le \h' = \h_j/2$, we can apply \refeq{scale-change} to the
left-hand side of \refeq{integration-property-a} to conclude that
\begin{align}
  \label{e:integration-property-pf}
     \|\Ex \delta I^X \theta F(Y) \|_{T_{\phi,j+1}(\h_{j+1})}
     &\leq
     \Econst^{|X|_j+|Y|_j}
     (C_{\delta V} \epdV)^{|X|_j}
     \| F(Y) \|_{\Gcal_j(\h_{j})}
    \Gcal_j(X\cup Y,\phi)^5.
\end{align}
For the norm pair \eqref{e:np1}, it suffices to consider the case
$\phi =0$, for which the regulator on the right-hand side of
\eqref{e:integration-property-pf} reduces to unity and the integration
property \refeq{integration-property} immediately follows in this
case.  For the norm pair \eqref{e:np2}, Lemma~\ref{lem:mart} gives
\begin{equation}
\label{e:Gtilmart}
    \tilde{G}_j(X,\cup Y,\phi)^5
    \le
    \tilde{G}_{j+1}^{\Gtilp}(X\cup Y,\phi),
\end{equation}
and with \eqref{e:integration-property-pf} this gives
\refeq{integration-property} in this case.  This completes the proof.
\end{proof}

\subsection{Proof of Proposition~\ref{prop:cl}}
\label{sec:contraction3-proof}

For convenience, we restate Proposition~\ref{prop:cl} as
Proposition~\ref{prop:cl-bis}.  Its proof uses
Proposition~\ref{prop:1-LTdefXY} in a crucial way.

\begin{prop}
\label{prop:cl-bis} Let $j<N$ and $V\in \bar\DV_j$.  Let $X \in
\Scal_j$ and $U = \overline X$.  Let $F(X) \in \Ncal(X^\Box)$ be such
that $\pi_\alpha F(X) =0$ when $X(\alpha)=\varnothing$.  We assume
that $\pi_{ab}V=\pi_{ab}F(X)=0$ unless $j \ge j_{ab}$ (recall
\eqref{e:Phi-def-jc}).  Then
\begin{align}
    \label{e:contraction3z}
    \|\Ipttil^{U\setminus X} \Ex_{C_{j+1}} \theta F (X) \|_{j+1}
    &
    \prec
    \cgam(X)
    \kappa_F
    + \kappa_{\LT F}
    ,
\end{align}
where
$\kappa_F=\|F (X)\|_{j}$,
$\kappa_{\LT F} =\|\Ipttil^X \LT_X \Ipttil^{-X} F(X) \|_j$, and
where the pair of norms is given by
either of \eqref{e:np1} or \eqref{e:np2}.
\end{prop}

\begin{proof}
We make the decomposition
\begin{equation}
\label{e:KLTdeca}
    F(X) = D(X)+E(X),
\end{equation}
with
\begin{align}
\label{e:EXdef}
    D(X) & = \Ipttil^X \LT_X \Ipttil^{-X} F(X),
    \quad\quad\quad
    E(X)  = \Ipttil^X (1-\LT_X) \Ipttil^{-X} F(X).
\end{align}
By the triangle inequality and the product property,
\begin{equation}
    \|\Ipttil^{U\setminus X} \Ex \theta F (X) \|_{j+1}
    \le
    \|\Ipttil^{U\setminus X} \|_{j+1} \|\Ex \theta D (X) \|_{j+1}
    +
    \|\Ipttil^{U\setminus X} \Ex \theta E (X) \|_{j+1}.
\end{equation}
Since $X\in\Scal_{j}$, its closure $U$ lies in  $\Scal_{j+1}$ and hence
consists of at most $2^d$ blocks.
Therefore, by the product property and \refeq{Iptass},
$\|\Ipttil^{U\setminus X} \|_{j+1}\le 2^{2^d}$.
By the integration property of Proposition~\ref{prop:ip},
\begin{align}
    \|\Ex \theta D (X) \|_{j+1}
    &\prec \|D(X)\|_j
    =
    \kappa_{\LT F}.
\label{e:Lkpfii-1zbis}
\end{align}
Thus the $D$ term in \eqref{e:KLTdeca}
leads to the final term of \eqref{e:contraction3z}.

For the term involving $E$,
we first apply the product property and
\cite[Proposition~\ref{norm-prop:EK}]{BS-rg-norm} (with its assumption given by
$h \ge \ell$ and \refeq{CLbd}) to obtain
\begin{equation}
\lbeq{cl-0}
    \|\Ipttil^{U\setminus X} \Ex \theta E (X) \|_{T_{\phi,j+1}(\h_{j+1})}
    \le
    \|\Ipttil^{U\setminus X} \|_{T_{\phi,j+1}(\h_{j+1})}
    \Ex \| E (X) \|_{T_{\phi+\xi,j+1}(2\h_{j+1})}.
\end{equation}
We recall the inequality
\begin{align}
\label{e:FXbdKz}
    \|F_1(1-\LT_X) F_2\|_{T_{\phi}'}
&\le
    \bar{C}
    \cgam(Y)
    \left(1 + \|\phi\|_{\Phi'}\right)^{A+d+1}
    \sup_{0\le t \le 1}
    \big(
    \|F_{1}F_{2}\|_{T_{t\phi}}
    +
    \|F_{1}\|_{T_{t\phi}}\|F_{2}\|_{T_{0}}\big)
\end{align}
from Proposition~\ref{prop:1-LTdefXY} (where its notation is defined).
To bound the semi-norm of $E(X)$, we apply \refeq{FXbdKz} (writing
$a=A+d+1$ and $\cgam=\cgam(X)$), to obtain
\begin{align}
\lbeq{cl-1}
    &
    \| E(X) \|_{T_{\phi+\xi,j+1}(2\h_{j+1})}
    \prec
    \gamma
    \left(1+\|\phi+\xi\|_{\Phi_{j+1}(X^\Box,2\h_{j+1})} \right)^{a}
    \\
    \nonumber
    & \quad \times
    \sup_{0 \le t\le 1}
    \left(
    \|F(X)\|_{T_{t(\phi+\xi),j}(\h_{j})}
    +
    \|\Ipttil^X\|_{T_{t(\phi+\xi),j}(\h_{j})}
    \|\Ipttil^{-X}\|_{T_{0,j}(\h_{j})}  \|F(X)\|_{T_{0,j}(\h_{j})}
    \right)
    .
\end{align}
Our assumption that $\pi_{ab}V=\pi_{ab}F(X)=0$ unless $j \ge j_{ab}$
provides a corresponding assumption for
Proposition~\ref{prop:1-LTdefXY}.  By the triangle inequality, the
polynomial factor can be bounded as
\begin{align}
    \left(1+\|\phi+ \xi\|_{\Phi_{j+1}(X^\Box)} \right)^{a}
    &\le
    \left(1+\|\phi\|_{\Phi_{j+1}(X^\Box )} \right)^{a}
    \left(1+\|\xi\|_{\Phi_{j+1}(X^\Box    )} \right)^{a}
    \nnb
    & \prec
     \left(1+\|\phi\|_{\Phi_{j+1}(X^\Box     )} \right)^{a}
    G_{j+1}(X,\xi),
\end{align}
where in the last step we used $\h_{j+1}\ge\ell_{j+1}$
to conclude the inequality
$\|\xi\|_{\Phi_{j+1}(2\h_{j+1})}\le \|\xi\|_{\Phi_{j+1}(\ell_{j+1})}$,
together with the fact that the regulator dominates polynomials by \refeq{exp1}.
Next, we apply \refeq{Iass}--\refeq{I-b:5} (the latter in
conjunction with the product property), together with the
definition of $\kappa_F$, to see that the quantity under the supremum
in \refeq{cl-1} is bounded above by a constant multiple of
$\kappa_F \Gcal_j(X,\phi+\xi)$.  Using $\|\phi+\xi\|^2 \le 2\|\phi\|^2 +
2\|\xi\|^2$ to estimate this last regulator, we obtain
\begin{align}
    \| E(X) \|_{T_{\phi+\xi,j+1}(2\h_{j+1})}
    & \prec
    \gamma
    \kappa_F
    \left(1+\|\phi \|_{\Phi_{j+1}(X^\Box,2\h_{j+1})} \right)^{a}
    \nnb & \quad \times
    \Gcal_j(X,\phi)^2
    \Gcal_j(X,\xi)^2
    G_{j+1}(X,\xi)
    .
\lbeq{cl-2}
\end{align}
Since $\Gcal \le G$,
we can then take the expectation using \refeq{EG}
(with Cauchy--Schwarz to separate the regulators at the two different scales), to obtain
\begin{align}
    \Ex \| E(X) \|_{T_{\phi+\xi,j+1}(2\h_{j+1})}
    & \prec
    \gamma
    \kappa_F
    \left(1+\|\phi \|_{\Phi_{j+1}(X^\Box,2\h_{j+1})} \right)^{a}
    \Gcal_j(X,\phi)^2
    .
\lbeq{cl-3}
\end{align}
With \refeq{cl-0}, this gives
\begin{align}
\lbeq{cl-4}
    \|\Ipttil^{U\setminus X} \Ex \theta E (X) \|_{T_{\phi,j+1}(\h_{j+1})}
    & \prec
    \gamma
    \kappa_F
    \|\Ipttil^{U\setminus X} \|_{T_{\phi,j+1}(\h_{j+1})}
    \nnb & \quad \times
    \left(1+\|\phi \|_{\Phi_{j+1}(X^\Box,2\h_{j+1})} \right)^{a}
    \Gcal_j(X,\phi)^2
    .
\end{align}
With an application of Proposition~\ref{prop:Iupper}, this gives
\begin{equation}
\lbeq{cl-5}
    \|\Ipttil^{U\setminus X} \Ex \theta E (X) \|_{T_{\phi,j+1}(\h_{j+1})}
    \prec
    \gamma
    \kappa_F
    \Gcal_{j+1}(U,\phi)^{\Gtilp/2}
    \Gcal_j(X,\phi)^2
    ,
\end{equation}
where the exponent $\Gtilp/2$ on $\Gcal_{j+1}$ is a convenient choice.

For the norm pair \refeq{np1} we set $\phi=0$, the regulators become
equal to $1$, and the desired result follows from \refeq{cl-5}.  For
the norm pair \refeq{np2}, we apply Lemma~\ref{lem:mart} and $X
\subset U$ to obtain
\begin{align}
    \tilde G_{j+1}(U,\phi)^{\Gtilp/2}
    \tilde G_j(X,\phi)^2
    & \prec
    \tilde G_{j+1}(U,\phi)^{\Gtilp/2}
    \tilde{G}_{j+1}(X,\phi)^{\Gtilp/2} \le \tilde G_{j+1}^{\Gtilp}(U,\phi)
    ,
\label{e:FXbdKzzz}
\end{align}
and the desired result
follows from \refeq{cl-5}.
This completes the proof.
\end{proof}

\setcounter{section}{0}
\renewcommand{\thesection}{\Alph{section}}

\section{\texorpdfstring{$L^p$}{Lp} norm estimates}
\label{sec:Lp}

Let $\phi : \Lambda \to \C$, and let $X \subset \Lambda$ be a subset
of cardinality $|X|$.  For $p\in [1,\infty )$, we define the $L^{p}$
norm
\begin{equation}\label{e:Lp-def}
    \|\phi\|_{L^{p} (X)}
=
     \frac{1}{\h}\left(\frac{1}{|X|}\sum_{x\in X} |\phi (x)|^{p}\right)^{1/p}.
\end{equation}
The weight $\h$ is included in the norm so that, according to
\eqref{e:phinorm} and \eqref{e:PhiXdef},
\begin{equation}\label{e:equivalent-norms4}
    \|\phi\|_{L^{p}(X)}^{p}
\le
    \| \phi\|_{\Phi (X)}^{p}.
\end{equation}
Proposition~\ref{prop:equivalent-norms} below provides a lattice Sobolev
inequality which shows that
\eqref{e:equivalent-norms4} can be reversed at the cost of an
additional term.
Our application of Proposition~\ref{prop:equivalent-norms} occurs in
\eqref{e:Sob2}, with $p=2$.

To prepare for the proposition, we first prove
a lemma which shows that the reversal is possible for
polynomials, even with an increase in the size of the domain of the
$\Phi$ norm (recall that the small set neighbourhood $X^\Box$ of $X$
was defined in \refeq{ssn}).
Throughout this appendix, we write $R=L^j$.
The hypothesis below, that $R \ge R_0$, can then be achieved uniformly in $j$
by taking $L$ sufficiently large.
Outside this appendix, we take the parameter $p_\Phi$
in the definition of the $\Phi$ norm to obey $p_\Phi \ge \frac{d+4}{2}$
(as mentioned in Section~\ref{sec:reg}),
but this restriction is unnecessary in the following lemma.

\begin{lemma}
\label{lem:PhiLp}
Let $p_\Phi, q\ge 0$ be integers.
Let $Q$ denote the vector space of complex-valued
polynomials defined on $\Rd$
and of degree at most $q$. Let $f$ be the
restriction of any polynomial in $Q$ to $\Zd$.
Let $B$ be a block of side $R$ in $\Zd$.
There exists $c_0 =
c_0(q,p)>0$ such that for $R\geq R_0(q,p)$
sufficiently large,
\begin{equation}
\label{e:fequiv}
    \| f\|_{\Phi (B^\Box)}
\le
    c_0 \|f\|_{L^{p} (B)}
.
\end{equation}
\end{lemma}

\begin{proof}
The inequality is homogeneous in $\h$ so without loss of generality
we take $\h =1$.
It suffices to consider the case where $p_\Phi=q$. In fact,
derivatives of $f \in Q$ having order higher than $q$ vanish
so the left-hand side of \refeq{fequiv} is constant in $q \ge p_\Phi$,
and the left-hand side is an increasing function of $p_\Phi$ so the
statement is strongest when $p_\Phi=q$.
Thus we take $p_\Phi=q$ throughout the proof.
Moreover, \refeq{fequiv} is trivial if $f$ is a constant, so we
consider the case $q \ge 1$.

Let $\Ccal^{q}$ denote
the space of $q$-times differentiable functions on $\R^{d}$ with norm
given by
\begin{equation}
    \|G\|_{\Ccal^{q}} = \sup_{x\in\Rd}\sup_{|\alpha| \leq q} |D^\alpha G (x)|,
\end{equation}
where $\alpha$ is a multi-index
and $D^\alpha$ is the derivative on
$\R^d$.
Without loss of generality, we assume that $B$ is centred at the origin
of $\Zd$.  We obtain a continuum version $\hat B^\Box\subset \Rd$
of $B^\Box$ by placing a
unit $\Rd$-cube centred at each point in $B^\Box$.  Let
$I^\Box = R^{-1}\hat B^\Box \subset \Rd$ be its rescaled version.
For $P \in Q$, let
\begin{equation}
    \|P\|_{\Ccal^{q}(I^\Box)} = \inf \{ \|P-G\|_{\Ccal^{q}} :
    G \in \Ccal^{q}, G|_{I^\Box}=0\}
.
\end{equation}
This defines a norm on $Q$.

Given $F \in Q$, let $f$ be the
restriction of $F$ to $\Zd$, and let $\hat F \in Q$
be defined by $\hat
F(x)=F(Rx)$ for $x \in \R^d$.  We prove that
\begin{equation}
\label{e:sob1}
    \|f\|_{\Phi (B^\Box)} \le \|\hat F\|_{\Ccal^{q}(I^\Box)},
\end{equation}
and that there is a $c_0(q,p)>0$ and an $R_0(q,p)$ such that for
$R \geq R_0$,
\begin{equation}
\label{e:sob2}
    \|\hat F\|_{\Ccal^{q}(I^\Box)} \le c_0 \|f\|_{L^{p} (B)}.
\end{equation}
Together, these two inequalities give \eqref{e:fequiv}.

We first prove \eqref{e:sob1}.
By Taylor's theorem, $R |\nabla^{e} f (x) |
\le \|D^{e} \hat F \|_{\Ccal^{0}}$.
By induction on $|\alpha |$, this gives
\begin{equation}
    \label{e:finite-diff}
    \sup_{x \in \R^d}  |\nabla_R^\alpha f(x)|
\le
    \|\hat F\|_{\Ccal^{q}},
\quad\quad
    |\alpha| \le q
,
\end{equation}
where $\nabla_R^\alpha = R^{|\alpha|}\nabla^\alpha$.
Given $\hat G \in \Ccal^{q}$, let $g(x)=
\hat G(R^{-1}x)$.
By definition, $f(x)-g(x) = \hat{F}(R^{-1}x)-
\hat{G}(R^{-1}x)$, so by \eqref{e:finite-diff} with $\hat F$ replaced
by $\hat{F}- \hat{G}$,
\begin{equation}
\lbeq{fghats}
    \sup_{x \in \Z^d} |\nabla_R^\alpha[f(x)-g(x)]| \le
    \|\hat F - \hat G\|_{\Ccal^{q}},
    \quad\quad
    |\alpha| \le q.
\end{equation}
Therefore,
\begin{equation}
\lbeq{fginf}
    \inf \left\{
    \|f -g \|_\Phi  :
    \hat G \in \Ccal^{q}, \hat G|_{I^\Box}=0 \right\}
    \le
    \|\hat F  \|_{\Ccal^{q}(I^\Box)}.
\end{equation}
The set of all lattice functions $g$ with $g|_{I^\Box}=0$ includes
all functions $g$ arising on the left-hand side, and the infimum over
this larger class is smaller that the infimum in \refeq{fginf}.
Thus the left-hand side of \refeq{fginf} is greater
than or equal to $\|f\|_{\Phi (B^\Box)}$.  This proves \eqref{e:sob1}.

To prove \eqref{e:sob2}, we define a second norm on $Q$, as follows.
For $P \in Q$, let
\begin{equation}
    \|P\|_{L^p(I)} = \left( \int_I |P(x)|^p dx \right)^{1/p}.
\end{equation}
Since all norms on the finite-dimensional vector space $Q$ are
equivalent, there exists a constant $c_1 = c_1(q,p)$ such
that, for all $P \in Q$,
\begin{equation}
\label{e:equivalent-norm1}
    \|P\|_{\Ccal^{q}(I^\Box)}^p  \leq c_1 \|P\|_{L^p(I)}^p.
\end{equation}
The difference
\begin{align}
    \|P\|_{L^p(I)}^p
    - \frac{1}{|B|} \sum_{x\in B} |P(R^{-1}x)|^p
    &=
    \int_I|P(x)|^p dx - \frac{1}{|B|} \sum_{x\in B} |P(R^{-1}x)|^p
\end{align}
is a Riemann sum approximation error.
It is therefore bounded in absolute value by
$R^{-1}$ times the maximum over $I$ of $|DP^{p}|$, which is
less than $pR^{-1} \|P\|_{\Ccal^{q}(I^\Box)}^p$
(here we use $q \ge 1$). Therefore,
\begin{equation}
\label{e:equivalent-norm2}
    \left(1 - \frac{p}{R}c_{1} \right)
    \|P\|_{\Ccal^{q}(I^\Box)}^{p}
    \leq
    c_1\frac{1}{|B|}
    \sum_{x\in B} |P(R^{-1}x)|^p
.
\end{equation}
We take $R$ large enough that $1 - \frac{p}{R}c_{1} \ge 1/2$,
and set $P= \hat F$ in \eqref{e:equivalent-norm2}, to conclude
\eqref{e:sob2} with $c_0 = (2c_1)^{1/p}$.  This completes the proof of
\eqref{e:sob2}, and hence of \eqref{e:fequiv}.
\end{proof}

\begin{prop}
\label{prop:equivalent-norms} Let $B$ be a block of side $R=L^j$ in the torus
$\Lambda$ of side length $L^N$, with $j \le N-1$. There are constants $c_{1}$, $c_2$ and
$R_{0}$ (depending on $p_{\Phi},p$) such that for $X \subset B$ with
$|X|\le c_{1} |B|$, and for $R \ge R_0$,
\begin{equation}
\label{e:fequiv0}
    \|\phi\|_{\Phi (B^\Box)}
\le
    c_2 \left( \|\phi\|_{L^{p} (B\setminus X)} +
    \|\phi\|_{\tilde{\Phi} (B^\Box)} \right).
\end{equation}
\end{prop}

\proof
The inequality
\eqref{e:fequiv0}
is homogeneous in $\h$ so we may assume that $\h=1$.
For any $f \in \C^\Lambda$,
\begin{align}
    \|f\|_{L^p(B\setminus X)}^p
    & \geq
    \frac{|B\setminus X|}{|B|}\|f\|_{L^{p} (B\setminus X)}^{p}
    =
    \|f \|_{L^{p} (B)}^{p}
    - \frac{|X|}{|B|}\, \|f\|_{L^p(X)}^{p}
    .
\end{align}
The restriction $j \le N-1$ is imposed
to ensure that the periodicity of $\Lambda$ plays no role, and
we may assume that we are working on $\Zd$ rather than on $\Lambda$.
We apply Lemma~\ref{lem:PhiLp} with
$q=1$.
With $c_{0}$ the constant of Lemma~\ref{lem:PhiLp},
let $c_{1}=(2c_0^p)^{-1}$.  By hypothesis,
$| X| \le (2c_0^p)^{-1}|B|$.
Let $f \in Q$, with $Q$ as in Lemma~\ref{lem:PhiLp}.  By
\eqref{e:equivalent-norms4} and the fact that $X\subset B^\Box$,
$\|f\|_{L^p(X)} \le \|f\|_{\Phi (X)} \le \|f\|_{\Phi (B^\Box)}$.
With Lemma~\ref{lem:PhiLp}, this gives
\begin{align}
    \|f\|_{L^p(B\setminus X)}^p
    &
    \ge
    \|f \|_{L^{p} (B)}^{p}
    - \frac{|X|}{|B|}\, \|f\|_{\Phi (B^\Box)}^{p}
    \ge
    \left[\frac {1}{c_0^p} - \frac{1}{2 c_0^p}\right]\|f  \|_{\Phi (B^\Box)}^{p}.
\end{align}
Therefore,
\begin{equation}
\label{e:fequiv2}
    \|f\|_{\Phi (B^\Box)}
\le
    2^{1/p}c_0\|f\|_{L^{p} (B\setminus X)}.
\end{equation}
Given $\phi :\Zd \to \C$ and $f \in Q$, we apply the triangle
inequality (twice), \eqref{e:fequiv2} and \eqref{e:equivalent-norms4} to see
that
\begin{align}
    \|\phi\|_{\Phi (B^\Box)}
&\le
    \|f\|_{\Phi (B^\Box)} + \|\phi -f\|_{\Phi (B^\Box)}\nnb
&\le
    2^{1/p}c_0\|f\|_{L^{p} (B\setminus X)} + \|\phi -f\|_{\Phi (B^\Box)}\nnb
&\le
    2^{1/p}c_0\|\phi\|_{L^{p} (B\setminus X)} +
    2^{1/p}c_0\|\phi-f\|_{L^{p} (B\setminus X)} +
    \|\phi -f\|_{\Phi (B^\Box)}\nnb
&\le
    2^{1/p}c_0\|\phi\|_{L^{p} (B\setminus X)} +
    \big(2^{1/p}c_0+1\big)\|\phi -f\|_{\Phi (B^\Box)}.
\end{align}
The desired inequality \eqref{e:fequiv0},
with $c_2=2^{1/p}c_0+1$, then follows by minimising
over $f \in Q$ once
we note that
\eq
    \inf\{\|\phi -f\|_{\Phi (B^\Box)}: f \in V \}
    =
    \|\phi\|_{\tilde\Phi(B^\Box)}
\en
by definition of the norms in \refeq{PhiXdef} and \refeq{Phipoltildef}.
\qed

\section{Further interaction estimates}
\label{sec:further-ie}

This section comprises estimates of a more specialised
nature, which are required in \cite{BS-rg-step}.  The estimates are
stated as three lemmas.
For the first lemma, for $B \in \Bcal_{j}$ we define
\begin{align}
\label{e:DeltaIdef}
    \Delta I(B)
    & =
    \tilde{I}(V,B)
    -
    I_{}(V,B)
    =
    e^{-V(B)}
    \left[
    \prod_{b \in \Bcal_{j-1}(B)}
    (1+W_{j}(V,b))
    -
    (1+W_{j}(V,B))
    \right]
    .
\end{align}

\begin{lemma}
\label{lem:DelIbd}
For $j \le N$,
for both choices
of $\|\cdot \|_j$ in \refeq{np1}--\refeq{np2},
for $B \in \Bcal_{j}$ and $V \in \bar\DV_j$,
\begin{align}
\label{e:DelIbd}
    \|\Delta I(B)\|_{j}
&\prec_{L}
    \epdV^{4}
    .
\end{align}
\end{lemma}

\begin{proof}
By \eqref{e:DeltaIdef}, together with the fact that
$W_j(V,B)=\sum_{b \in \Bcal_{j-1}(B)}  W_{j}(V,b)$ by
\eqref{e:WLTF},
\begin{align}
    \Delta I(B)
    &
    =
    e^{-V(B)}
    \!\!\!\!\!\!\!\!
    \sumtwo{X \in \Pcal_{j-1}(B) :}{|X|_{j-1} \ge 2}
    \prod_{b \in \Bcal_{j-1}(X)}
    W_{j}(V_{j},b)
    .
\end{align}
Then \refeq{Wbomega} gives a bound of order $(\epdV^2)^2$ for the $T_{0}$
semi-norm of the above sum, and the desired estimate follows from this together
with \refeq{IF}.
\end{proof}

\begin{lemma}
\label{lem:JCK-app-2}
For $V \in \bar\DV$,  $X \in \Scal$ and $F \in\Ncal(X^\Box)$,
\begin{equation}
\label{e:JCK3-app}
    \left\|
    \LT_X \left( (I^{-X}- \Itilde_{\pt}^{-X} )F\right)
    \right\|_{T_{0} }
    \prec \;
    C_{\delta V}\epdV
    \|F\|_{T_0 }
    .
\end{equation}
All quantities and norms are at scale $j<N$, and norms are computed with either $\h=\ell$ or
$\h=h$.
\end{lemma}

\begin{proof}
It follows from
\cite[Proposition~\ref{loc-prop:opLTdefXY}]{BS-rg-loc} that
\begin{equation}
    \left\|
    \LT_X \left( (I^{-X}- \Itilde_{\pt}^{-X} )F\right)
    \right\|_{T_{0} }
    \prec
    \left\|
    \left( (I^{-X}- \Itilde_{\pt}^{-X} )F\right)
    \right\|_{T_{0} }
\end{equation}
To estimate the right-hand side, we use the identity
\begin{equation}
    \prod_{i}a_{i}^{-1} - \prod_{i}b_{i}^{-1}
=
     \sum_{k}\Big(\prod_{i\le k}a_{i}^{-1}\Big)
     (a_{k}-b_{k})
     \Big(\prod_{i\ge k}b_{i}^{-1}\Big)
     ,
\end{equation}
the triangle inequality, the product property of the norm, and
\eqref{e:I-b:5}, to obtain
\begin{equation}
    \label{e:JCK3-1}
    \left\|
    \LT_X \left( (I^{-X}- \Itilde_{\pt}^{-X} )F\right)
    \right\|_{T_{0}}
\prec
    \sup_{B \in \Bcal (X)}
    \|
    I (B) - \Itilde_{\pt} (B)
    \|_{T_{0}}
    \|F\|_{T_{0}}
.
\end{equation}
We are thus reduced to estimates on a single block, and we henceforth omit $B$ arguments.

To account for the fact
that $I$ involves $W_j$ whereas $\Itilde_{\pt}$ involves $W_{j+1}$, we
define $\Ipt = I(\Vpt) = I_j(\Vpt)$.
Then
\eq
\lbeq{I-Itil}
    \|
    I   - \Itilde_{\pt}
    \|_{T_{0}}
    \le
    \|
    I   - \Ipt
    \|_{T_{0}}
    +
        \|
    \Ipt   - \Itilde_{\pt}
    \|_{T_{0}}.
\en
By \refeq{IF} and \eqref{e:Wbomega}, the second term on the right-hand side obeys
\begin{align}
\label{e:JCK3-2}
    \|\Ipt -  \Itilde_{\pt}\|_{T_{0}}
&=
    \|e^{-\Vpt} (W_{j}-W_{j+1})\|_{T_{0}}
\prec_{L}
    \epdV^2
    .
\end{align}

To estimate the first term on the right-hand side of \refeq{I-Itil},
we proceed as in the proof of \refeq{JCK1-app} and now define
$V_s= V + s(\Vpt -V)$, $I_s=I(V_s)$,
$\Ical_s = e^{-V_{s}}$, and $W_s=W(V_{s})$.  The steps leading to \refeq{IIhatdif}
give
\begin{equation}
    \|I - \Ipttil\|_{T_0}
    \le
    \sup_{s\in [0,1]}
    \left( \|I_s\|_{T_0} \| (\Vpt -V)\|_{T_0} + \|\Ical_s\|_{T_0} \| W_s'\|_{T_0} \right).
\end{equation}
The norms of $I_s$ and $\Ical_s$ are bounded by $2$, by \refeq{Iupper-a}.
 Also, $ \| \Vpt -V\|_{T_0}$ was encountered in
\refeq{dVbd1} and proved to be at most $C_{\delta V}\epdV$.
With \refeq{Wprimebd-app}, we then obtain $\| W_s'\|_{T_0} \prec_l \epdV^2$.
This completes the proof.
\end{proof}

The next lemma is applied in
\cite[Lemmas~\ref{step-lem:K4}--\ref{step-lem:K7a}]{BS-rg-step}.  To
prepare for its statement, given $V' \in \Qcal$ we define a new
element $V'' \in \Qcal$ by
\begin{equation}
\label{e:VptVplus}
    V''
    =
    V'
    + y    (\tau_{\Delta} -  \tau_{\nabla\nabla}),
\end{equation}
where $y$ is the coefficient of $\tau_{\nabla\nabla}$ in $V'$.  Thus
the term $y\tau_{\nabla\nabla}$ in $V'$ is replaced by $y\tau_\Delta$
to produce $V''$.  We also define
\begin{align}
    \delta I^{+}(B)
    &=
    e^{-V'(B)}\big(W_{j+1}(V',B) -  W_{j+1}(V'',B)\big)
    .
\end{align}
The definition of $V''$ is motivated by the fact that, for a polymer
$X$, $V''(X)$ and $V' (X)$ are equal up to a polynomial in the fields
that is supported on the boundary of $X$. To see this let $\chi$, $f$
and $g$ be functions on $\Lambda$. Then
\begin{align}
    -
    \sum_{x \in \Lambda,e \in \Ucal}
    (\nabla^{e} \chi)_{x} \big(\nabla^{e} (fg)\big)_{x}
    =
    \sum_{x \in \Lambda}\chi_{x}
    \Big((\Delta f)_{x}g_{x} + f_{x} (\Delta g)_{x}\Big)
    \nonumber
    \\+
    \sum_{x \in \Lambda,e \in \Ucal} \chi_{x}
    (\nabla^{e}f)_{x}(\nabla^{e}g)_{x}
.
\end{align}
This is proved by using summation by parts ($\nabla^{e}$ and
$\nabla^{-e}$ are adjoints) to rewrite the summand on the left as
$\chi_{x} (\Delta fg)_{x}$, followed by writing $\Delta (fg)_{x}$ as
the sum over $e\in\Ucal$ of $f_{x+e} g_{x+e} - f_{x}g_{x}$ and
using simple algebra.  Choosing $f = \phi,\psi$ and $g=
\bar{\phi},\bar{\psi}$ and referring to \eqref{e:addDelta} we obtain
\begin{equation}
    -
    \sum_{x \in \Lambda,e \in \Ucal}
    (\nabla^{e} \chi)_{x} (\nabla^{e}\tau)_{x}
    =
    2\sum_{x \in \Lambda}\chi_{x}
    \Big(- \tau_{\Delta ,x} + \tau_{\nabla\nabla,x}\Big)
.
\end{equation}
For a polymer $X$ in $\Pcal_{j+1}$ let $\chi$ be the indicator
function of $X$. Then from \eqref{e:VptVplus},
\begin{equation}
    V''(X) - V'(X)
    =
    \frac{1}{2}y
    \sum_{x \in \Lambda,e \in \Ucal}
    (\nabla^{e} \chi)_{x} (\nabla^{e}\tau)_{x}
.
\end{equation}
Let $\partial X$ denote the points in $X$ with a neighbour in $\Lambda
\setminus X$. The right hand side is a sum of $\tau_{z'}-\tau_{z}$
over nearest neighbours $z',z$ where $z$ is in $X$ and $z'$ is not in
$X$. By rewriting the fields in $\tau_{z'}$ using $f_{z'} = f_{z}+
(\nabla^{e}f)_{z}$ we find that there exists a polynomial $V_\partial$
which is quadratic in the fields and their derivatives such that
\begin{equation}
     V''(X) - V'(X) = \sum_{z \in \partial X} V_{\partial,z}
\end{equation}
and every term in $V_{\partial,z}$ has at least one derivative.

For $X \in \Pcal_{j+1}$ and $B \in \Bcal_{j+1} (\Lambda \setminus X)$,
we set $R_{X} (B)=\delta I_{X}^{(6)} (B)=0$ if $B$ does not have a
neighbour in $\partial X$, and otherwise define
\begin{align}
    R_{X} (B)
    & =
    e^{- V_{\partial} (\partial X \cap B^1)}-1
    ,
    \quad\quad
    \delta I_{X} (B)
    =
    R_{X} (B) I(V'',B)
,
\end{align}
where $B^1 = B \cup \partial(\Lambda \setminus B)$.

\begin{lemma}
\label{lem:sbp-bds}
Let $j<N$, and $B \in \Bcal_{j+1}$.  Suppose that
$V' \in \bar\DV_{j+1}$ has $y \tau_{\nabla\nabla}$ term which obeys
$\|y \tau_{\nabla\nabla}(b)\|_{T_{0,j}} \prec \epdV$ when $b \in \Bcal_j$.
Let  $X \in \Pcal_{j+1}$.  Then
for both choices of $\|\cdot \|_{j+1}$ in \refeq{np1}--\refeq{np2},
\begin{align}
\label{e:map6-bd0}
    \|\delta I^+(B)\|_{j+1}
&
\prec_{L}
    \epdV^2
, \\
\lbeq{delI6bd}
    \|\delta I_X (B)\|_{j+1}
&
    \prec  \;
    \epdV
   .
\end{align}
\end{lemma}

\begin{proof}
By direct calculation, $\|\tau_{\nabla\nabla}(b)\|_{T_{0,j}} \asymp L^{(d-2)(j)}\h_{j}^2$,
and the right-hand side is $\ell_0^2$ for $\h=\ell$ and $k_0^2 \ggen_{j}^{-1/2}$
for $\h=h$.  Therefore, by hypothesis and by definition of $\epdV$, we have
\begin{equation}
    |y| \prec
    \begin{cases}
    \ell_0^{-2} \epdV & \h=\ell
    \\
    k_0^{-2}  \ggen_j^{1/2} \epdV & \h=h.
    \end{cases}
\end{equation}
Since $V_{\partial}$ is given by a
sum over $O(L^{(d-1)(j+1)})$ boundary
points of terms containing at least one gradient and two fields, this gives
\begin{align}
\lbeq{Vpartialbd}
    \| V_\partial (\partial X \cap B)\|_{T_0}
    &\prec
    \begin{cases}
    \ell_0^{-2}\epdV
    L^{(d-1)(j+1)}
    L^{- (j+1)} \ell^{2}_{j+1} & \h=\ell
    \\
    k_0^{-2}\ggen^{1/2} \epdV
    L^{(d-1)(j+1)}
    L^{- (j+1)} h^{2}_{j+1}
    & \h=h
    \end{cases}
    \nnb &
    =
    \epdV.
\end{align}

To prove \refeq{map6-bd0}, we apply \refeq{IF} to obtain
\begin{equation}
    \|\delta I^+(B)\|_{j+1}
    \prec
     \|W_{j+1}(V',B) -  W_{j+1}(V'',B)\|_{T_0 (\h)}
    ,
\end{equation}
and then use \refeq{Wprimebd-app}--\refeq{Wprimebd-app2} to see that the
right-hand side is $\prec_l \, \epdV^2$ as required.
(In fact we use a small variation of \refeq{Wprimebd-app}--\refeq{Wprimebd-app2}
in which we regard $V'-V''$ as supported on $\partial X \cap B$, with
\refeq{Vpartialbd}.)

To prove \refeq{delI6bd},
we set $I_\partial(t) = I(V'',B) e^{- tV_{\partial}}$,
with $V_{\partial}=V_{\partial}(\partial X \cap B)$.
By the Fundamental Theorem of Calculus,
\begin{equation}
    \delta I_{X} (B)
    =
    V_{\partial}
    \int_0^1
    I_\partial(t)
     dt
    ,
\end{equation}
and hence
\begin{equation}
    \| \delta I_{X} (B) \|_{j+1}
    \le \sup_{t \in [0,1]}
    \|I_\partial(t)
    V_{\partial} \|_{j+1}
    .
\end{equation}
The polynomial $V''$
obeys our stability estimates since compared to $V'$ its $z\tau_\Delta$ term is
modified by $z \mapsto z+y$ and this change is such that $\epsilon_{V''} \le \epsilon_{V'}$,
and hence
$V''\in\bar\DV$.
By
\cite[\eqref{norm-e:apos-e}]{BS-rg-norm} and
\cite[Proposition~\ref{norm-prop:T0K}]{BS-rg-norm},
$\|e^{-tV_\partial}\|_{T_\phi} \le e^{\|V_\partial\|_{T_\phi}}
\le e^{\|V_\partial\|_{T_0}(1+\|\phi\|_{\Phi}^2)}$.
The bound on
$\|e^{-tV_\partial}\|_{T_\phi}$ is no larger than the effect of $Q$ handled in
\refeq{apos-e}, and thus
$ e^{- tV_{\partial}}$ is a negligible perturbation of $I(V'',B)$,
and $I_\partial(t)$ also obeys the stability bounds.
Thus we obtain from \refeq{IF} and \refeq{Vpartialbd} that
\begin{equation}
    \| \delta I_{X} (B) \|_{j+1}
    \prec
    \| V_{\partial}  \|_{T_0}
    \prec \epdV
    ,
\end{equation}
and the proof is complete.
\end{proof}

%%%%%%%%%%%%%%%%%%%%%%%%%%%%%%%%%%%%%%%%%%%%%%%%%%%%%%%%%%%%%%%%%%%%%%
%%%%%%%%%%%%%%%%%%%%%%%%%%%%%%%%%%%%%%%%%%%%%%%%%%%%%%%%%%%%%%%%%%%%%%
\section*{Acknowledgements}

The work of both authors was supported in part by NSERC of Canada.
DB gratefully acknowledges the support and hospitality of
the Institute for Advanced Study at Princeton and of Eurandom during part
of this work.
GS gratefully acknowledges the support and hospitality of
the Institut Henri Poincar\'e, and of the Mathematical Institute of Leiden
University, where part of this work was done.
We thank Roland Bauerschmidt for numerous helpful discussions.

\bibliography{../../bibdef/bib}
\bibliographystyle{plain}

\end{document}